\documentclass[longauth]{aaEC}

\usepackage{graphicx}
\usepackage{natbib}
\usepackage{scalerel}
\usepackage{lastpage}
\usepackage{mathtools}
\usepackage{multirow}
\usepackage{sidecap}
\usepackage[table]{xcolor}

\DeclareMathOperator{\asinh}{asinh}

\usepackage{txfonts}
\usepackage{hyperref}
\hypersetup{
   colorlinks=true,
   linkcolor=blue,
   filecolor=magenta,    
   urlcolor=blue,
   citecolor=blue
}

\makeatletter
\renewcommand*\aa@pageof{, page \thepage{} of \pageref*{LastPage}}
\makeatother

\usepackage[utf8]{inputenc}

\usepackage[switch, modulo]{lineno}

\usepackage{euclid}

\usepackage[nameinlink,capitalise]{cleveref}

\crefname{section}{Sect.}{Sects.}
\Crefname{section}{Section}{Sections}
\crefname{figure}{Fig.}{Figs.}
\Crefname{figure}{Figure}{Figures}
\crefname{equation}{Eq.}{Eqs.}
\Crefname{equation}{Equation}{Equations}
\crefname{table}{Table}{Tables}
\crefname{appendix}{Appendix}{Appendices}

\usepackage{orcidlink}
\newcommand{\orcid}[1]{\orcidlink{#1}}

\defcitealias{Q1-SP027}{MZ25}
\defcitealias{Q1-SP015}{MB25}
\defcitealias{Q1-TP004}{R25}

\begin{document} 

\title{Euclid Quick Data Release (Q1)}
\subtitle{Active galactic nuclei identification using diffusion-based inpainting of \Euclid VIS images}  
		   
\author{Euclid Collaboration: G.~Stevens\orcid{0000-0002-8885-4443}\thanks{\email{grant.stevens@bristol.ac.uk}}\inst{\ref{aff1}}
\and S.~Fotopoulou\orcid{0000-0002-9686-254X}\inst{\ref{aff1}}
\and M.~N.~Bremer\inst{\ref{aff1}}
\and T.~Matamoro~Zatarain\orcid{0009-0007-2976-293X}\inst{\ref{aff1}}
\and K.~Jahnke\orcid{0000-0003-3804-2137}\inst{\ref{aff2}}
\and B.~Margalef-Bentabol\orcid{0000-0001-8702-7019}\inst{\ref{aff3}}
\and M.~Huertas-Company\orcid{0000-0002-1416-8483}\inst{\ref{aff4},\ref{aff5},\ref{aff6},\ref{aff7}}
\and M.~J.~Smith\orcid{0000-0003-0220-5125}\inst{\ref{aff8},\ref{aff9}}
\and M.~Walmsley\orcid{0000-0002-6408-4181}\inst{\ref{aff10},\ref{aff11}}
\and M.~Salvato\orcid{0000-0001-7116-9303}\inst{\ref{aff12}}
\and M.~Mezcua\orcid{0000-0003-4440-259X}\inst{\ref{aff13},\ref{aff14}}
\and A.~Paulino-Afonso\orcid{0000-0002-0943-0694}\inst{\ref{aff15},\ref{aff16}}
\and M.~Siudek\orcid{0000-0002-2949-2155}\inst{\ref{aff5},\ref{aff13}}
\and M.~Talia\orcid{0000-0003-4352-2063}\inst{\ref{aff17},\ref{aff18}}
\and F.~Ricci\orcid{0000-0001-5742-5980}\inst{\ref{aff19},\ref{aff20}}
\and W.~Roster\orcid{0000-0002-9149-6528}\inst{\ref{aff12}}
\and N.~Aghanim\orcid{0000-0002-6688-8992}\inst{\ref{aff21}}
\and B.~Altieri\orcid{0000-0003-3936-0284}\inst{\ref{aff22}}
\and S.~Andreon\orcid{0000-0002-2041-8784}\inst{\ref{aff23}}
\and H.~Aussel\orcid{0000-0002-1371-5705}\inst{\ref{aff24}}
\and C.~Baccigalupi\orcid{0000-0002-8211-1630}\inst{\ref{aff25},\ref{aff26},\ref{aff27},\ref{aff28}}
\and M.~Baldi\orcid{0000-0003-4145-1943}\inst{\ref{aff29},\ref{aff18},\ref{aff30}}
\and S.~Bardelli\orcid{0000-0002-8900-0298}\inst{\ref{aff18}}
\and P.~Battaglia\orcid{0000-0002-7337-5909}\inst{\ref{aff18}}
\and A.~Biviano\orcid{0000-0002-0857-0732}\inst{\ref{aff26},\ref{aff25}}
\and A.~Bonchi\orcid{0000-0002-2667-5482}\inst{\ref{aff31}}
\and E.~Branchini\orcid{0000-0002-0808-6908}\inst{\ref{aff32},\ref{aff33},\ref{aff23}}
\and M.~Brescia\orcid{0000-0001-9506-5680}\inst{\ref{aff34},\ref{aff35}}
\and J.~Brinchmann\orcid{0000-0003-4359-8797}\inst{\ref{aff16},\ref{aff36}}
\and S.~Camera\orcid{0000-0003-3399-3574}\inst{\ref{aff37},\ref{aff38},\ref{aff39}}
\and G.~Ca\~nas-Herrera\orcid{0000-0003-2796-2149}\inst{\ref{aff40},\ref{aff41},\ref{aff42}}
\and V.~Capobianco\orcid{0000-0002-3309-7692}\inst{\ref{aff39}}
\and C.~Carbone\orcid{0000-0003-0125-3563}\inst{\ref{aff43}}
\and J.~Carretero\orcid{0000-0002-3130-0204}\inst{\ref{aff44},\ref{aff45}}
\and M.~Castellano\orcid{0000-0001-9875-8263}\inst{\ref{aff20}}
\and G.~Castignani\orcid{0000-0001-6831-0687}\inst{\ref{aff18}}
\and S.~Cavuoti\orcid{0000-0002-3787-4196}\inst{\ref{aff35},\ref{aff46}}
\and K.~C.~Chambers\orcid{0000-0001-6965-7789}\inst{\ref{aff47}}
\and A.~Cimatti\inst{\ref{aff48}}
\and C.~Colodro-Conde\inst{\ref{aff4}}
\and G.~Congedo\orcid{0000-0003-2508-0046}\inst{\ref{aff49}}
\and C.~J.~Conselice\orcid{0000-0003-1949-7638}\inst{\ref{aff11}}
\and L.~Conversi\orcid{0000-0002-6710-8476}\inst{\ref{aff50},\ref{aff22}}
\and Y.~Copin\orcid{0000-0002-5317-7518}\inst{\ref{aff51}}
\and A.~Costille\inst{\ref{aff52}}
\and F.~Courbin\orcid{0000-0003-0758-6510}\inst{\ref{aff53},\ref{aff54}}
\and H.~M.~Courtois\orcid{0000-0003-0509-1776}\inst{\ref{aff55}}
\and M.~Cropper\orcid{0000-0003-4571-9468}\inst{\ref{aff56}}
\and A.~Da~Silva\orcid{0000-0002-6385-1609}\inst{\ref{aff57},\ref{aff58}}
\and H.~Degaudenzi\orcid{0000-0002-5887-6799}\inst{\ref{aff59}}
\and G.~De~Lucia\orcid{0000-0002-6220-9104}\inst{\ref{aff26}}
\and C.~Dolding\orcid{0009-0003-7199-6108}\inst{\ref{aff56}}
\and H.~Dole\orcid{0000-0002-9767-3839}\inst{\ref{aff21}}
\and M.~Douspis\orcid{0000-0003-4203-3954}\inst{\ref{aff21}}
\and F.~Dubath\orcid{0000-0002-6533-2810}\inst{\ref{aff59}}
\and X.~Dupac\inst{\ref{aff22}}
\and S.~Dusini\orcid{0000-0002-1128-0664}\inst{\ref{aff60}}
\and S.~Escoffier\orcid{0000-0002-2847-7498}\inst{\ref{aff61}}
\and M.~Farina\orcid{0000-0002-3089-7846}\inst{\ref{aff62}}
\and S.~Ferriol\inst{\ref{aff51}}
\and K.~George\orcid{0000-0002-1734-8455}\inst{\ref{aff63}}
\and C.~Giocoli\orcid{0000-0002-9590-7961}\inst{\ref{aff18},\ref{aff30}}
\and B.~R.~Granett\orcid{0000-0003-2694-9284}\inst{\ref{aff23}}
\and A.~Grazian\orcid{0000-0002-5688-0663}\inst{\ref{aff64}}
\and F.~Grupp\inst{\ref{aff12},\ref{aff63}}
\and S.~V.~H.~Haugan\orcid{0000-0001-9648-7260}\inst{\ref{aff65}}
\and I.~M.~Hook\orcid{0000-0002-2960-978X}\inst{\ref{aff66}}
\and F.~Hormuth\inst{\ref{aff67}}
\and A.~Hornstrup\orcid{0000-0002-3363-0936}\inst{\ref{aff68},\ref{aff69}}
\and P.~Hudelot\inst{\ref{aff70}}
\and M.~Jhabvala\inst{\ref{aff71}}
\and E.~Keih\"anen\orcid{0000-0003-1804-7715}\inst{\ref{aff72}}
\and S.~Kermiche\orcid{0000-0002-0302-5735}\inst{\ref{aff61}}
\and A.~Kiessling\orcid{0000-0002-2590-1273}\inst{\ref{aff73}}
\and M.~Kilbinger\orcid{0000-0001-9513-7138}\inst{\ref{aff24}}
\and B.~Kubik\orcid{0009-0006-5823-4880}\inst{\ref{aff51}}
\and M.~K\"ummel\orcid{0000-0003-2791-2117}\inst{\ref{aff63}}
\and H.~Kurki-Suonio\orcid{0000-0002-4618-3063}\inst{\ref{aff74},\ref{aff75}}
\and Q.~Le~Boulc'h\inst{\ref{aff76}}
\and A.~M.~C.~Le~Brun\orcid{0000-0002-0936-4594}\inst{\ref{aff77}}
\and D.~Le~Mignant\orcid{0000-0002-5339-5515}\inst{\ref{aff52}}
\and P.~B.~Lilje\orcid{0000-0003-4324-7794}\inst{\ref{aff65}}
\and V.~Lindholm\orcid{0000-0003-2317-5471}\inst{\ref{aff74},\ref{aff75}}
\and I.~Lloro\orcid{0000-0001-5966-1434}\inst{\ref{aff78}}
\and G.~Mainetti\orcid{0000-0003-2384-2377}\inst{\ref{aff76}}
\and D.~Maino\inst{\ref{aff79},\ref{aff43},\ref{aff80}}
\and E.~Maiorano\orcid{0000-0003-2593-4355}\inst{\ref{aff18}}
\and O.~Marggraf\orcid{0000-0001-7242-3852}\inst{\ref{aff81}}
\and M.~Martinelli\orcid{0000-0002-6943-7732}\inst{\ref{aff20},\ref{aff82}}
\and N.~Martinet\orcid{0000-0003-2786-7790}\inst{\ref{aff52}}
\and F.~Marulli\orcid{0000-0002-8850-0303}\inst{\ref{aff17},\ref{aff18},\ref{aff30}}
\and R.~Massey\orcid{0000-0002-6085-3780}\inst{\ref{aff83}}
\and S.~Maurogordato\inst{\ref{aff84}}
\and H.~J.~McCracken\orcid{0000-0002-9489-7765}\inst{\ref{aff70}}
\and E.~Medinaceli\orcid{0000-0002-4040-7783}\inst{\ref{aff18}}
\and S.~Mei\orcid{0000-0002-2849-559X}\inst{\ref{aff85},\ref{aff86}}
\and M.~Melchior\inst{\ref{aff87}}
\and M.~Meneghetti\orcid{0000-0003-1225-7084}\inst{\ref{aff18},\ref{aff30}}
\and E.~Merlin\orcid{0000-0001-6870-8900}\inst{\ref{aff20}}
\and G.~Meylan\inst{\ref{aff88}}
\and A.~Mora\orcid{0000-0002-1922-8529}\inst{\ref{aff89}}
\and M.~Moresco\orcid{0000-0002-7616-7136}\inst{\ref{aff17},\ref{aff18}}
\and L.~Moscardini\orcid{0000-0002-3473-6716}\inst{\ref{aff17},\ref{aff18},\ref{aff30}}
\and R.~Nakajima\orcid{0009-0009-1213-7040}\inst{\ref{aff81}}
\and C.~Neissner\orcid{0000-0001-8524-4968}\inst{\ref{aff90},\ref{aff45}}
\and S.-M.~Niemi\orcid{0009-0005-0247-0086}\inst{\ref{aff40}}
\and C.~Padilla\orcid{0000-0001-7951-0166}\inst{\ref{aff90}}
\and S.~Paltani\orcid{0000-0002-8108-9179}\inst{\ref{aff59}}
\and F.~Pasian\orcid{0000-0002-4869-3227}\inst{\ref{aff26}}
\and K.~Pedersen\inst{\ref{aff91}}
\and W.~J.~Percival\orcid{0000-0002-0644-5727}\inst{\ref{aff92},\ref{aff93},\ref{aff94}}
\and V.~Pettorino\inst{\ref{aff40}}
\and G.~Polenta\orcid{0000-0003-4067-9196}\inst{\ref{aff31}}
\and M.~Poncet\inst{\ref{aff95}}
\and L.~A.~Popa\inst{\ref{aff96}}
\and L.~Pozzetti\orcid{0000-0001-7085-0412}\inst{\ref{aff18}}
\and F.~Raison\orcid{0000-0002-7819-6918}\inst{\ref{aff12}}
\and R.~Rebolo\orcid{0000-0003-3767-7085}\inst{\ref{aff4},\ref{aff97},\ref{aff98}}
\and A.~Renzi\orcid{0000-0001-9856-1970}\inst{\ref{aff99},\ref{aff60}}
\and J.~Rhodes\orcid{0000-0002-4485-8549}\inst{\ref{aff73}}
\and G.~Riccio\inst{\ref{aff35}}
\and E.~Romelli\orcid{0000-0003-3069-9222}\inst{\ref{aff26}}
\and M.~Roncarelli\orcid{0000-0001-9587-7822}\inst{\ref{aff18}}
\and R.~Saglia\orcid{0000-0003-0378-7032}\inst{\ref{aff63},\ref{aff12}}
\and A.~G.~S\'anchez\orcid{0000-0003-1198-831X}\inst{\ref{aff12}}
\and D.~Sapone\orcid{0000-0001-7089-4503}\inst{\ref{aff100}}
\and J.~A.~Schewtschenko\orcid{0000-0002-4913-6393}\inst{\ref{aff49}}
\and M.~Schirmer\orcid{0000-0003-2568-9994}\inst{\ref{aff2}}
\and P.~Schneider\orcid{0000-0001-8561-2679}\inst{\ref{aff81}}
\and T.~Schrabback\orcid{0000-0002-6987-7834}\inst{\ref{aff101}}
\and A.~Secroun\orcid{0000-0003-0505-3710}\inst{\ref{aff61}}
\and S.~Serrano\orcid{0000-0002-0211-2861}\inst{\ref{aff14},\ref{aff102},\ref{aff13}}
\and P.~Simon\inst{\ref{aff81}}
\and C.~Sirignano\orcid{0000-0002-0995-7146}\inst{\ref{aff99},\ref{aff60}}
\and G.~Sirri\orcid{0000-0003-2626-2853}\inst{\ref{aff30}}
\and J.~Skottfelt\orcid{0000-0003-1310-8283}\inst{\ref{aff103}}
\and L.~Stanco\orcid{0000-0002-9706-5104}\inst{\ref{aff60}}
\and J.~Steinwagner\orcid{0000-0001-7443-1047}\inst{\ref{aff12}}
\and P.~Tallada-Cresp\'{i}\orcid{0000-0002-1336-8328}\inst{\ref{aff44},\ref{aff45}}
\and A.~N.~Taylor\inst{\ref{aff49}}
\and I.~Tereno\orcid{0000-0002-4537-6218}\inst{\ref{aff57},\ref{aff104}}
\and S.~Toft\orcid{0000-0003-3631-7176}\inst{\ref{aff105},\ref{aff106}}
\and R.~Toledo-Moreo\orcid{0000-0002-2997-4859}\inst{\ref{aff107}}
\and F.~Torradeflot\orcid{0000-0003-1160-1517}\inst{\ref{aff45},\ref{aff44}}
\and I.~Tutusaus\orcid{0000-0002-3199-0399}\inst{\ref{aff108}}
\and L.~Valenziano\orcid{0000-0002-1170-0104}\inst{\ref{aff18},\ref{aff109}}
\and J.~Valiviita\orcid{0000-0001-6225-3693}\inst{\ref{aff74},\ref{aff75}}
\and T.~Vassallo\orcid{0000-0001-6512-6358}\inst{\ref{aff63},\ref{aff26}}
\and G.~Verdoes~Kleijn\orcid{0000-0001-5803-2580}\inst{\ref{aff110}}
\and A.~Veropalumbo\orcid{0000-0003-2387-1194}\inst{\ref{aff23},\ref{aff33},\ref{aff32}}
\and Y.~Wang\orcid{0000-0002-4749-2984}\inst{\ref{aff111}}
\and J.~Weller\orcid{0000-0002-8282-2010}\inst{\ref{aff63},\ref{aff12}}
\and A.~Zacchei\orcid{0000-0003-0396-1192}\inst{\ref{aff26},\ref{aff25}}
\and G.~Zamorani\orcid{0000-0002-2318-301X}\inst{\ref{aff18}}
\and F.~M.~Zerbi\inst{\ref{aff23}}
\and I.~A.~Zinchenko\orcid{0000-0002-2944-2449}\inst{\ref{aff63}}
\and E.~Zucca\orcid{0000-0002-5845-8132}\inst{\ref{aff18}}
\and V.~Allevato\orcid{0000-0001-7232-5152}\inst{\ref{aff35}}
\and M.~Ballardini\orcid{0000-0003-4481-3559}\inst{\ref{aff112},\ref{aff113},\ref{aff18}}
\and M.~Bolzonella\orcid{0000-0003-3278-4607}\inst{\ref{aff18}}
\and E.~Bozzo\orcid{0000-0002-8201-1525}\inst{\ref{aff59}}
\and C.~Burigana\orcid{0000-0002-3005-5796}\inst{\ref{aff114},\ref{aff109}}
\and R.~Cabanac\orcid{0000-0001-6679-2600}\inst{\ref{aff108}}
\and A.~Cappi\inst{\ref{aff18},\ref{aff84}}
\and J.~A.~Escartin~Vigo\inst{\ref{aff12}}
\and L.~Gabarra\orcid{0000-0002-8486-8856}\inst{\ref{aff115}}
\and W.~G.~Hartley\inst{\ref{aff59}}
\and J.~Mart\'{i}n-Fleitas\orcid{0000-0002-8594-569X}\inst{\ref{aff89}}
\and S.~Matthew\orcid{0000-0001-8448-1697}\inst{\ref{aff49}}
\and R.~B.~Metcalf\orcid{0000-0003-3167-2574}\inst{\ref{aff17},\ref{aff18}}
\and A.~Pezzotta\orcid{0000-0003-0726-2268}\inst{\ref{aff116},\ref{aff12}}
\and M.~P\"ontinen\orcid{0000-0001-5442-2530}\inst{\ref{aff74}}
\and I.~Risso\orcid{0000-0003-2525-7761}\inst{\ref{aff117}}
\and V.~Scottez\inst{\ref{aff118},\ref{aff119}}
\and M.~Sereno\orcid{0000-0003-0302-0325}\inst{\ref{aff18},\ref{aff30}}
\and M.~Tenti\orcid{0000-0002-4254-5901}\inst{\ref{aff30}}
\and M.~Wiesmann\orcid{0009-0000-8199-5860}\inst{\ref{aff65}}
\and Y.~Akrami\orcid{0000-0002-2407-7956}\inst{\ref{aff120},\ref{aff121}}
\and S.~Alvi\orcid{0000-0001-5779-8568}\inst{\ref{aff112}}
\and I.~T.~Andika\orcid{0000-0001-6102-9526}\inst{\ref{aff122},\ref{aff123}}
\and S.~Anselmi\orcid{0000-0002-3579-9583}\inst{\ref{aff60},\ref{aff99},\ref{aff124}}
\and M.~Archidiacono\orcid{0000-0003-4952-9012}\inst{\ref{aff79},\ref{aff80}}
\and F.~Atrio-Barandela\orcid{0000-0002-2130-2513}\inst{\ref{aff125}}
\and D.~Bertacca\orcid{0000-0002-2490-7139}\inst{\ref{aff99},\ref{aff64},\ref{aff60}}
\and M.~Bethermin\orcid{0000-0002-3915-2015}\inst{\ref{aff126}}
\and L.~Bisigello\orcid{0000-0003-0492-4924}\inst{\ref{aff64}}
\and A.~Blanchard\orcid{0000-0001-8555-9003}\inst{\ref{aff108}}
\and L.~Blot\orcid{0000-0002-9622-7167}\inst{\ref{aff127},\ref{aff77}}
\and S.~Borgani\orcid{0000-0001-6151-6439}\inst{\ref{aff128},\ref{aff25},\ref{aff26},\ref{aff27},\ref{aff129}}
\and M.~L.~Brown\orcid{0000-0002-0370-8077}\inst{\ref{aff11}}
\and S.~Bruton\orcid{0000-0002-6503-5218}\inst{\ref{aff130}}
\and A.~Calabro\orcid{0000-0003-2536-1614}\inst{\ref{aff20}}
\and F.~Caro\inst{\ref{aff20}}
\and T.~Castro\orcid{0000-0002-6292-3228}\inst{\ref{aff26},\ref{aff27},\ref{aff25},\ref{aff129}}
\and F.~Cogato\orcid{0000-0003-4632-6113}\inst{\ref{aff17},\ref{aff18}}
\and S.~Davini\orcid{0000-0003-3269-1718}\inst{\ref{aff33}}
\and G.~Desprez\orcid{0000-0001-8325-1742}\inst{\ref{aff110}}
\and A.~D\'iaz-S\'anchez\orcid{0000-0003-0748-4768}\inst{\ref{aff131}}
\and J.~J.~Diaz\orcid{0000-0003-2101-1078}\inst{\ref{aff4}}
\and S.~Di~Domizio\orcid{0000-0003-2863-5895}\inst{\ref{aff32},\ref{aff33}}
\and J.~M.~Diego\orcid{0000-0001-9065-3926}\inst{\ref{aff132}}
\and P.-A.~Duc\orcid{0000-0003-3343-6284}\inst{\ref{aff126}}
\and A.~Enia\orcid{0000-0002-0200-2857}\inst{\ref{aff29},\ref{aff18}}
\and Y.~Fang\inst{\ref{aff63}}
\and A.~G.~Ferrari\orcid{0009-0005-5266-4110}\inst{\ref{aff30}}
\and A.~Finoguenov\orcid{0000-0002-4606-5403}\inst{\ref{aff74}}
\and A.~Fontana\orcid{0000-0003-3820-2823}\inst{\ref{aff20}}
\and A.~Franco\orcid{0000-0002-4761-366X}\inst{\ref{aff133},\ref{aff134},\ref{aff135}}
\and J.~Garc\'ia-Bellido\orcid{0000-0002-9370-8360}\inst{\ref{aff120}}
\and T.~Gasparetto\orcid{0000-0002-7913-4866}\inst{\ref{aff26}}
\and V.~Gautard\inst{\ref{aff136}}
\and E.~Gaztanaga\orcid{0000-0001-9632-0815}\inst{\ref{aff13},\ref{aff14},\ref{aff137}}
\and F.~Giacomini\orcid{0000-0002-3129-2814}\inst{\ref{aff30}}
\and F.~Gianotti\orcid{0000-0003-4666-119X}\inst{\ref{aff18}}
\and M.~Guidi\orcid{0000-0001-9408-1101}\inst{\ref{aff29},\ref{aff18}}
\and C.~M.~Gutierrez\orcid{0000-0001-7854-783X}\inst{\ref{aff138}}
\and A.~Hall\orcid{0000-0002-3139-8651}\inst{\ref{aff49}}
\and S.~Hemmati\orcid{0000-0003-2226-5395}\inst{\ref{aff139}}
\and H.~Hildebrandt\orcid{0000-0002-9814-3338}\inst{\ref{aff140}}
\and J.~Hjorth\orcid{0000-0002-4571-2306}\inst{\ref{aff91}}
\and J.~J.~E.~Kajava\orcid{0000-0002-3010-8333}\inst{\ref{aff141},\ref{aff142}}
\and Y.~Kang\orcid{0009-0000-8588-7250}\inst{\ref{aff59}}
\and V.~Kansal\orcid{0000-0002-4008-6078}\inst{\ref{aff143},\ref{aff144}}
\and D.~Karagiannis\orcid{0000-0002-4927-0816}\inst{\ref{aff112},\ref{aff145}}
\and C.~C.~Kirkpatrick\inst{\ref{aff72}}
\and S.~Kruk\orcid{0000-0001-8010-8879}\inst{\ref{aff22}}
\and L.~Legrand\orcid{0000-0003-0610-5252}\inst{\ref{aff146},\ref{aff147}}
\and M.~Lembo\orcid{0000-0002-5271-5070}\inst{\ref{aff112},\ref{aff113}}
\and F.~Lepori\orcid{0009-0000-5061-7138}\inst{\ref{aff148}}
\and G.~Leroy\orcid{0009-0004-2523-4425}\inst{\ref{aff149},\ref{aff83}}
\and J.~Lesgourgues\orcid{0000-0001-7627-353X}\inst{\ref{aff150}}
\and L.~Leuzzi\orcid{0009-0006-4479-7017}\inst{\ref{aff17},\ref{aff18}}
\and T.~I.~Liaudat\orcid{0000-0002-9104-314X}\inst{\ref{aff151}}
\and J.~Macias-Perez\orcid{0000-0002-5385-2763}\inst{\ref{aff152}}
\and M.~Magliocchetti\orcid{0000-0001-9158-4838}\inst{\ref{aff62}}
\and F.~Mannucci\orcid{0000-0002-4803-2381}\inst{\ref{aff153}}
\and R.~Maoli\orcid{0000-0002-6065-3025}\inst{\ref{aff154},\ref{aff20}}
\and C.~J.~A.~P.~Martins\orcid{0000-0002-4886-9261}\inst{\ref{aff15},\ref{aff16}}
\and L.~Maurin\orcid{0000-0002-8406-0857}\inst{\ref{aff21}}
\and M.~Miluzio\inst{\ref{aff22},\ref{aff155}}
\and P.~Monaco\orcid{0000-0003-2083-7564}\inst{\ref{aff128},\ref{aff26},\ref{aff27},\ref{aff25}}
\and G.~Morgante\inst{\ref{aff18}}
\and K.~Naidoo\orcid{0000-0002-9182-1802}\inst{\ref{aff137}}
\and A.~Navarro-Alsina\orcid{0000-0002-3173-2592}\inst{\ref{aff81}}
\and F.~Passalacqua\orcid{0000-0002-8606-4093}\inst{\ref{aff99},\ref{aff60}}
\and K.~Paterson\orcid{0000-0001-8340-3486}\inst{\ref{aff2}}
\and L.~Patrizii\inst{\ref{aff30}}
\and A.~Pisani\orcid{0000-0002-6146-4437}\inst{\ref{aff61}}
\and D.~Potter\orcid{0000-0002-0757-5195}\inst{\ref{aff148}}
\and S.~Quai\orcid{0000-0002-0449-8163}\inst{\ref{aff17},\ref{aff18}}
\and M.~Radovich\orcid{0000-0002-3585-866X}\inst{\ref{aff64}}
\and P.-F.~Rocci\inst{\ref{aff21}}
\and G.~Rodighiero\orcid{0000-0002-9415-2296}\inst{\ref{aff99},\ref{aff64}}
\and S.~Sacquegna\orcid{0000-0002-8433-6630}\inst{\ref{aff134},\ref{aff133},\ref{aff135}}
\and M.~Sahl\'en\orcid{0000-0003-0973-4804}\inst{\ref{aff156}}
\and D.~B.~Sanders\orcid{0000-0002-1233-9998}\inst{\ref{aff47}}
\and E.~Sarpa\orcid{0000-0002-1256-655X}\inst{\ref{aff28},\ref{aff129},\ref{aff27}}
\and A.~Schneider\orcid{0000-0001-7055-8104}\inst{\ref{aff148}}
\and M.~Schultheis\inst{\ref{aff84}}
\and D.~Sciotti\orcid{0009-0008-4519-2620}\inst{\ref{aff20},\ref{aff82}}
\and E.~Sellentin\inst{\ref{aff157},\ref{aff42}}
\and F.~Shankar\orcid{0000-0001-8973-5051}\inst{\ref{aff158}}
\and L.~C.~Smith\orcid{0000-0002-3259-2771}\inst{\ref{aff159}}
\and K.~Tanidis\orcid{0000-0001-9843-5130}\inst{\ref{aff115}}
\and G.~Testera\inst{\ref{aff33}}
\and R.~Teyssier\orcid{0000-0001-7689-0933}\inst{\ref{aff160}}
\and S.~Tosi\orcid{0000-0002-7275-9193}\inst{\ref{aff32},\ref{aff33},\ref{aff23}}
\and A.~Troja\orcid{0000-0003-0239-4595}\inst{\ref{aff99},\ref{aff60}}
\and M.~Tucci\inst{\ref{aff59}}
\and C.~Valieri\inst{\ref{aff30}}
\and D.~Vergani\orcid{0000-0003-0898-2216}\inst{\ref{aff18}}
\and G.~Verza\orcid{0000-0002-1886-8348}\inst{\ref{aff161}}
\and N.~A.~Walton\orcid{0000-0003-3983-8778}\inst{\ref{aff159}}}

\institute{School of Physics, HH Wills Physics Laboratory, University of Bristol, Tyndall Avenue, Bristol, BS8 1TL, UK\label{aff1}
\and
Max-Planck-Institut f\"ur Astronomie, K\"onigstuhl 17, 69117 Heidelberg, Germany\label{aff2}
\and
SRON Netherlands Institute for Space Research, Landleven 12, 9747 AD, Groningen, The Netherlands\label{aff3}
\and
Instituto de Astrof\'{\i}sica de Canarias, V\'{\i}a L\'actea, 38205 La Laguna, Tenerife, Spain\label{aff4}
\and
Instituto de Astrof\'isica de Canarias (IAC); Departamento de Astrof\'isica, Universidad de La Laguna (ULL), 38200, La Laguna, Tenerife, Spain\label{aff5}
\and
Universit\'e PSL, Observatoire de Paris, Sorbonne Universit\'e, CNRS, LERMA, 75014, Paris, France\label{aff6}
\and
Universit\'e Paris-Cit\'e, 5 Rue Thomas Mann, 75013, Paris, France\label{aff7}
\and
School of Physics, Astronomy and Mathematics, University of Hertfordshire, College Lane, Hatfield AL10 9AB, UK\label{aff8}
\and
Aspia Space, Falmouth, TR10 9TA, UK\label{aff9}
\and
David A. Dunlap Department of Astronomy \& Astrophysics, University of Toronto, 50 St George Street, Toronto, Ontario M5S 3H4, Canada\label{aff10}
\and
Jodrell Bank Centre for Astrophysics, Department of Physics and Astronomy, University of Manchester, Oxford Road, Manchester M13 9PL, UK\label{aff11}
\and
Max Planck Institute for Extraterrestrial Physics, Giessenbachstr. 1, 85748 Garching, Germany\label{aff12}
\and
Institute of Space Sciences (ICE, CSIC), Campus UAB, Carrer de Can Magrans, s/n, 08193 Barcelona, Spain\label{aff13}
\and
Institut d'Estudis Espacials de Catalunya (IEEC),  Edifici RDIT, Campus UPC, 08860 Castelldefels, Barcelona, Spain\label{aff14}
\and
Centro de Astrof\'{\i}sica da Universidade do Porto, Rua das Estrelas, 4150-762 Porto, Portugal\label{aff15}
\and
Instituto de Astrof\'isica e Ci\^encias do Espa\c{c}o, Universidade do Porto, CAUP, Rua das Estrelas, PT4150-762 Porto, Portugal\label{aff16}
\and
Dipartimento di Fisica e Astronomia "Augusto Righi" - Alma Mater Studiorum Universit\`a di Bologna, via Piero Gobetti 93/2, 40129 Bologna, Italy\label{aff17}
\and
INAF-Osservatorio di Astrofisica e Scienza dello Spazio di Bologna, Via Piero Gobetti 93/3, 40129 Bologna, Italy\label{aff18}
\and
Department of Mathematics and Physics, Roma Tre University, Via della Vasca Navale 84, 00146 Rome, Italy\label{aff19}
\and
INAF-Osservatorio Astronomico di Roma, Via Frascati 33, 00078 Monteporzio Catone, Italy\label{aff20}
\and
Universit\'e Paris-Saclay, CNRS, Institut d'astrophysique spatiale, 91405, Orsay, France\label{aff21}
\and
ESAC/ESA, Camino Bajo del Castillo, s/n., Urb. Villafranca del Castillo, 28692 Villanueva de la Ca\~nada, Madrid, Spain\label{aff22}
\and
INAF-Osservatorio Astronomico di Brera, Via Brera 28, 20122 Milano, Italy\label{aff23}
\and
Universit\'e Paris-Saclay, Universit\'e Paris Cit\'e, CEA, CNRS, AIM, 91191, Gif-sur-Yvette, France\label{aff24}
\and
IFPU, Institute for Fundamental Physics of the Universe, via Beirut 2, 34151 Trieste, Italy\label{aff25}
\and
INAF-Osservatorio Astronomico di Trieste, Via G. B. Tiepolo 11, 34143 Trieste, Italy\label{aff26}
\and
INFN, Sezione di Trieste, Via Valerio 2, 34127 Trieste TS, Italy\label{aff27}
\and
SISSA, International School for Advanced Studies, Via Bonomea 265, 34136 Trieste TS, Italy\label{aff28}
\and
Dipartimento di Fisica e Astronomia, Universit\`a di Bologna, Via Gobetti 93/2, 40129 Bologna, Italy\label{aff29}
\and
INFN-Sezione di Bologna, Viale Berti Pichat 6/2, 40127 Bologna, Italy\label{aff30}
\and
Space Science Data Center, Italian Space Agency, via del Politecnico snc, 00133 Roma, Italy\label{aff31}
\and
Dipartimento di Fisica, Universit\`a di Genova, Via Dodecaneso 33, 16146, Genova, Italy\label{aff32}
\and
INFN-Sezione di Genova, Via Dodecaneso 33, 16146, Genova, Italy\label{aff33}
\and
Department of Physics "E. Pancini", University Federico II, Via Cinthia 6, 80126, Napoli, Italy\label{aff34}
\and
INAF-Osservatorio Astronomico di Capodimonte, Via Moiariello 16, 80131 Napoli, Italy\label{aff35}
\and
Faculdade de Ci\^encias da Universidade do Porto, Rua do Campo de Alegre, 4150-007 Porto, Portugal\label{aff36}
\and
Dipartimento di Fisica, Universit\`a degli Studi di Torino, Via P. Giuria 1, 10125 Torino, Italy\label{aff37}
\and
INFN-Sezione di Torino, Via P. Giuria 1, 10125 Torino, Italy\label{aff38}
\and
INAF-Osservatorio Astrofisico di Torino, Via Osservatorio 20, 10025 Pino Torinese (TO), Italy\label{aff39}
\and
European Space Agency/ESTEC, Keplerlaan 1, 2201 AZ Noordwijk, The Netherlands\label{aff40}
\and
Institute Lorentz, Leiden University, Niels Bohrweg 2, 2333 CA Leiden, The Netherlands\label{aff41}
\and
Leiden Observatory, Leiden University, Einsteinweg 55, 2333 CC Leiden, The Netherlands\label{aff42}
\and
INAF-IASF Milano, Via Alfonso Corti 12, 20133 Milano, Italy\label{aff43}
\and
Centro de Investigaciones Energ\'eticas, Medioambientales y Tecnol\'ogicas (CIEMAT), Avenida Complutense 40, 28040 Madrid, Spain\label{aff44}
\and
Port d'Informaci\'{o} Cient\'{i}fica, Campus UAB, C. Albareda s/n, 08193 Bellaterra (Barcelona), Spain\label{aff45}
\and
INFN section of Naples, Via Cinthia 6, 80126, Napoli, Italy\label{aff46}
\and
Institute for Astronomy, University of Hawaii, 2680 Woodlawn Drive, Honolulu, HI 96822, USA\label{aff47}
\and
Dipartimento di Fisica e Astronomia "Augusto Righi" - Alma Mater Studiorum Universit\`a di Bologna, Viale Berti Pichat 6/2, 40127 Bologna, Italy\label{aff48}
\and
Institute for Astronomy, University of Edinburgh, Royal Observatory, Blackford Hill, Edinburgh EH9 3HJ, UK\label{aff49}
\and
European Space Agency/ESRIN, Largo Galileo Galilei 1, 00044 Frascati, Roma, Italy\label{aff50}
\and
Universit\'e Claude Bernard Lyon 1, CNRS/IN2P3, IP2I Lyon, UMR 5822, Villeurbanne, F-69100, France\label{aff51}
\and
Aix-Marseille Universit\'e, CNRS, CNES, LAM, Marseille, France\label{aff52}
\and
Institut de Ci\`{e}ncies del Cosmos (ICCUB), Universitat de Barcelona (IEEC-UB), Mart\'{i} i Franqu\`{e}s 1, 08028 Barcelona, Spain\label{aff53}
\and
Instituci\'o Catalana de Recerca i Estudis Avan\c{c}ats (ICREA), Passeig de Llu\'{\i}s Companys 23, 08010 Barcelona, Spain\label{aff54}
\and
UCB Lyon 1, CNRS/IN2P3, IUF, IP2I Lyon, 4 rue Enrico Fermi, 69622 Villeurbanne, France\label{aff55}
\and
Mullard Space Science Laboratory, University College London, Holmbury St Mary, Dorking, Surrey RH5 6NT, UK\label{aff56}
\and
Departamento de F\'isica, Faculdade de Ci\^encias, Universidade de Lisboa, Edif\'icio C8, Campo Grande, PT1749-016 Lisboa, Portugal\label{aff57}
\and
Instituto de Astrof\'isica e Ci\^encias do Espa\c{c}o, Faculdade de Ci\^encias, Universidade de Lisboa, Campo Grande, 1749-016 Lisboa, Portugal\label{aff58}
\and
Department of Astronomy, University of Geneva, ch. d'Ecogia 16, 1290 Versoix, Switzerland\label{aff59}
\and
INFN-Padova, Via Marzolo 8, 35131 Padova, Italy\label{aff60}
\and
Aix-Marseille Universit\'e, CNRS/IN2P3, CPPM, Marseille, France\label{aff61}
\and
INAF-Istituto di Astrofisica e Planetologia Spaziali, via del Fosso del Cavaliere, 100, 00100 Roma, Italy\label{aff62}
\and
Universit\"ats-Sternwarte M\"unchen, Fakult\"at f\"ur Physik, Ludwig-Maximilians-Universit\"at M\"unchen, Scheinerstrasse 1, 81679 M\"unchen, Germany\label{aff63}
\and
INAF-Osservatorio Astronomico di Padova, Via dell'Osservatorio 5, 35122 Padova, Italy\label{aff64}
\and
Institute of Theoretical Astrophysics, University of Oslo, P.O. Box 1029 Blindern, 0315 Oslo, Norway\label{aff65}
\and
Department of Physics, Lancaster University, Lancaster, LA1 4YB, UK\label{aff66}
\and
Felix Hormuth Engineering, Goethestr. 17, 69181 Leimen, Germany\label{aff67}
\and
Technical University of Denmark, Elektrovej 327, 2800 Kgs. Lyngby, Denmark\label{aff68}
\and
Cosmic Dawn Center (DAWN), Denmark\label{aff69}
\and
Institut d'Astrophysique de Paris, UMR 7095, CNRS, and Sorbonne Universit\'e, 98 bis boulevard Arago, 75014 Paris, France\label{aff70}
\and
NASA Goddard Space Flight Center, Greenbelt, MD 20771, USA\label{aff71}
\and
Department of Physics and Helsinki Institute of Physics, Gustaf H\"allstr\"omin katu 2, 00014 University of Helsinki, Finland\label{aff72}
\and
Jet Propulsion Laboratory, California Institute of Technology, 4800 Oak Grove Drive, Pasadena, CA, 91109, USA\label{aff73}
\and
Department of Physics, P.O. Box 64, 00014 University of Helsinki, Finland\label{aff74}
\and
Helsinki Institute of Physics, Gustaf H{\"a}llstr{\"o}min katu 2, University of Helsinki, Helsinki, Finland\label{aff75}
\and
Centre de Calcul de l'IN2P3/CNRS, 21 avenue Pierre de Coubertin 69627 Villeurbanne Cedex, France\label{aff76}
\and
Laboratoire d'etude de l'Univers et des phenomenes eXtremes, Observatoire de Paris, Universit\'e PSL, Sorbonne Universit\'e, CNRS, 92190 Meudon, France\label{aff77}
\and
SKA Observatory, Jodrell Bank, Lower Withington, Macclesfield, Cheshire SK11 9FT, UK\label{aff78}
\and
Dipartimento di Fisica "Aldo Pontremoli", Universit\`a degli Studi di Milano, Via Celoria 16, 20133 Milano, Italy\label{aff79}
\and
INFN-Sezione di Milano, Via Celoria 16, 20133 Milano, Italy\label{aff80}
\and
Universit\"at Bonn, Argelander-Institut f\"ur Astronomie, Auf dem H\"ugel 71, 53121 Bonn, Germany\label{aff81}
\and
INFN-Sezione di Roma, Piazzale Aldo Moro, 2 - c/o Dipartimento di Fisica, Edificio G. Marconi, 00185 Roma, Italy\label{aff82}
\and
Department of Physics, Institute for Computational Cosmology, Durham University, South Road, Durham, DH1 3LE, UK\label{aff83}
\and
Universit\'e C\^{o}te d'Azur, Observatoire de la C\^{o}te d'Azur, CNRS, Laboratoire Lagrange, Bd de l'Observatoire, CS 34229, 06304 Nice cedex 4, France\label{aff84}
\and
Universit\'e Paris Cit\'e, CNRS, Astroparticule et Cosmologie, 75013 Paris, France\label{aff85}
\and
CNRS-UCB International Research Laboratory, Centre Pierre Binetruy, IRL2007, CPB-IN2P3, Berkeley, USA\label{aff86}
\and
University of Applied Sciences and Arts of Northwestern Switzerland, School of Engineering, 5210 Windisch, Switzerland\label{aff87}
\and
Institute of Physics, Laboratory of Astrophysics, Ecole Polytechnique F\'ed\'erale de Lausanne (EPFL), Observatoire de Sauverny, 1290 Versoix, Switzerland\label{aff88}
\and
Aurora Technology for European Space Agency (ESA), Camino bajo del Castillo, s/n, Urbanizacion Villafranca del Castillo, Villanueva de la Ca\~nada, 28692 Madrid, Spain\label{aff89}
\and
Institut de F\'{i}sica d'Altes Energies (IFAE), The Barcelona Institute of Science and Technology, Campus UAB, 08193 Bellaterra (Barcelona), Spain\label{aff90}
\and
DARK, Niels Bohr Institute, University of Copenhagen, Jagtvej 155, 2200 Copenhagen, Denmark\label{aff91}
\and
Waterloo Centre for Astrophysics, University of Waterloo, Waterloo, Ontario N2L 3G1, Canada\label{aff92}
\and
Department of Physics and Astronomy, University of Waterloo, Waterloo, Ontario N2L 3G1, Canada\label{aff93}
\and
Perimeter Institute for Theoretical Physics, Waterloo, Ontario N2L 2Y5, Canada\label{aff94}
\and
Centre National d'Etudes Spatiales -- Centre spatial de Toulouse, 18 avenue Edouard Belin, 31401 Toulouse Cedex 9, France\label{aff95}
\and
Institute of Space Science, Str. Atomistilor, nr. 409 M\u{a}gurele, Ilfov, 077125, Romania\label{aff96}
\and
Consejo Superior de Investigaciones Cientificas, Calle Serrano 117, 28006 Madrid, Spain\label{aff97}
\and
Universidad de La Laguna, Departamento de Astrof\'{\i}sica, 38206 La Laguna, Tenerife, Spain\label{aff98}
\and
Dipartimento di Fisica e Astronomia "G. Galilei", Universit\`a di Padova, Via Marzolo 8, 35131 Padova, Italy\label{aff99}
\and
Departamento de F\'isica, FCFM, Universidad de Chile, Blanco Encalada 2008, Santiago, Chile\label{aff100}
\and
Universit\"at Innsbruck, Institut f\"ur Astro- und Teilchenphysik, Technikerstr. 25/8, 6020 Innsbruck, Austria\label{aff101}
\and
Satlantis, University Science Park, Sede Bld 48940, Leioa-Bilbao, Spain\label{aff102}
\and
Centre for Electronic Imaging, Open University, Walton Hall, Milton Keynes, MK7~6AA, UK\label{aff103}
\and
Instituto de Astrof\'isica e Ci\^encias do Espa\c{c}o, Faculdade de Ci\^encias, Universidade de Lisboa, Tapada da Ajuda, 1349-018 Lisboa, Portugal\label{aff104}
\and
Cosmic Dawn Center (DAWN)\label{aff105}
\and
Niels Bohr Institute, University of Copenhagen, Jagtvej 128, 2200 Copenhagen, Denmark\label{aff106}
\and
Universidad Polit\'ecnica de Cartagena, Departamento de Electr\'onica y Tecnolog\'ia de Computadoras,  Plaza del Hospital 1, 30202 Cartagena, Spain\label{aff107}
\and
Institut de Recherche en Astrophysique et Plan\'etologie (IRAP), Universit\'e de Toulouse, CNRS, UPS, CNES, 14 Av. Edouard Belin, 31400 Toulouse, France\label{aff108}
\and
INFN-Bologna, Via Irnerio 46, 40126 Bologna, Italy\label{aff109}
\and
Kapteyn Astronomical Institute, University of Groningen, PO Box 800, 9700 AV Groningen, The Netherlands\label{aff110}
\and
Infrared Processing and Analysis Center, California Institute of Technology, Pasadena, CA 91125, USA\label{aff111}
\and
Dipartimento di Fisica e Scienze della Terra, Universit\`a degli Studi di Ferrara, Via Giuseppe Saragat 1, 44122 Ferrara, Italy\label{aff112}
\and
Istituto Nazionale di Fisica Nucleare, Sezione di Ferrara, Via Giuseppe Saragat 1, 44122 Ferrara, Italy\label{aff113}
\and
INAF, Istituto di Radioastronomia, Via Piero Gobetti 101, 40129 Bologna, Italy\label{aff114}
\and
Department of Physics, Oxford University, Keble Road, Oxford OX1 3RH, UK\label{aff115}
\and
INAF - Osservatorio Astronomico di Brera, via Emilio Bianchi 46, 23807 Merate, Italy\label{aff116}
\and
INAF-Osservatorio Astronomico di Brera, Via Brera 28, 20122 Milano, Italy, and INFN-Sezione di Genova, Via Dodecaneso 33, 16146, Genova, Italy\label{aff117}
\and
Institut d'Astrophysique de Paris, 98bis Boulevard Arago, 75014, Paris, France\label{aff118}
\and
ICL, Junia, Universit\'e Catholique de Lille, LITL, 59000 Lille, France\label{aff119}
\and
Instituto de F\'isica Te\'orica UAM-CSIC, Campus de Cantoblanco, 28049 Madrid, Spain\label{aff120}
\and
CERCA/ISO, Department of Physics, Case Western Reserve University, 10900 Euclid Avenue, Cleveland, OH 44106, USA\label{aff121}
\and
Technical University of Munich, TUM School of Natural Sciences, Physics Department, James-Franck-Str.~1, 85748 Garching, Germany\label{aff122}
\and
Max-Planck-Institut f\"ur Astrophysik, Karl-Schwarzschild-Str.~1, 85748 Garching, Germany\label{aff123}
\and
Laboratoire Univers et Th\'eorie, Observatoire de Paris, Universit\'e PSL, Universit\'e Paris Cit\'e, CNRS, 92190 Meudon, France\label{aff124}
\and
Departamento de F{\'\i}sica Fundamental. Universidad de Salamanca. Plaza de la Merced s/n. 37008 Salamanca, Spain\label{aff125}
\and
Universit\'e de Strasbourg, CNRS, Observatoire astronomique de Strasbourg, UMR 7550, 67000 Strasbourg, France\label{aff126}
\and
Center for Data-Driven Discovery, Kavli IPMU (WPI), UTIAS, The University of Tokyo, Kashiwa, Chiba 277-8583, Japan\label{aff127}
\and
Dipartimento di Fisica - Sezione di Astronomia, Universit\`a di Trieste, Via Tiepolo 11, 34131 Trieste, Italy\label{aff128}
\and
ICSC - Centro Nazionale di Ricerca in High Performance Computing, Big Data e Quantum Computing, Via Magnanelli 2, Bologna, Italy\label{aff129}
\and
California Institute of Technology, 1200 E California Blvd, Pasadena, CA 91125, USA\label{aff130}
\and
Departamento F\'isica Aplicada, Universidad Polit\'ecnica de Cartagena, Campus Muralla del Mar, 30202 Cartagena, Murcia, Spain\label{aff131}
\and
Instituto de F\'isica de Cantabria, Edificio Juan Jord\'a, Avenida de los Castros, 39005 Santander, Spain\label{aff132}
\and
INFN, Sezione di Lecce, Via per Arnesano, CP-193, 73100, Lecce, Italy\label{aff133}
\and
Department of Mathematics and Physics E. De Giorgi, University of Salento, Via per Arnesano, CP-I93, 73100, Lecce, Italy\label{aff134}
\and
INAF-Sezione di Lecce, c/o Dipartimento Matematica e Fisica, Via per Arnesano, 73100, Lecce, Italy\label{aff135}
\and
CEA Saclay, DFR/IRFU, Service d'Astrophysique, Bat. 709, 91191 Gif-sur-Yvette, France\label{aff136}
\and
Institute of Cosmology and Gravitation, University of Portsmouth, Portsmouth PO1 3FX, UK\label{aff137}
\and
Instituto de Astrof\'\i sica de Canarias, c/ Via Lactea s/n, La Laguna 38200, Spain. Departamento de Astrof\'\i sica de la Universidad de La Laguna, Avda. Francisco Sanchez, La Laguna, 38200, Spain\label{aff138}
\and
Caltech/IPAC, 1200 E. California Blvd., Pasadena, CA 91125, USA\label{aff139}
\and
Ruhr University Bochum, Faculty of Physics and Astronomy, Astronomical Institute (AIRUB), German Centre for Cosmological Lensing (GCCL), 44780 Bochum, Germany\label{aff140}
\and
Department of Physics and Astronomy, Vesilinnantie 5, 20014 University of Turku, Finland\label{aff141}
\and
Serco for European Space Agency (ESA), Camino bajo del Castillo, s/n, Urbanizacion Villafranca del Castillo, Villanueva de la Ca\~nada, 28692 Madrid, Spain\label{aff142}
\and
ARC Centre of Excellence for Dark Matter Particle Physics, Melbourne, Australia\label{aff143}
\and
Centre for Astrophysics \& Supercomputing, Swinburne University of Technology,  Hawthorn, Victoria 3122, Australia\label{aff144}
\and
Department of Physics and Astronomy, University of the Western Cape, Bellville, Cape Town, 7535, South Africa\label{aff145}
\and
DAMTP, Centre for Mathematical Sciences, Wilberforce Road, Cambridge CB3 0WA, UK\label{aff146}
\and
Kavli Institute for Cosmology Cambridge, Madingley Road, Cambridge, CB3 0HA, UK\label{aff147}
\and
Department of Astrophysics, University of Zurich, Winterthurerstrasse 190, 8057 Zurich, Switzerland\label{aff148}
\and
Department of Physics, Centre for Extragalactic Astronomy, Durham University, South Road, Durham, DH1 3LE, UK\label{aff149}
\and
Institute for Theoretical Particle Physics and Cosmology (TTK), RWTH Aachen University, 52056 Aachen, Germany\label{aff150}
\and
IRFU, CEA, Universit\'e Paris-Saclay 91191 Gif-sur-Yvette Cedex, France\label{aff151}
\and
Univ. Grenoble Alpes, CNRS, Grenoble INP, LPSC-IN2P3, 53, Avenue des Martyrs, 38000, Grenoble, France\label{aff152}
\and
INAF-Osservatorio Astrofisico di Arcetri, Largo E. Fermi 5, 50125, Firenze, Italy\label{aff153}
\and
Dipartimento di Fisica, Sapienza Universit\`a di Roma, Piazzale Aldo Moro 2, 00185 Roma, Italy\label{aff154}
\and
HE Space for European Space Agency (ESA), Camino bajo del Castillo, s/n, Urbanizacion Villafranca del Castillo, Villanueva de la Ca\~nada, 28692 Madrid, Spain\label{aff155}
\and
Theoretical astrophysics, Department of Physics and Astronomy, Uppsala University, Box 515, 751 20 Uppsala, Sweden\label{aff156}
\and
Mathematical Institute, University of Leiden, Einsteinweg 55, 2333 CA Leiden, The Netherlands\label{aff157}
\and
School of Physics \& Astronomy, University of Southampton, Highfield Campus, Southampton SO17 1BJ, UK\label{aff158}
\and
Institute of Astronomy, University of Cambridge, Madingley Road, Cambridge CB3 0HA, UK\label{aff159}
\and
Department of Astrophysical Sciences, Peyton Hall, Princeton University, Princeton, NJ 08544, USA\label{aff160}
\and
Center for Computational Astrophysics, Flatiron Institute, 162 5th Avenue, 10010, New York, NY, USA\label{aff161}}    

\abstract 
{
Light emission from galaxies exhibit diverse brightness profiles, influenced by factors such as galaxy type, structural features, and interactions with other galaxies. Elliptical galaxies feature more uniform light distributions, while spiral and irregular galaxies have complex, varied light profiles due to their structural heterogeneity and star-forming activity. In addition, galaxies with active galactic nuclei (AGN) feature intense, concentrated emission from gas accretion around supermassive black holes, superimposed on regular galactic light, while quasi-stellar objects (QSOs) represent extreme cases in which AGN emissions dominate their host galaxies. The challenge of identifying AGN and QSOs has been discussed many times in the literature, often requiring multi-wavelength observations. This paper introduces a novel approach to identify AGN and QSOs from a single image.
Diffusion models have recently been developed in the machine-learning literature to generate realistic-looking images of everyday objects. Utilising the spatial resolving power of the \Euclid VIS images, we created a diffusion model trained on one million sources, without using any source pre-selection or labels. The model learns to reconstruct light distributions of normal galaxies, since the population is dominated by them. We conditioned the prediction of the central light distribution by masking the central few pixels of each source and reconstructed the light according to the diffusion model. We further used this prediction to identify sources that deviate from this profile by examining the reconstruction error of the few central pixels regenerated in each source's core.
Our approach, solely using VIS imaging, features high completeness compared to traditional methods of AGN and QSO selection, including optical, near-infrared, mid-infrared, and X-rays. 
Our study offers practical insights for refining diffusion models and broadening their applications throughout the \Euclid survey area, underscoring the utility of this approach in diverse astronomical contexts beyond just AGN identification.
}

   \keywords{Galaxies: active -- techniques: image processing -- methods: data analysis -- methods: observational}

   \titlerunning{AGN identification using diffusion-based inpainting of \Euclid VIS images}
   \authorrunning{Euclid Collaboration: G. Stevens et al.}

   \maketitle

\section{Introduction}

Ever since the discovery that the mass or dynamics of a spheroid within a galaxy correlates with the mass of any detectable supermassive black hole (SMBH) at its centre (e.g.\
\citealt{1989IAUS..134..217D, 1995ARA&A..33..581K, 1998AJ....115.2285M, 2000ApJ...539L...9F}, and more recently, \citealt{Sahu2019ApJ...876..155S,Davis2018ApJ...869..113D,Davis2019ApJ...873...85D} and references therein), it has been believed that most -- if not all -- galaxies with a spheroidal (or pseudo-spheroidal) component contain a SMBH at their centre. If a galaxy's emission contains evidence of its central SMBH being actively fuelled, it will emit non-thermal radiation, leading it to be classified as an active galactic nucleus (AGN). Depending upon the luminosity of that emission and its contrast with the stellar emission from the surrounding galaxy, it could have a classification ranging from a Seyfert galaxy (for the weakest emission) through to a quasi-stellar object or QSO (for the most luminous).

For imaging in wavebands such as the optical and near-IR, the presence and detectability of emission from the area directly around a SMBH depends on multiple factors. This can include the rate of infall of material into the region around the SMBH, the amount of obscuration (absorption or scattering) towards that region, the contrast between the SMBH-related emission and that of the surrounding galaxy, the redshift, and the intrinsic compactness of the stellar distribution of the host galaxy. The growth and mass of a central SMBH is intimately related to the evolution of its host galaxy, not least through the action of energy, momentum, and radiative feedback from outflows originating from the area around the SMBH on the properties of the stellar populations and ISM (see \citealt{harrison2024observational} for a recent review). Consequently, identifying and characterising SMBHs, particularly during periods of active feedback, is central to the study of SMBH and galaxy evolution.

A more precise classification of a source as an AGN is possible when a much wider range of multi-wavelength data is available. This can include spectroscopy and considering the much broader spectral energy distribution of the galaxy and/or the candidate AGN. However, it is still worth considering the efficient selection of AGN candidates from single-band imaging as a first step in identifying large samples of AGN for further study. In particular, with the advent of the uniform, deep, and high-spatial-resolution optical imaging dataset from the \Euclid mission, covering a large fraction of the extragalactic sky, there is the potential to derive unprecedentedly large samples of AGN across a wide range of the (optical) luminosity and redshift parameter space with the use of an appropriate and sufficiently efficient technique. The technique can then subsequently be refined by feeding back the result of using other multi-wavelength data, allowing us to confirm or reject candidates. With sufficient refinement, this further assessment step eventually becomes either unnecessary, or at the very least, an efficient stage in compiling reliable samples. This then opens the door to the production of unprecedentedly large and reliable samples of optically selected AGN.

\begin{figure*}[!ht]
   \centering
   \includegraphics[width=\linewidth]{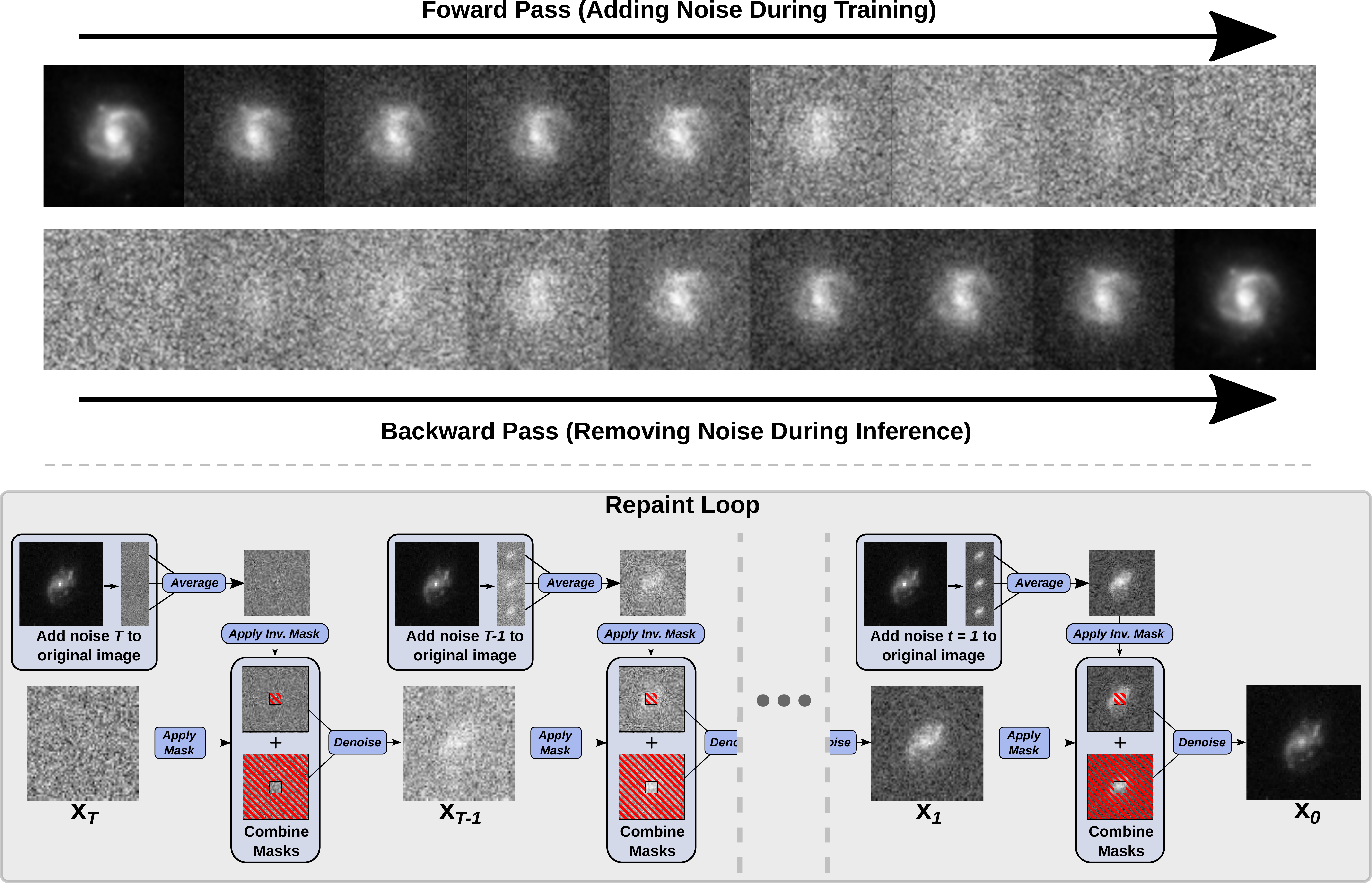}
   \caption{Diffusion pipeline (\textit{top}) that progressively adds noise to images, training the model to predict what noise was added from the previous step. Once trained and during inference, the model takes pure Gaussian noise as input and can iteratively remove the noise until a realistic galaxy image remains. Repeat inference runs will provide a different and unique galaxy from those it was trained on. The Repaint pipeline (\textit{bottom}) takes the trained diffusion model and enables conditioning to allow parts of an existing image to be preserved by masking. At each denoising step, noise levels in the preserved pixels are adjusted to ensure they integrate correctly with the newly generated sections. After $T$ iterations, the output includes the retained pixels and newly generated areas, creating a different yet plausible final image.}
   \label{fig:repaint_loop}
\end{figure*}

This work presents a technique that can achieve the above goal. In the following, we describe the use of diffusion models for generating samples of AGN candidates using \Euclid VIS \IE band \citep{EuclidSkyOverview, EuclidSkyVIS} imaging alone. Diffusion-based methods are machine-learning models that generate or reconstruct data by evolving a random distribution of pixels into a structured output over several steps. This technique outperforms traditional methods such as median pixel substitution or interpolation, which often struggle to recreate complex structures within images. Using these models for `inpainting', whereby we adaptively recreate parts of the image, we can measure the discrepancies between the original and generated pixels. Examining these errors allows us to separate AGN, QSOs, and other sources from the broader galaxy population. 

Our method\footnote{The code is publicly available here: \url{https://github.com/grant-m-s/Astro-Repaint}.} identifies potential AGN by treating the task as an unsupervised anomaly detection problem. It begins with an image of a galaxy, potentially containing AGN emission at its centre, and masks out the central pixels that might be influenced by such emission. A diffusion model, trained on a large, unlabeled dataset of galaxies -- most of which are normal, non-active -- then predicts or `inpaints' the masked region based on the learned prior distribution of typical galaxy morphologies. Because AGN are rare, there is no requirement to remove them from the training sample; the model is inherently biased towards reconstructing the central region as if it belonged to a typical, non-active galaxy. As a result, any significant discrepancy between the inpainted and actual central region suggests the presence of an AGN. This reconstruction error thus serves as a powerful anomaly score for identifying candidate AGN.

While this paper details the fundamental technique applied to VIS imaging data, as this approach utilises machine learning, future improvements could incorporate findings from a detailed assessment that involves additional datasets. These could include multi-wavelength cross-matching or spectroscopic validation, from which we can ultimately reduce the need for further time-expensive refinement steps. Such enhancements would help to further improve the accuracy and reliability of the AGN candidate samples generated by the technique. Additionally, the technique presented here uses a standard treatment of diffusion models, which in principle can be significantly further optimised for the specific characteristics of the \Euclid dataset, thereby improving further the efficiency of the processing of the \Euclid data. Such work would be the subject of future papers.

\section{Diffusion-based inpainting}
\subsection{Generative models}

Generative models are a class of machine learning systems designed to learn an underlying data distribution, $p(x)$, from a finite set of independent and identically distributed (i.i.d.) samples, $\{x_{i}\}^N_{i=1} \sim p(x)$. The primary goal of these models is to develop a tractable method for generating new, synthetic samples that are statistically indistinguishable from the training data. Over the years, several distinct families of generative models have emerged, each with unique architectural and training paradigms. A key distinction lies in their optimisation objectives, which often seek to minimise a measure of divergence between the true data distribution, $p(x)$, and the model's learned distribution, $p_{\theta}(x)$. The choice of divergence has significant implications for model behavior.
Prominent examples include variational autoencoders \citep[VAEs,][]{kingma2013auto}, which learn a latent variable model of the data, and generative adversarial networks \citep[GANs,][]{goodfellow2014generative,goodfellow2020generative}, which involve a contest between a generator and a discriminator. In practice, GANs implicitly optimise an objective related to the reverse Kullback-Leibler (KL) divergence, $D_{\mathrm{KL}}[p_{\theta}(x)||p(x)]$, which exhibits a `mode-seeking' property. This encourages the generation of high-fidelity samples but can lead to mode collapse, whereby the model learns only a subset of the data distribution~\citep{thanh2020catastrophic}. In contrast, VAEs and diffusion models optimise objectives related to the forward KL divergence, $D_{\mathrm{KL}}[p(x)||p_{\theta}(x)]$, which is `mass-covering'. This property encourages the model to account for all modes present in the data, leading to greater sample diversity and more comprehensive coverage of the data distribution.

\subsection{Diffusion models}

Diffusion models \citep{sohl2015deep,ho2020denoising}, the focus of this work, systematically corrupt data with noise and then learn to reverse the process. They have been shown to achieve superior sample quality and mode coverage compared to other classes of models~\citep{dhariwal2021diffusion}, making them a compelling choice for modelling the complex and diverse morphologies of galaxies observed by Euclid.

For both GANs and VAEs, their significant advantage compared to diffusion models is that they offer faster sample generation and lower computational costs. Conversely, diffusion models trade off speed for superior sample quality, mode coverage, and flexibility in controllable generation, making them our preferred choice in the task of AGN identification.

This paper makes use of one such implementation, denoising diffusion probabilistic models \citep[DDPMs,][]{ho2020denoising}. Diffusion works on the following premise: by continually adding a small amount of noise to an image, eventually, one is left with an image of complete noise without any remnants of the original input. If the noise is generated in a stochastic but consistent way, such as from a Gaussian distribution, we can use a network to learn the dynamics of the noise and perform a reverse process. By predicting the noise that was added to an image, $x$, at timestep $t$, we can remove this noise to produce image $x_{t-1}$. This forms an iterative process originating from some final timestep, $T$, an image of pure noise, to a realistic image from the data distribution at timestep $0$.

The formal definition of this diffusion process, following \citet{ho2020denoising} and \citet{nichol2021improved}, is expressed as $q(x_{1},\dots,x_{T})$, which represents the joint distribution of a sequence in which each image, $x_{t}$, is progressively noised.
\begin{equation}
   \label{eq:diffusion_full}
   q(x_{1} , \dots , x_{T} | x_{0}) := \prod_{t=1}^{T} q(x_{t} | x_{t-1})\;,
\end{equation}

\noindent where $q(x_{t} | x_{t-1})$ specifies the conditional distribution at each timestep, $t$, modelling the incremental addition of noise:
\begin{equation}
   \label{eq:diffusion_step}
   q(x_{t} | x_{t-1}) := \mathcal{N}(x_{t}; \sqrt{1-\beta_{t}}x_{t-1},\beta_{t}I)\;.
\end{equation}

The $\beta_{t}$ term defines the variance of the Gaussian noise added at each timestep. By ensuring a gradual increase in noise, the model can more easily learn the transition from the data distribution to the noise distribution.

With a large enough $T$ and an adjusted time-dependent variance, $\beta_{t}$, $x_{T}$ will approach a Gaussian distribution that is isotropic, meaning uniform in all directions. For the denoising process, starting at a $x_{T} \sim \mathcal{N}(0,I)$, accurately modelling $q(x_{t-1}|x_{t})$ is not tractable, requiring a neural network to approximate it. An example of the full diffusion process can be seen in the top panel of \cref{fig:repaint_loop}.

\subsubsection{Background noise versus diffusion noise}

It is important to clarify the distinctions between the definition and use of `noise' in diffusion models compared to the traditional understanding of noise in astronomy-based data. In the context of astronomical imaging, noise refers to the random fluctuations inherent in observational data, often caused by, for example, background emission, electronics read-out noise, and atmospheric disturbances. This `background noise' reduces the signal-to-noise ratio (S/N) of the image, affecting the ability to detect subtle details in any analysis.

In contrast, the diffusion process uses an intentional integration of Gaussian noise to facilitate the training of the generative model. The pixels of image $x_{T}$ follow the distribution $\mathcal{N}(0,I)$, rather than following the distribution of the background noise. The goal of the reverse diffusion process is to remove the artificial, high-variance Gaussian noise that was intentionally added during the forward process, thereby transforming a sample from a simple prior distribution [$\mathcal{N}(0,I)$] into a realistic galaxy image. A realistic galaxy image, as defined by our training dataset, inherently includes the source's light profile superimposed on the characteristic background noise of the Euclid instrument. Therefore, the model learns to generate images that replicate this entire distribution, including the statistical properties of the background noise. The diffusion pipeline does not aim to produce a `denoised' image in the traditional astronomical sense (i.e. a background-subtracted image).

\subsubsection{Use in astronomy applications}
\citet{2022MNRAS.511.1808S} also utilised the DDPM framework to generate a dataset of realistic galaxies. The model was trained using the Photometry and Rotation curve OBservations from Extragalactic Surveys (PROBES) dataset \citep{stone2019intrinsic,stone2021intrinsic}, a collection of large, well-resolved objects that feature significant internal structure. They provide analysis of the quality of their generated images, showcasing the diffusion model's ability to create visually realistic images that feature similar physical property distributions such as the half-light radius and flux-space colour values. They also briefly explore the use of inpainting with their model to remove satellite trails.

When researching diffusion models, papers utilising score-based models are often used and spoken of interchangeably, as they are different implementations of the same generative process \citep{2022MNRAS.511.1808S}. Rather than the fixed sequence of timesteps to denoise data, score-based methods apply stochastic differential equations (SDEs) to estimate the data distribution's gradient (score). Working with SDEs and gradients allows for a continuous range of possible diffusion paths. It is for this reason that score-based models can be classed as the more general framework for diffusion-based generative models. The use of score-based generative models within astronomical applications includes galaxy image deconvolution \citep{adam2023echoes,spagnoletti2024bayesian}, gravitation lensing analysis \citep{adam2022posterior,remy2023probabilistic}, and deblending \citep{2024A&C....4900875S}.

\subsection{Training objective}
\label{sec:Training_objective}

For training, we used 64$\times$64 pixel VIS cutouts for each of the selected sources that are discussed in \cref{sec:sample_selection}. We used the training pipeline described in \citet{dhariwal2021diffusion}\footnote{Available at \href{https://github.com/openai/guided-diffusion}{https://github.com/openai/guided-diffusion}} and adopted a hybrid loss function that combines a simple mean squared error (MSE) term with the full variational lower bound (VLB).

The MSE objective, proposed by \citet{ho2020denoising}, is an MSE loss between the true and predicted Gaussian noise, $\epsilon$,
\begin{equation}
   L_\mathrm{MSE} = \mathbb{E}_{t, x_{0}, \epsilon} \left[ \left\| \epsilon - \epsilon_{\theta} \left( \sqrt{\bar{\alpha}_t} x_{0} + \sqrt{1 - \bar{\alpha}_t} \epsilon, t \right) \right\|_{2}^{2} \right]
   \label{eq:MSE}\;,
\end{equation}
where $x_{0}$ is an original image, $t$ is a uniformly sampled timestep, and $\epsilon\sim N(0,I)$ is the true noise. For our implementation $\bar{\alpha}_{t}$ is described below (the reader is directed to \citet{nichol2021improved} for specifics on its derivation):
\begin{equation}
   \centering
   \bar{\alpha}_{t} = \cos\left(\frac{t/T + 0.008}{1.008} \frac{\pi}{2}\right)^{2}\;.
   \label{eq:alpha_bar}
\end{equation}

While this objective is effective for training the mean of the reverse process transition, $\mu_{\theta}(x_{t},t)$, it provides no learning signal for its variance, $\Sigma_{\theta}(x_{t},t)$. To address this, \citet{nichol2021improved} proposed learning this variance by incorporating the full VLB of the log-likelihood into the training objective. The VLB is given by
\begin{gather}
    L_\mathrm{VLB} = L_{0} + L_{1} + \dots + L_{T-1} + L_{T}, \label{eq:VLB} \\
    L_{t-1} = D_\mathrm{KL}\left[q(x_{t-1} \mid x_t, x_0) \parallel p(x_{t-1} \mid x_t)\right], \label{eq:VLB_step} \\
    L_{T} = D_\mathrm{KL}\left[q(x_{T} \mid x_0) \parallel p(x_{t})\right]. \label{eq:VLB_Final}
\end{gather}

Since $L_{\mathrm{VLB}}$ explicitly depends on the parameters of $\Sigma_{\theta}$, it can be used to train them. We adopted the hybrid objective
\begin{equation}
   L_{\mathrm{hybrid}} = L_{\mathrm{MSE}} + \lambda L_{\mathrm{VLB}}\;.
   \label{eq:default_loss}
\end{equation}

We used a small weighting constant ($\lambda=0.001$) to prevent the $L_\mathrm{VLB}$ term from overpowering the $L_\mathrm{MSE}$ term, which remains the primary driver of sample quality. This hybrid loss allows the model to learn an optimal reverse process variance, which can improve model likelihoods and enable more efficient sampling.  
A key challenge with astronomical data is its high dynamic range. A standard MSE-based loss is not scale-invariant and will be dominated by the few brightest pixels or sources in a training batch. This can lead to poor performance on the much larger population of fainter objects. To address this, we introduced a normalised version of the MSE loss term. For each image, $x_{i}$, we divided its error by its respective maximum pixel value, denoted as $\max(x_{i})$. This normalisation ensures that the relative error is penalised, allowing for more equitable training across both bright and faint sources. The final loss used for our primary model is thus
\begin{equation}
   L_{\mathrm{normalised}} = \frac{L_{\mathrm{MSE}}}{\max(x_{i})} + \lambda L_{\mathrm{VLB}}\;.
   \label{eq:normalised_loss}
\end{equation}

Throughout the remainder of the paper, we refer to this variant of the loss as the normalised loss. The training curves of this scale-invariant loss compared with the original loss used in \citet{nichol2021improved} are shown in \cref{fig:mse_max_pixel_losses}. By incorporating this normalised loss, we enable more equitable training across the full range of galaxy brightnesses, rather than being dominated by the brightest objects, which occurs when using the default loss.

\subsection{Noise scheduling}
\label{sec:scheduler} 
The total number of timesteps, $T$, is a critical hyperparameter in the diffusion process. A larger $T$ corresponds to smaller, more refined steps in both the forward (noising) and reverse (denoising) processes. While this can, in principle, allow the model to learn a more accurate representation of the data distribution, it comes with a significant trade-off. The computational cost of inference (i.e. generating a sample) is directly proportional to $T$, as the reverse process must be iterated $T$ times. The cost of a single training step, however, is not directly dependent on $T$, because at each step a random timestep, $t$, is sampled and the loss is evaluated only for that level of noise.

\begin{figure}[!t]
   \centering
   \includegraphics[width=\linewidth]{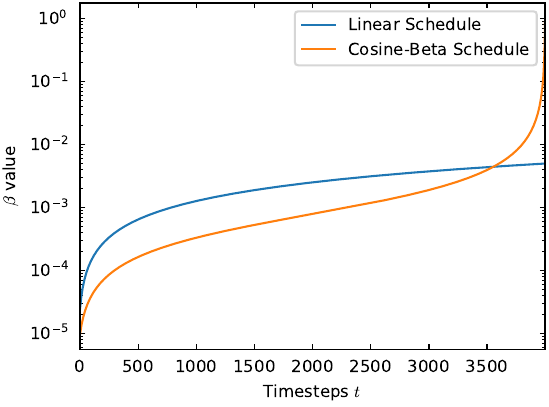}
   \caption{Linear schedule from the original diffusion implementation, which causes the parameters to converge early in the timesteps, resulting in training images becoming pure noise too soon and leading to suboptimal performance. The switch to the cosine-beta schedule adds noise at a much slower rate, prioritising smaller updates in the early stages, leading to more unique noised images throughout training.}
   \label{fig:scheduler_beta}
\end{figure}

In the original DDPM paper, \citet{ho2020denoising} made use of a linear schedule for the amount of noise added to the image at each $t$. Although the produced images competed with state-of-the-art methods, \citet{nichol2021improved} found that a linear schedule was not well suited to low-resolution images and proposed using a cosine-beta noise schedule as a replacement. By stretching the noise levels to add less noise in early iterations and more in the latter, the process overcame the issue of images becoming too noisy too quickly. \Cref{fig:scheduler_beta} shows how the scheduler impacts the $\beta$ parameter used in \cref{eq:diffusion_step}. \Cref{eq:beta_t} shows how $\beta_{t}$ is calculated (the reader is directed to \citealt{nichol2021improved} for specifics on its derivation):
\begin{equation}
   \centering
   \label{eq:beta_t}
   \beta_{t} = 1-\frac{\bar{\alpha}_{t+1}}{\bar{\alpha}_{t}}\;,
\end{equation}

\noindent where $\bar{\alpha}_{t}$ is shown in \cref{eq:alpha_bar}.

All models in this paper make use of the cosine-beta scheduler. The impact of this scheduler on the high dynamic range nature of astronomy data is explored in detail in \cref{sec:Astro_noise_vs_diffusion}.

\subsection{Conditioning the model}

Using the pipeline of DDPMs allows us to impose conditioning on our input space. Throughout training, the model is already conditioned on the timestep, $t$. This conditioning is crucial, as it informs the neural network of the noise level present in the input image, $x_{t}$. The network learns a single function, $\epsilon_{\theta}(x_{t},t)$, which is trained to predict the noise component, $\epsilon$, from a noised image, $x_{t}$, at any given timestep, $t$. The characteristics of the noise to be removed differ substantially between a timestep close to $T$ (where the image is almost pure noise) and a timestep close to $0$ (where the image is almost fully reconstructed). Providing $t$ as an input allows the single network to adapt its predictions accordingly.

However, we are not limited to only conditioning by timestep; often, multiple parameters are provided alongside each image. Typically, this is to allow `class-conditioning', whereby users can tune the expected output of the diffusion model to a specific class or category \citep{radford2021learning,zhang2023adding}. By providing an embedding space during training, the model can learn not only the dynamics of the noise but also the inherent differences between noise and non-noise pixel values for each of the class representations. This conditioning provides small nudges to the random walk, guiding the denoising to the respective population of the class. Providing rich representations to the model has led to the impressive text-to-image capabilities of popular diffusion-based implementations, such as Dall-E \citep{ramesh2022hierarchical} or Imagen \citep{saharia2022photorealistic}.

This type of conditioning allows the user to direct the image generation to a particular class or style of image. However, with diffusion's ability to cover large areas of the search space, a repeat sample of the same prompt will likely produce an image very different from the first, even if the generated image belongs to the requested class. Even with the possibility of detailed prompts and refined classes, which can guide the generation towards a narrower or more specific region of the learned space, it may be difficult to know exactly what the output image will look like.  This is due to the one-to-many relationship between classes or prompts and their subsequent generated outputs. 

However, for our task, we require a different form of conditioning. Instead of generating an entire image based on a class label, we need to reconstruct a specific, masked region of an existing image, conditioned on the surrounding, unmasked pixels. This task is known as inpainting~\citep{pathak2016context,yeh2017semantic,yu2018generative,wang2018image}. To achieve this, we do not train a new, specialised conditional model. Instead, we employ a `plug-and-play' (PnP) approach, whereby conditioning is introduced during the inference phase of an unconditionally trained generative model \citep{nguyen2017plug,graikos2022diffusion}. Our diffusion model, trained on a vast sample of galaxies, serves as a powerful, non-parametric prior for galaxy morphology. The PnP approach allows us to apply this prior to the inpainting task with great flexibility, as the model does not need to be retrained for different images or mask geometries. This is particularly advantageous for large-scale surveys like Euclid. Recent works in astrophysics have successfully employed similar PnP or posterior sampling strategies for complex inference tasks \citep{feng2023score,dia2025iris}.

\subsection{Repainting}
For this work, we used the Repaint algorithm \citep{lugmayr2022repaint}, a PnP method specifically designed for diffusion models. The core challenge in diffusion-based inpainting is to ensure semantic and structural consistency between the known (unmasked) and unknown (masked) regions. A naive approach of simply replacing the masked region with noise and then running the reverse process would fail, as the denoising network would alter the known, unmasked pixels, corrupting the conditioning information.

Repaint overcomes this by iteratively enforcing the data constraint at each step, $t$, of the reverse diffusion process. Given an image, $x_{0}$, and a binary mask, $M$ (where $1$ indicates the region to be inpainted and $0$ indicates the known region), the algorithm proceeds as follows. At each timestep, $t$, of the reverse process, starting from a noisy sample, $x_{t}$:

\begin{enumerate}
   \item Sample the known region: The original image, $x_{0}$, is noised to the appropriate level for timestep $t$ using the forward process, $q(x_{t}|x_{0})$, resulting in a sample of the known part, $x^\mathrm{known}_{t}$.
   \item Sample the unknown region: The output image from the previous, slightly noisier step, $x_{t+1}$, is denoised for one step using the unconditional diffusion model to get a sample of the unknown part, $x^\mathrm{unknown}_{t}$.
   \item Combine and denoise: The known and unknown parts are combined: $x'_{t} = (1-M)\odot x^\mathrm{known}_{t} + M\odot x^\mathrm{unknown}_{t}$. This composite image is then passed to the next iteration of the repaint loop to produce the estimate for step $t-1$. Once this image has been passed through one step of the learned reverse diffusion process $p_{\theta}(x_{t-1}|x'_{t})$, it will produce the slightly less noisy image $x_{t-1}$.
\end{enumerate}

This procedure, illustrated in \cref{fig:repaint_loop}, ensures that the generated region is coherently conditioned on the surrounding galaxy structure at every stage of the generation. The `denoise' operation in this context refers to the application of one step of the learned reverse diffusion process, $p_{\theta}(x_{t-1}|x_{t})$, which involves using the trained neural network, $\epsilon_{\theta}$, to predict the noise in $x_t$ and deriving an estimate for $x_{t-1}$. By denoising an image composed of both fixed and generated pixels -- each matched to the appropriate noise level $t$ -- we ensure that the generated region is properly conditioned on the information from the rest of the galaxy, resulting in a final image in which all pixels are consistent and coherent.

While alternative conditioning methods based on guiding the reverse process with a likelihood score function exist~\citep{graikos2022diffusion}, we adopted the Repaint algorithm for its straightforward implementation and proven effectiveness in enforcing data consistency in a PnP setting.

Incorporating the Repaint method for AGN identification has the following benefits:
\begin{itemize}
   \item Diffusion methods have proven their ability in many applications to accurately recreate complex datasets, allowing for the generation of realistic galaxy morphology images.
   \item A large imbalance of normal galaxies exists compared to AGN, allowing the model to learn a bias. Such a bias can be exploited to create an outlier detection-based classifier without explicitly requiring labels during training.
   \item There are no constraints on the size and shape of masks used, and since Repainting is only applied at inference, no additional retraining is required.
   
\end{itemize}

\section{Training and inference pipeline}
\label{sec:Training}

The diffusion and inpainting methods are two distinct pipelines, one for the initial training of the model and a second for the actual inpainting. Training the diffusion model uses the whole image and learns to predict what noise to remove at each timestep{, allowing for} novel examples to be generated from the trained distribution.

\subsection{Mask creation}
\label{sec:mask_creation}
Due to the nature of the application, our aim is to mask out any suspected AGN without the need to rely on labels ahead of time. Each mask is positioned so that the brightest pixel in the centre of the source becomes the centre of the mask. Although each image is centred on the source, various interactions and dynamics allow suspected AGN to be offset from the centre. To allow for some deviation, a 9$\times$9 pixel window around the centre of the image is used to detect the brightest pixel. This offset allows for the correct masking of less symmetric sources, whilst limiting the likelihood of masking an adjacent source by mistake. The masks for each image are assigned automatically throughout the data-loading pipeline.

Throughout this paper, we use a 5$\times$5 pixel mask, unless explicitly stated otherwise, centred on the calculated brightest pixel. A 5$\times$5 pixel mask, which covers 0\farcs5$\times$0\farcs5, allows us to effectively cover the AGN contribution, whilst minimising loss of the surrounding galaxy structure. Given that the VIS PSF has a full width half maximum (FWHM) of 0\farcs13 \citep{EuclidSkyOverview}, the AGN's light is spread over multiple pixels. Taking into account the redshift of the galaxy, an AGN within nearby sources ($z=0.01$) will cover 4.8 pixels, ensuring that we can fully encapsulate the core without unnecessary overlap of the host morphology. By retaining as much of the host galaxy information as possible for inpainting, we also reduce the risk of introducing artificial features into the reconstructed image.

\subsection{Reconstruction rescaling}
Unlike some image processing pipelines that rescale image data to a limited range, such as when working with JPG images that are binned to the standard 0--255 pixel scale range and converted to floats between 0--1 \citep{walmsley2020galaxy,EP-Aussel}, our approach utilises the raw values direct from the sensor as input to our network. As comparisons of individual pixels are being made on such a small mask, the relative differences become much more impactful and would likely be lost if values were binned.

One potential issue with this set-up is that the libraries used for training are optimised for images within the standard pixel range. Although this did not prevent the model from effectively learning the distribution, it does require a rescaling of the generated output as the pixels are mostly between $-$1 and $+$1 to improve training stability.

We experimented with simply normalising the input data so that all our images were in the range 0--1; however, due to the large dynamic range of the data, the faintest pixels ended up being $10^{-4}$. Given the previous discussions on how MSE does not penalise smaller errors effectively, the model could not accurately recreate sources, outputting a near-empty image of very low pixel values.

To allow the inpainted region to be meaningfully compared (e.g. via MSE) to the original pixel values in that masked region, a post-inference rescaling must be applied. The vast majority of pixels in the image (the unmasked region, which constitutes 99.5\% of a 64$\times$64 image when a 5$\times$5 central mask is used) correspond to known pixel values from the original astronomical image. We leveraged these known, unmasked pixels to perform a linear rescaling of the diffusion model's initial output (which is in a normalised range) back to the original image's dynamic range. This procedure ensures consistency in the unmasked regions and appropriately scales the newly inpainted pixels in the masked region. The following scaling can be applied instantly on output of the repaint procedure using each image's respective values to allow the non-masked pixels of the input and output images to be consistent:
\begin{equation}
\label{eq:output_scaling}
   \mathrm{output_{scaled}} = \mathrm{output} \times \mathrm{scale} + \mathrm{offset}\;,
\end{equation}

where
\[
\mathrm{scale} = \frac{\mathrm{max}(\mathrm{input_{nonmasked}})-\mathrm{min}(\mathrm{input_{nonmasked}})}{\mathrm{max}(\mathrm{output_{nonmasked}})-\mathrm{min}(\mathrm{output_{nonmasked}})}
\]

and
\[\mathrm{offset} = \mathrm{min}(\mathrm{input_{nonmasked}}) - \mathrm{scale}\times\mathrm{min}(\mathrm{output_{nonmasked}})\;.\]

By ensuring that the shared pixels are consistent, the inpainted pixels can also be correctly adjusted.

\begin{figure*}[!ht]
   \centering
   \includegraphics[width=\textwidth]{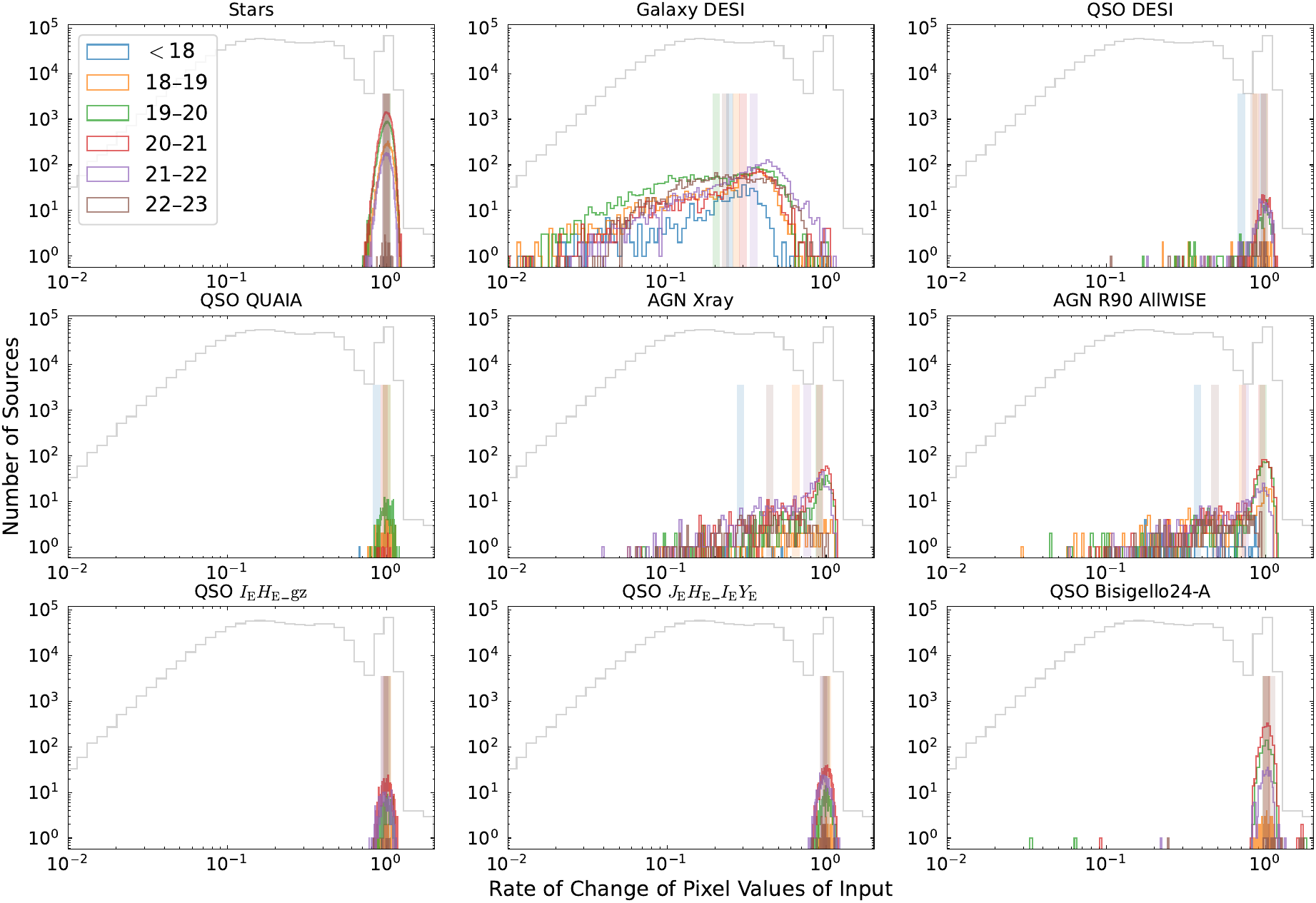}
   \caption{Initial results for pixel value differences across various selections. Comparing the ratios of the centre's brightest pixel with the means of the surrounding 1- and 2-pixel-wide regions shows a clear distribution difference between galaxy and non-galaxy classes. The grey histogram shows the distribution of the whole dataset, showing how the images not captured in these selections compare. The median value for each \IE magnitude bin is shown in the respective vertical line.}
   \label{fig:ratios_12}
\end{figure*}

\section{Data}\label{sec:Data}
\subsection{Sample selection}\label{sec:sample_selection}
Covering 63.1\,deg$^{2}$, the Euclid Q1 Data Release \citep{Q1cite} contains just under 30 million catalogued sources \citep{Q1-TP004} in the Euclid Deep Fields North (EDF-N), South (EDF-S), and Fornax (EDF-F). In this work, we use exclusively the VIS images (\Euclid \IE band) because we aim to identify AGN candidates on a single image. However, due to the varying distances and sizes of sources, an appropriate subset must be chosen to ensure the model is presented with a sufficiently broad selection of resolved galaxies. The following criteria are applied to the 30 million source catalogue of Q1:

\begin{itemize}
   \item The source must be detected in the VIS band (\texttt{vis\_det\_euclid$>$0});
   \item the source must have a \IE magnitude brighter than $22.5$ [\texttt{23.5 $-$ 2.5log(flux\_detection\_total\_euclid) $<$ 22.5}];
   \item the source must not have any defects or abnormalities as signified by the detection flags (\texttt{det\_quality\_flag\_euclid = 0} and \texttt{flag\_vis\_euclid = 0});
   \item the source must have a low probability of being spurious (\texttt{spurious\_prob\_euclid $<$ 0.2});
   \item the source must have a large segmentation area (\texttt{segmentation\_area\_euclid$>$200});
   \item the S/N of the total flux measurement must be sufficiently high ($\frac{\texttt{flux\_detection\_total\_euclid}}{\texttt{fluxerr\_detection\_total\_euclid}}$ $>$ \texttt{15}); and
   \item if the S/N criterion is not met, then the segmentation area of the source must be very large (\texttt{segmentation\_area\_euclid $>$ 1200}), as defined in \cite{Q1-SP047}.
\end{itemize}

It is crucial to note that because AGN are intrinsically rare compared to the general galaxy population, the training set does not need to be pre-filtered to remove them. The diffusion model, in learning the data distribution, naturally develops a prior that is heavily biased towards the `typical' light profiles of non-active galaxies. This inherent bias is the key mechanism that enables the subsequent outlier detection.

After this sample selection, we are left with 1\,142\,606 sources across the three Q1 fields. VIS image cutouts (64$\times$64 pixels) were created for each source. Problematic images, such as those at the very edge of tiles where a full cutout could not be created, were removed. This provided us with images and catalogue data for 1\,060\,864 sources. We then applied a 80\,:\,20 train and test split. Even though the inpainting and its results are a separate pipeline to the diffusion training, all of the decision boundaries created are based solely on the training set to prevent any potential data leakage. When showing classifier performance, all results are from the unseen 20\% of the data. This prevents any potential memorisation of the training images from impacting the results. However, in our analysis, no obvious differences have been found in the inpainting results between the two sets, showing that the model is not simply memorising the training set.

To explore the impact point-like sources have on training and the inpainting process, a second smaller selection was created that combined all the selections above with the following:

\begin{itemize}
   \item The source must have a low probability of being a star. (\texttt{phz\_star\_prob\_phz\_class} $<$ \texttt{0.1}); and
   \item the source must have a low probability of being point-like. (\texttt{point\_like\_prob\_euclid} $<$ \texttt{0.2}).
\end{itemize}
After removing problematic images as before, this smaller dataset contained 794\,624 sources.

\subsection{Comparisons with other classifications}\label{sec:comparing_selections}
Although our diffusion model is trained in an unsupervised manner without labels, it is essential to compare the distributions of metrics over different source selection criteria. This allows us to compare the performance of detecting particular sources and determine what sources or factors may hinder the pipeline's ability to classify objects sufficiently.

In particular, we use some of the sample definitions discussed in \citet{Q1-SP027}. A selection of stars are used from DESI \citep{DESI-2024AJ....168...58D} and \textit{Gaia} \citep{GaiaCollaboration_2023_2023A&A...674A...1G,GaiaCollaboration_2023_2023A&A...674A..41G}. The normal galaxy sample is also from DESI \citep{DESI-2024AJ....168...58D}. Comparisons are made to selections of AGN and quasars using \Euclid photometry with the $I_{\text{E}}H$\_$gz$ and $JH$\_$I_{\text{E}}Y$ selections from \cite{Q1-SP027} and the two-colour selection from \citet[][hereafter Bisigello24-A]{EP-Bisigello}. Additional AGN and QSO selections are used from DESI \citep{DESI-2024AJ....168...58D}, \textit{Gaia}/Quaia \citep{Storey-Fisher2024}, and the 90\% reliability WISE sample from \cite{Assef_2018_2018ApJS..234...23A}, which is referred to as R90 for the remainder of the paper. The reader is directed to \citet{Q1-SP027} for a comprehensive overview of each of the selections used throughout our paper.

In this work, when referring to completeness against a specific reference sample (e.g. a catalogue of selected AGN candidates), we mean the fraction of objects in that reference sample that are successfully identified by our method. This is due to the true astrophysical completeness being unknown and extremely difficult to measure from the observational data alone. Our context is equivalent to the standard machine learning metric of recall for the positive class, which is defined in \cref{eq:prec_rec}.

When developing a method for initial candidate selection, whereby the goal is to flag objects for follow-up or multi-wavelength analysis, high recall is often prioritised. The aim is to minimise the loss of potentially interesting or rare sources (i.e. reduce false negatives), even if this comes at the cost of lower precision (i.e. including a larger fraction of contaminants or false positives in the selected sample). These contaminants can often be filtered out through subsequent analysis stages or by incorporating additional data. Our primary goal here is to demonstrate the method's capability to recover a wide variety of known AGN and QSO types, making recall a key indicator of its potential utility in broad searches.

As is particularly true for AGN and QSOs, it is impossible to create a selection criterion that will identify a single population of objects with high completeness and high purity. Instead, we shall use the aforementioned samples, many of which are only candidates themselves, to assess the contamination and purity of our method to the extent that is possible.

\subsection{Data expectations}
\label{sec:Data_expectations}

As a baseline for our morphology-based approach, it is first necessary to verify that the \Euclid VIS images contain sufficient information to resolve the contrast between the compact central emission characteristic of an AGN and the broader light profiles of typical galaxies. We analysed the relationship between the brightest central pixel and its surrounding pixels to quantify a difference between normal galaxies and AGN. By producing ratios comparing the central pixel to 1 and 2-pixels-wide regions around the central pixel, we formed a rate of change (ROC) metric over sources, which is described by
\begin{equation}
   \label{eq:ROC}
   \mathrm{ROC} = R_{1}^{2} + R_{2}^{2}\;,
\end{equation}

\noindent where
\[
   R_{1} = \frac{\mathrm{Input}_\mathrm{max} - \mathrm{Input}_\mathrm{1s}}{\mathrm{Input}_\mathrm{max}}\\
   R_{2} = \frac{\mathrm{Input}_\mathrm{1s} - \mathrm{Input}_\mathrm{2s}}{\mathrm{Input}_\mathrm{1s}},
\]
and 
$\mathrm{Input}_\mathrm{1s}$ and $\mathrm{Input}_\mathrm{2s}$ are the mean values of the surrounding rings of 1 (total 8 pixels) and 2 pixels (total 16 pixels), respectively. $\mathrm{Input}_\mathrm{max}$ is the pixel value of the brightest pixel within the centre 9$\times$9 pixels.

Pixels surrounding the brightest pixel, which share values similar to the maximum, will produce a ROC value closer to zero, implying that the source does not feature a prominent active component. Sources featuring a single very bright pixel surrounded by fainter pixels will have a ROC value closer to its maximum of 2. The produced distributions of sources, analysed according to both \IE magnitude and various selection criteria (\cref{sec:comparing_selections}), are shown in \cref{fig:ratios_12}. As was expected, sources with a high-flux point spread function (PSF) light profile show the highest ROC values. This effect is seen in the star sample and the QSO-dominated selections (R90 WISE, Quaia, DESI-QSO, and Euclid-QSO). On the other hand, spectroscopically confirmed normal galaxies show very broad ROC distributions.

\subsection{Addressing the use of asinh for image scaling}

\begin{figure}
   \centering
   \includegraphics[width=\linewidth]{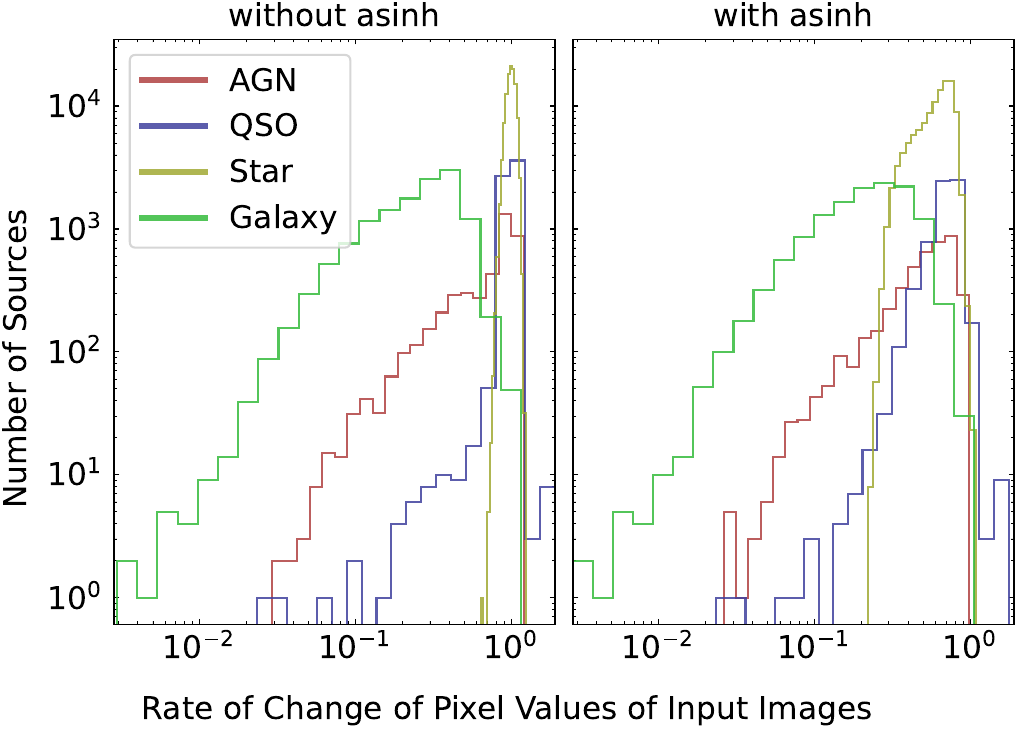}
   \caption{Distribution differences of the ROC value over each class. The narrow peaks from the non-galaxy class are significantly widened after applying the asinh transformation, causing a large overlap between classes.}
   \label{fig:asinh_diff}
\end{figure}

\begin{table}
\centering
\caption{Percentage of galaxies that have a ROC value higher than the mean ROC of the other classes.}
\label{tab:asinh_impact}
\begin{tabular}{ccc}
\hline\hline
\noalign{\vskip 1pt}
 Galaxy overlap & Without asinh& With asinh\\ 
 \hline
 \noalign{\vskip 1pt}
 AGN & 0.92\% & 3.98\% \\
 QSO & 0.13\% & 0.72\% \\
 Stars & 0.12\% & 1.80\% \\ \hline
 
\end{tabular}

\end{table}

Given its prevalence in galaxy morphology-based research \citep{lupton2004preparing}, the effect of using an $\asinh$ transformation on the data is also investigated. The function is often used to benefit visual inspections by reducing the difference between the brightest and faintest pixels. In our approach, however, when attempting to differentiate galaxy and non-galaxy sources, the $\asinh$ function becomes a hindrance, causing the distributions to more significantly overlap, as is shown in \cref{fig:asinh_diff}. To generate a measure for class overlap, we measured the ROC for each image, taking the mean of each non-galaxy class. We used the percentage of our true galaxy sample with a ROC value over the respective class mean as our overlap metric. The extent of the overlap is showcased in \cref{tab:asinh_impact}, where galaxy overlap increases by at least a factor of 4 and in the extreme case, nearly 15 times more galaxies overlap with stars.

In the following sections of the paper, we discuss the results of the diffusion model and analyse the performance of the inpainting pipeline that automatically learns to reconstruct the central emission expected for the normal galaxy population, allowing us to identify stars and AGN as outliers.

\begin{figure*}
   \centering
   \includegraphics[width=\linewidth]{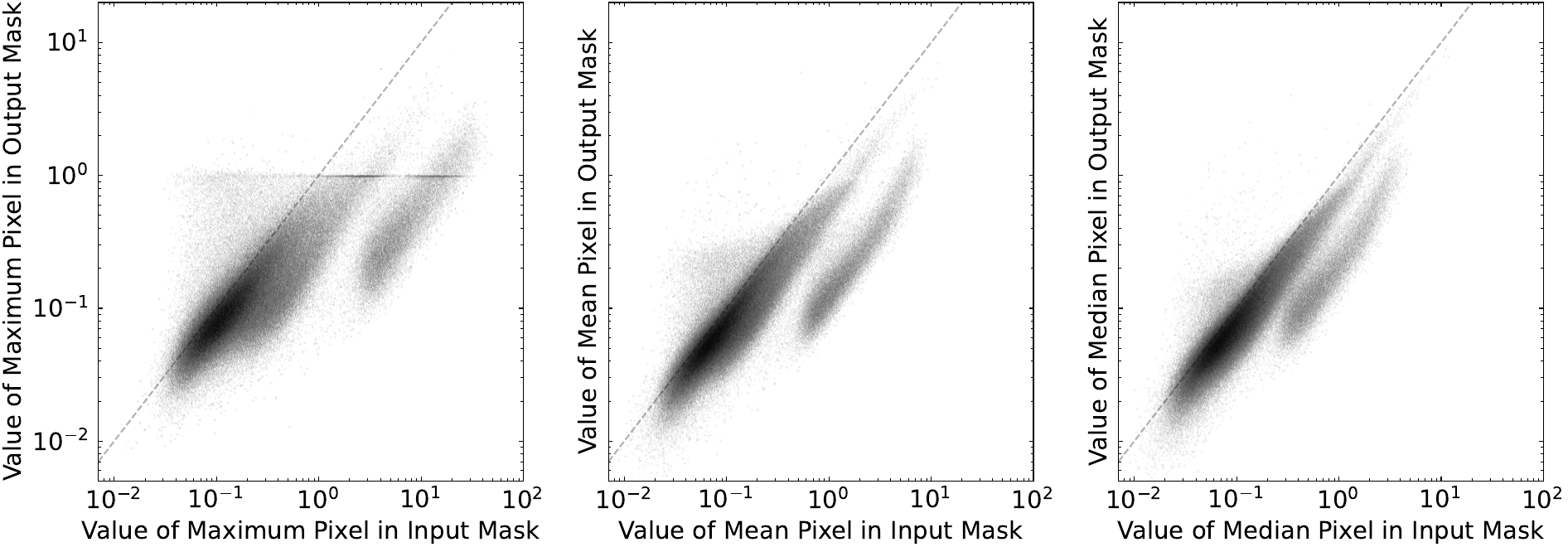}
   \caption{Comparison between the original masked pixels and the pixels of the generated output. The model's outputs demonstrated a consistent prediction along the gradient of $y = x$ (dashed line), with a bias for underpredicting the true value. A secondary cluster, also following the same gradient, is shown to reduce the pixel values by a factor of 10. This implies an inherent difference in the input image causing the model to behave differently. This collection of sources could be prime AGN candidates using our model.}
   \label{fig:performance_blanks}
\end{figure*}

\section{Results}

Although the focus of the paper is on the model's performance on small areas of a galaxy's core, Repaint's adaptability allows the pipeline to handle a variety of mask shapes and sizes. More general examples of the inpainting performance can be seen in \cref{fig:masks_output}. With larger masks, fewer pixels remain for conditioning, making the output more susceptible to artefacts. However, \citet{lugmayr2022repaint} demonstrate that this can be mitigated by increasing the number of resamples during the Repaint loop, at the cost of longer inference times.

\subsection{Inpainting of AGN}
For our inpainting, we used a 5$\times$5 pixel mask centred on the brightest pixel, following the constraints described in \cref{sec:mask_creation}. Comparing the expected pixel values with the generated outputs, \cref{fig:performance_blanks} shows the model's performance. The assumption that the model will predict a fainter core than the original does appear to be true, with a near-global bias for underpredicting, as is shown by the dashed $y=x$ line. The majority of sources behave in a consistent way, with the exception of a smaller cluster of sources that predict a core more than an order of magnitude fainter than the original. This distinct cluster could provide sufficient candidates for AGN.

The horizontal bar in the leftmost panel of \cref{fig:performance_blanks}, where the maximum output value appears to equal one, is a side effect of \cref{eq:output_scaling}. The values themselves are not exactly one, with $<$1\% of the data falling within $\epsilon=0.01$ of having a maximum output of one. There does exist a convergence point in \cref{eq:output_scaling}, where when the maximum and minimum pixel values of an input image equal one, the output of the rescaling always equals one; however, such an image would have all its pixels valued one, and so is not present in the data. Although no specific convergence point explains this trend, the majority of the images that fall on this line occur when the network outputs a maximum pixel value close to one. However, many more images have a network output close to one that is not close to this line, as well as images close to the line without a network output close to one. We have verified that both point-like and extended sources appear within this line; therefore, due to the low number of sources impacted, we leave more in-depth exploration to future work.

To further validate the model's output and show it is producing realistic outputs, \cref{fig:ROC_input_output} shows how the ROC metric compares between input and output images. The significant overlap, especially with respect to the peak of each distribution, shows that the model is accurately recreating the dynamics of the data.

\begin{figure*}[!ht]
   \centering
   \includegraphics[width=\linewidth]{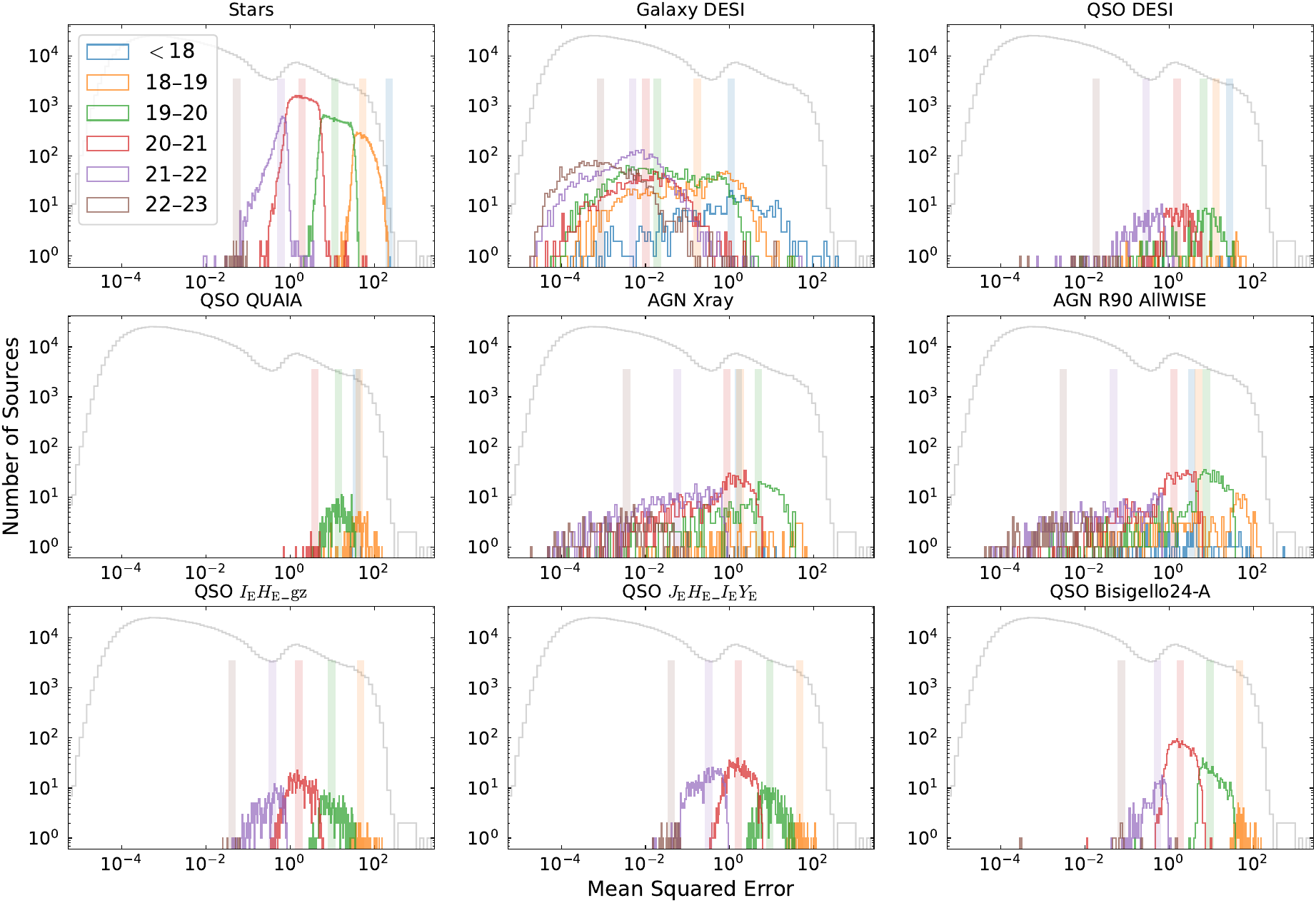}
   \caption{MSE of the 5$\times$5 pixel mask from the input and generated output. AGN selection show a skew toward higher errors, especially for lower-magnitude sources. The grey histogram shows the distribution of the whole dataset, showing how the errors of images not captured in these selections compare. The median value for each \IE magnitude bin is shown in the respective vertical line.}
   \label{fig:mse_all_mags}
\end{figure*}

\begin{figure*}[ht!]
   \centering
   \includegraphics[width=0.75\linewidth]{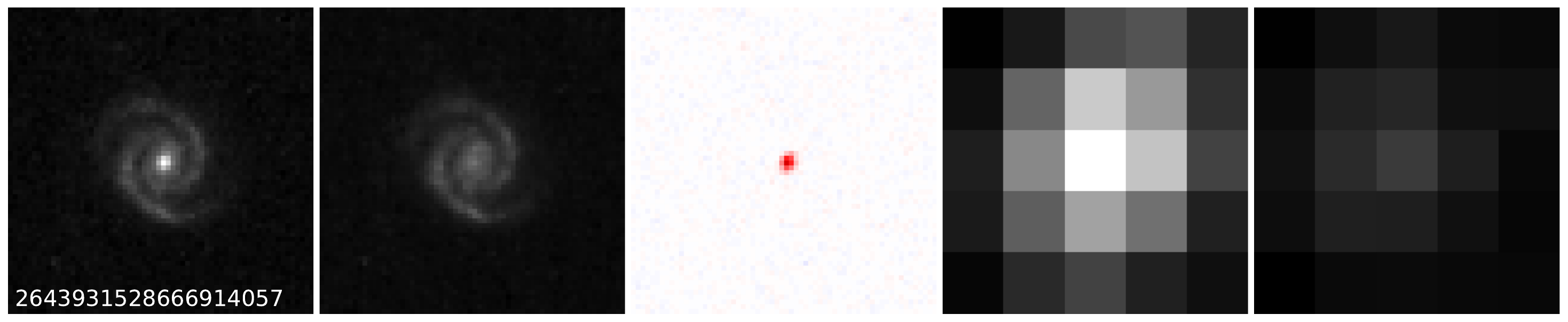}
   \includegraphics[width=0.75\linewidth]{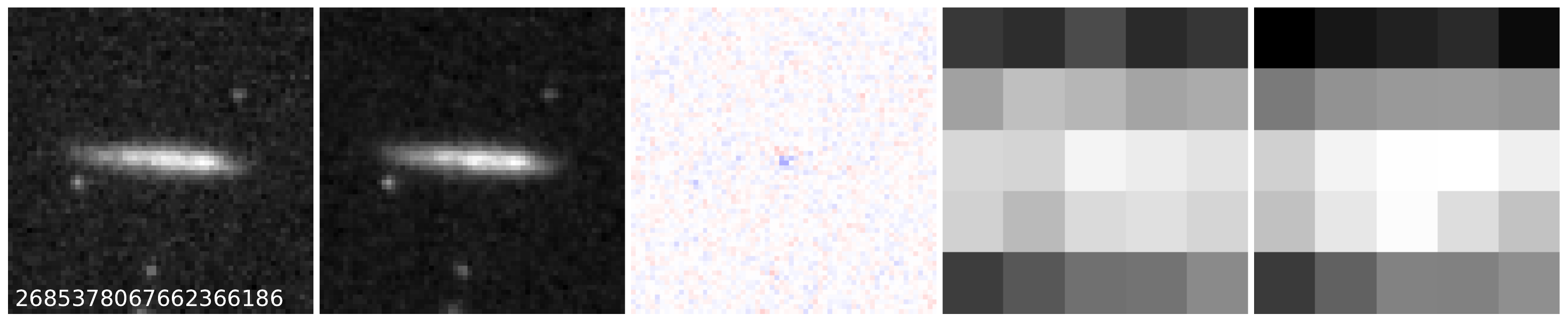}
   \includegraphics[width=0.75\linewidth]{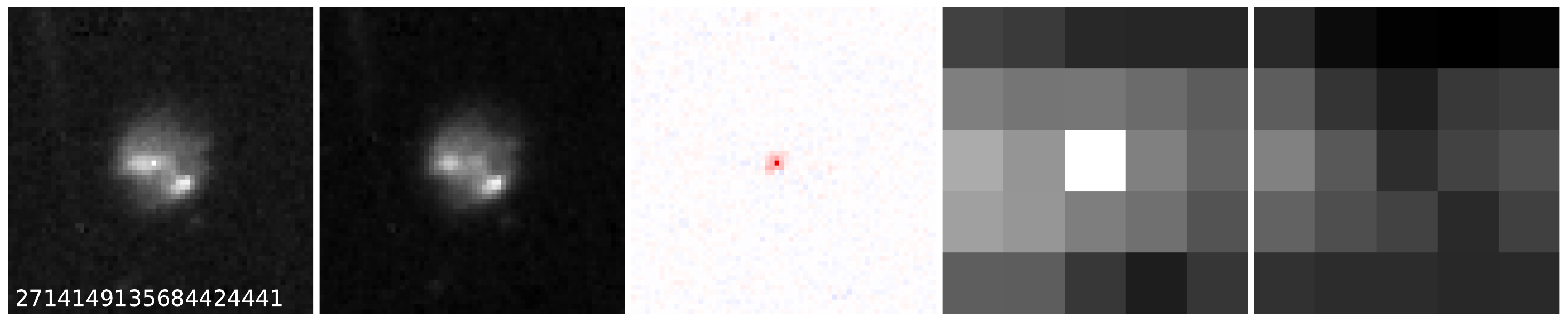}
   \includegraphics[width=0.75\linewidth]{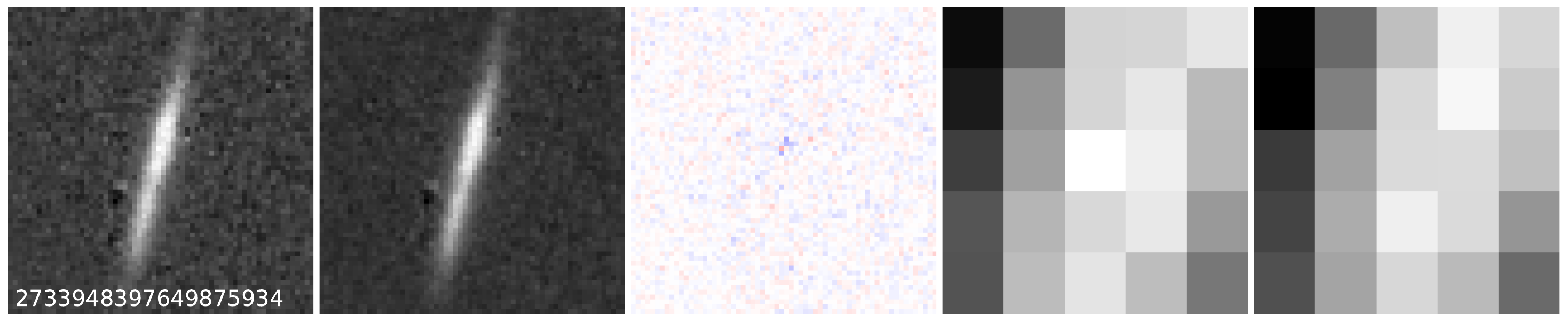}
   \includegraphics[width=0.75\linewidth]{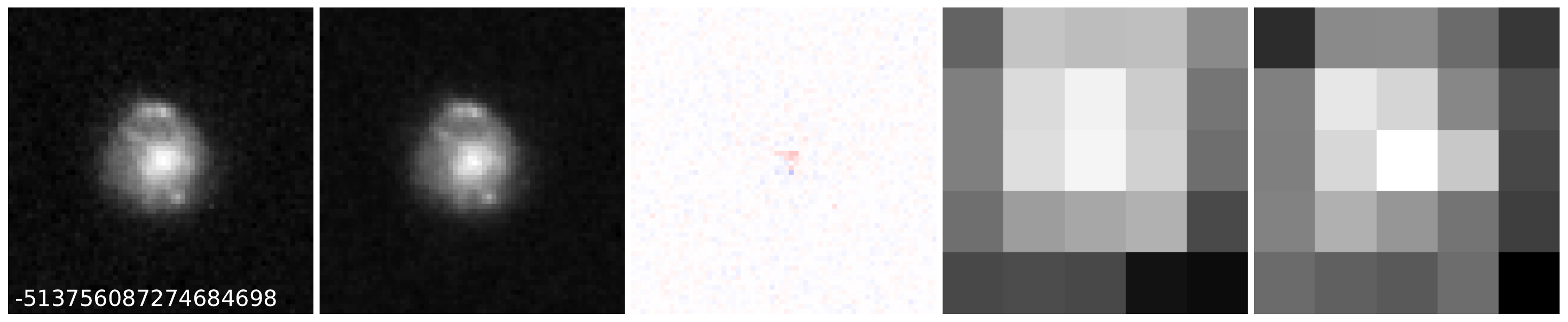}
   \includegraphics[width=0.75\linewidth]{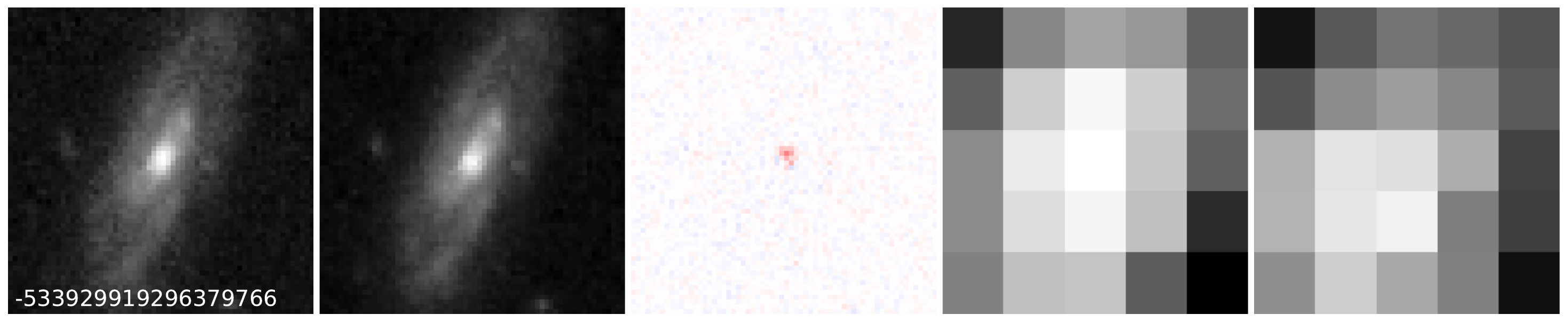}
   \includegraphics[width=0.75\linewidth]{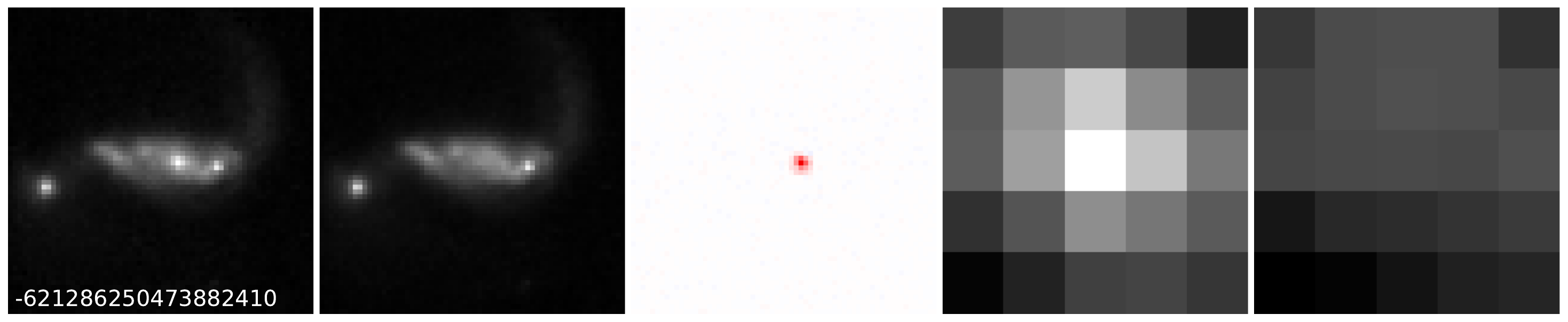}
   \caption{Example outputs from the repainting model. A 5$\times$5 pixel mask, centred on the brightest pixel within the central region of the input image, is repainted. The unmasked region is used to condition the generation of the new pixels. Each example shows (from left to right): the input image with its corresponding \Euclid ID; the output image (same scale as input); the residuals of the generated image; the masked pixels from the original image; and the inpainted pixels of the output image. The first three columns are the original 64$\times$64 cutout size, while the last two columns are the 5$\times$5 masked region.}
   \label{fig:outputs}
\end{figure*}

\begin{figure*}
   \centering
   \includegraphics[width=\linewidth]{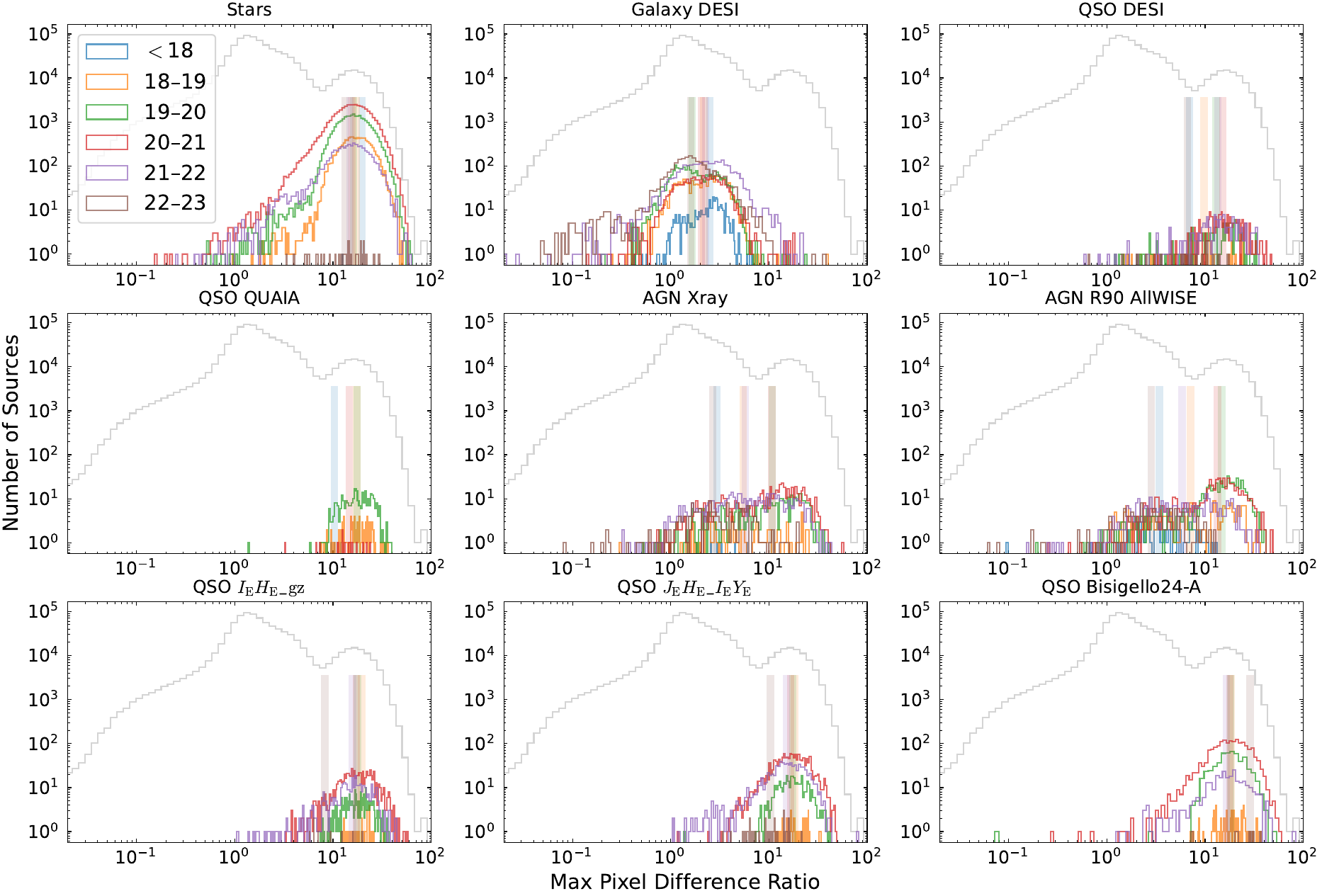}
   \caption{Measured ratio between the maximum pixel within the masked pixels of the input image and the max of the pixels in the output. The galaxy sample is the only sample that does not feature a right skew, allowing a threshold between this class and the rest. The brightness differences caused by the varying magnitudes do not significantly alter the ratios within each class selection. The grey histogram shows the distribution of the whole dataset, showing how the ratio of images not captured in these selections compares. The median value for each \IE magnitude bin is shown in the respective vertical line.}
   \label{fig:max_ratio_all_mags}
\end{figure*}

To analyse the reconstruction error for each subset of class candidates, comparisons were made between the original pixels and inpainted reconstruction for each image. The MSE of the pixel values are shown in \cref{fig:mse_all_mags}. Each subset was binned according to its \IE magnitude, providing a distinction between the relative brightness of each image.

There is a clear distribution difference between the respective class candidates. Brighter sources, particularly stars and luminous QSOs, produce higher errors, followed by fainter QSOs and the lowest-magnitude AGN candidates. Galaxies and the faintest AGN candidates show significantly lower errors, showing the model's tendancy to reconstruct sources without strong AGN features more effectively. This indicates that the model has learned the anticipated bias introduced by the uneven distribution within the dataset. While MSE is scale-dependent and influenced by the source's VIS \IE magnitude, the separation between the errors from the galaxy class and those of the other classes suggests it remains useful as a selection metric. The presence of low-error AGN candidates may also indicate contamination within the AGN labels.

It is acknowledged that because the model is trained on the full, unlabelled galaxy population, its prior distribution does contain information about AGN, even if they are rare. This could, in principle, lead the model to occasionally reconstruct AGN-like features, resulting in false negatives. However, the strong performance in separating known AGN and QSO populations, as is shown in \cref{fig:mse_all_mags}, suggests that the prior is sufficiently dominated by non-active galaxies for this outlier detection framework to be effective.

To showcase how the performance of the inpainting method impacts the overall image, \cref{fig:outputs} presents the input, output, and residuals for a selection of sources, including a close-up of the masked area both before and after inpainting. It is clear how bright central cores can be entirely removed with inpainting and replaced by regions consistent with the surrounding pixels. The model can also accurately recreate images that do not feature any significant core. This shows that the model has a good internal representation of galaxies and how they should behave. 

\begin{figure*}
   \centering
   \includegraphics[width=\linewidth]{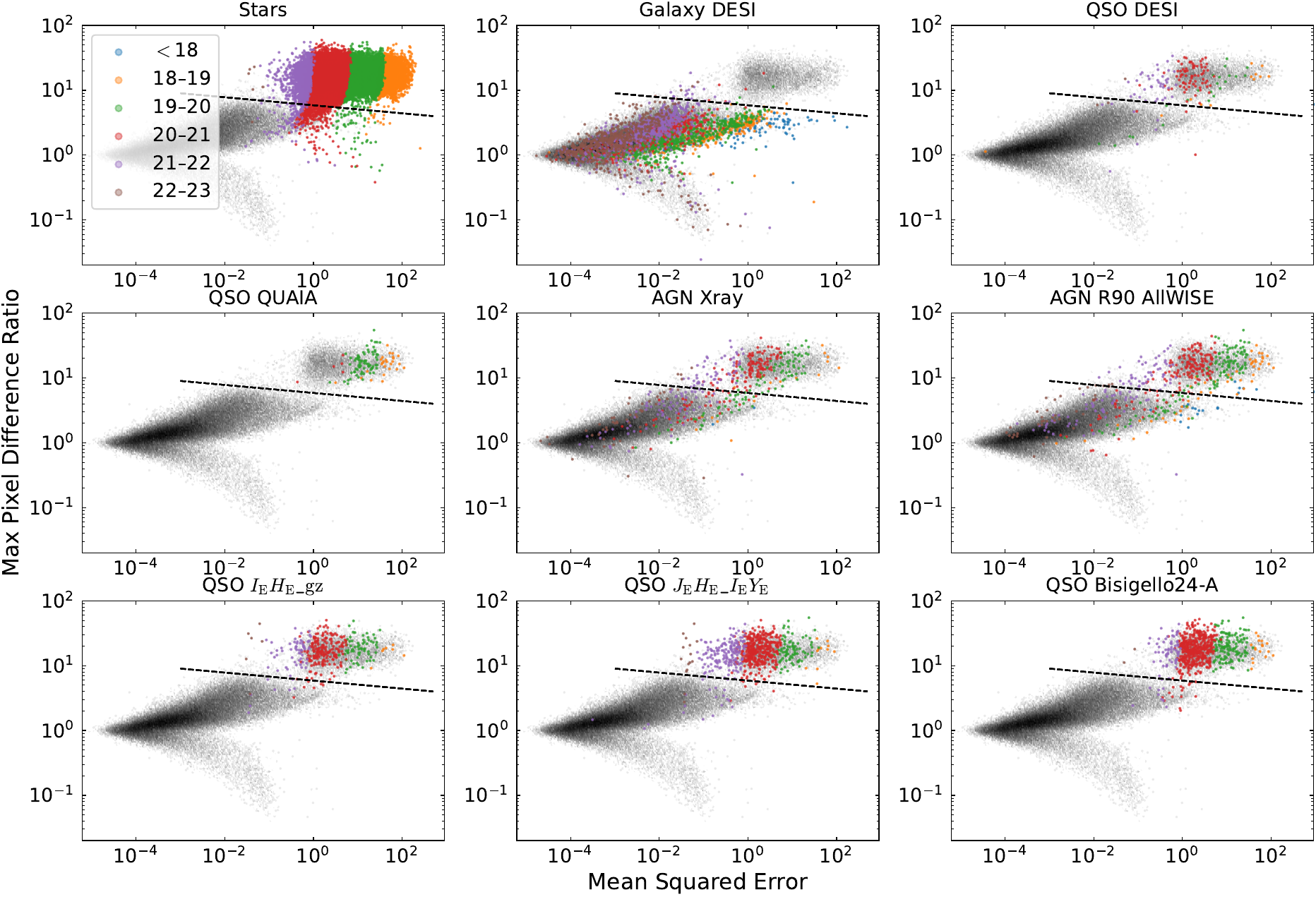}
   \caption{Threshold produced between the top right cluster of points and the largely galaxy-dominated cluster. The coloured dots represent the VIS \IE magnitude for each positive candidate for the respective selections. The grey dots show the values of all the images in the dataset. The majority of the sources within each non-galaxy class are captured with this simple linear boundary, allowing for high recall across all classes. The sources not captured are typically those of a higher VIS magnitude, and therefore fainter. The tail formed, where sources exhibit a maximum pixel difference ratio of $<10^0$, shows outlier sources with brighter predictions after inpainting. }
   \label{fig:scatter_mse_maxratio_train}
\end{figure*}

\subsection{Max pixel difference ratio}
\label{sec:max_pixel_difference}

Although the reconstruction error allows stars and a subset of AGN to stand out, a substantial overlap between suspected AGN and DESI-classified galaxies remains, especially at higher magnitudes. Even though measured on a pixel-by-pixel scale, MSE averages away specific pixel differences, leading to less nuanced differences between input and output. For example, if the reconstructed image ended up being the correct values but shifted by one pixel, this would be penalised heavily under MSE. From the perspective of this paper, such a scenario would represent a desirable outcome, as the model would correctly predict the presence of a bright, unresolved component, even if its exact location were shifted by a few pixels within the mask. Another example would be if only the brightest pixel were different in the reconstruction. The MSE would significantly reduce the impact of this measurement, potentially scoring similarly to a source that gave minor errors in every pixel within the mask.

To ensure such cases are not overlooked, we measured the pixel value ratio between the brightest pixel from the input mask and the brightest pixel from the reconstruction. As is shown in \cref{fig:max_ratio_all_mags}, the produced distributions show a much tighter spread of values for many of the AGN selections, with a skew towards higher scores. The majority of the galaxy subset produces a ratio of less than 10, whereas a large population of suspected AGN from each selection features a larger ratio.

Another important difference between this metric and MSE is how the results are relative, reducing the impact of magnitude on the recorded values. Many of the high-density peaks remain consistent, even as the magnitude of the sources increases. This implies that this metric is more robust at detecting fainter AGN than the MSE.

The distribution also highlights the scale of the extreme values, with many sources showing pixel values approaching 100 times greater than their counterparts. Exploring these large disparities can provide insights into the types of sources that are significantly sensitive to the model's capabilities. Sources producing the highest values for the ratio are those for which the reconstruction produces a cutout significantly fainter than the original image. As is seen in \cref{fig:outputs_removed}, this space is dominated by point-like objects. Having the sharp fall-off in pixel intensity in such a small area leads to the mask covering the majority of the source, if not the entirety. When the model attempts to reconstruct the missing pixels, it generates values that are coherent with the surrounding pixels. In these cases, the point-like source contributes very little to the surrounding pixels, leading to the generation of only background noise without any source.

The behaviour of the previous examples matches the model's expected behaviour and the paper's initial assumptions. Applying inpainting to this task is expected to produce outputs that are similar to, or produce a flattened light profile of, the original pixels. However, a subset of sources exists that exhibit the opposite behaviour and generate a much brighter collection of pixels. \Cref{fig:outputs_brighter} shows examples of such sources. When analysing the sources that exhibit this behaviour, we found that they have overwhelmingly low-S/N images. Further exploration of the impact of S/N on the diffusion process is shown in \cref{sec:Astro_noise_vs_diffusion}.

\begin{table}
   \centering
   \caption{Number of sources that produced brighter pixels in the masked area after repainting.}
   \label{tab:outputs_brighter}
   \begin{tabular}{ccc}
   \hline\hline
    \noalign{\vskip 1pt}
   Dataset& \multicolumn{2}{c}{Loss function} \\
    & Default & Normalised \\
   \hline
    \noalign{\vskip 1pt}
   All data& 55\,008 (6.49\%) & 33\,075 (3.90\%) \\
   
   No point-like & 52\,313 (6.17\%) & 39\,507 (4.66\%) \\
   \hline
   \end{tabular}

   \end{table}

The number of sources whose max pixel difference ratio is less than one for models trained on the two loss functions discussed in \cref{sec:Training_objective} is shown in \cref{tab:outputs_brighter}. The normalised loss significantly reduces the number of sources exhibiting this behaviour. The normalised loss trained on the complete dataset was selected as our applied model because it further reduced the number of erroneous predictions.

\renewcommand{\creflastconjunction}{, and }

\subsection{Morphology-based performance}

Given the variety of shapes and sizes of objects that will be witnessed using \Euclid, it is important to verify that the model can handle the complexities of different galaxy morphologies. Utilising the refined morphology classifications provided from \citet{Q1-SP047}, we provide examples of edge-on galaxies (\texttt{disk-edge-on\_yes\_fraction} $>0.5$), spirals (\texttt{has-spiral-arms\_yes\_fraction} $>0.7$), and mergers (\texttt{merging\_merger\_fraction} $>0.3$) in \cref{fig:morphology_edge_on,fig:morphology_spiral,,fig:morphology_merger}, respectively.

These images show how our diffusion model and the inpainting pipeline can effectively adapt to different morphologies. For edge-on in particular, the region of bright pixels within the masked area is often narrower than the mask itself. Despite this, the inpainting correctly preserves the orientation of the features rather than uniformly filling the entire masked area with bright pixels.

\subsection{Creating a classifier}

Due to the difference in distributions between the two metrics, any non-overlapping cases can be found when used in combination. As is seen in \cref{fig:scatter_mse_maxratio_train}, there is a cluster of sources in the top right. The key aspect of this cluster is its clear boundary from the DESI galaxy selection. This is vital for ensuring high reliability in classifications. We created a simple decision boundary between the two clusters of points, as is shown in \cref{fig:scatter_mse_maxratio_train}. Although a more fine-tuned boundary would be possible, especially when incorporating a secondary machine-learning model, it is important to show the benefits of utilising the inpainting method through only its error metrics. The equation of our linear boundary is $y=5.8x^{ -0.06}$.

\subsection{Variance of reconstructions}
Given the stochastic nature of the diffusion model pipeline, it is important to verify that any produced results can remain consistent if applied to the wider dataset. Although the nonmasked region of the images will remain fixed over repeat samples, the random walk performed in the inpainting procedure could lead the samples to a different area of the search space. This would result in a different, but possibly equally plausible, output image.

\begin{figure}
   \centering
   \includegraphics[width=\linewidth]{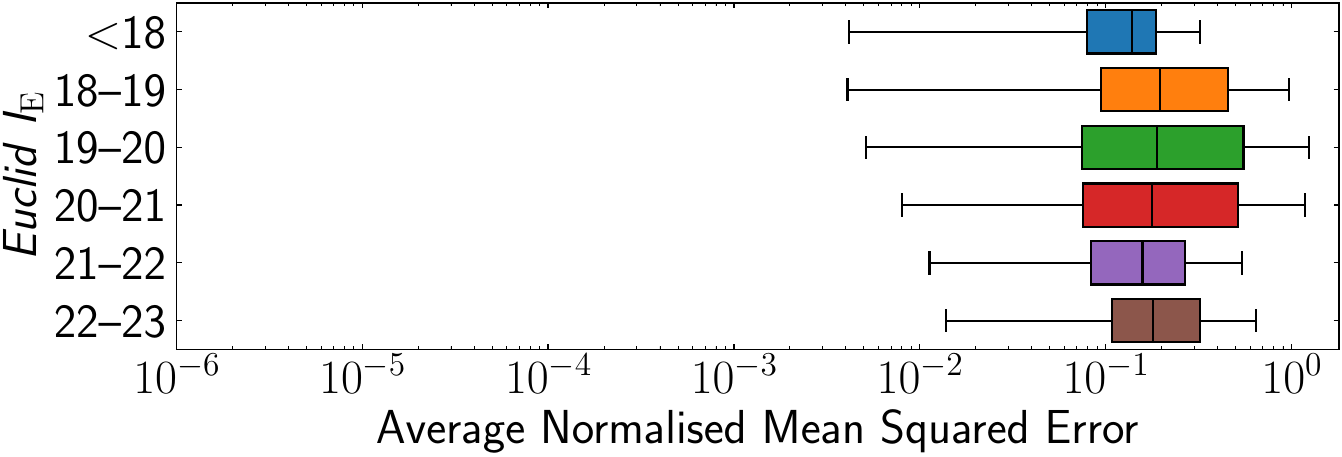}

    \vspace{3mm}
   
   \includegraphics[width=\linewidth]{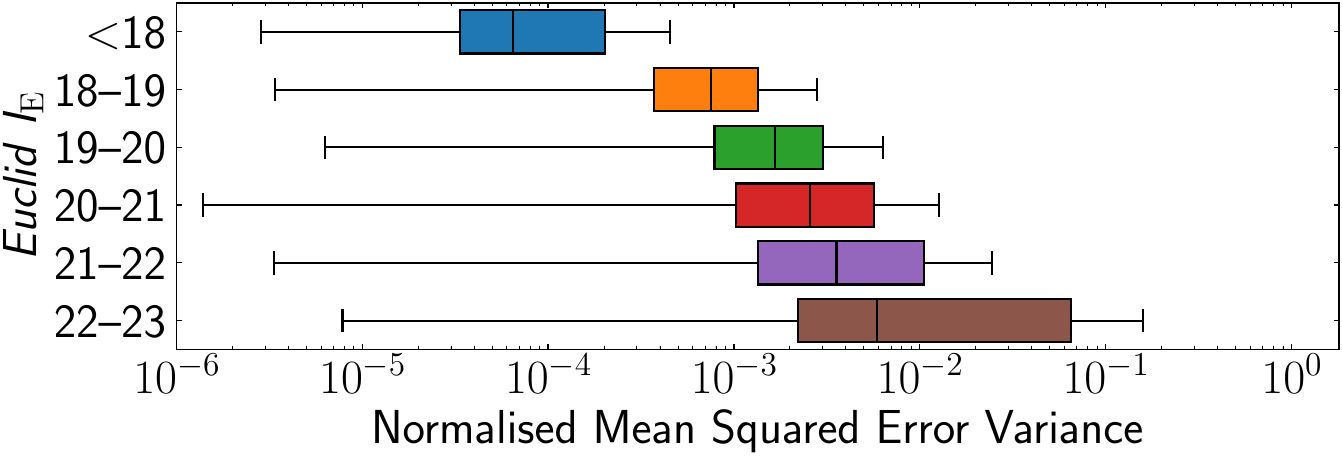}
   
   \caption{Variations in the output error from repeat samples of the same fixed pixels. Over seven separate inpainting runs, the model shows a consistent median error when normalised by each respective magnitude (top panel). The variance in the output errors is less consistent with the brighter, lower magnitudes, which provide more robust inpainting than those of fainter images.}
   \label{fig:boxplots_mse}
\end{figure}

As part of a single application of the pipeline, seven samples are generated. To ensure that any low variance in samples is not caused by the model simply `remembering' the true image from its training examples, all resampling is applied on images the model has never seen before. The mean and variance over a normalised-by-brightness MSE are shown in \cref{fig:boxplots_mse}.

The top figure shows that, irrespective of magnitude, the model achieves a similar relative median error across all images, highlighting the impact of the normalised loss throughout training. However, the difference in the scheduler over images of different quality and brightness is evident in the bottom figure. The consistency of generated outputs varies drastically as fainter (and therefore typically lower S/N) images produce widely different pixels when resampled.

\begin{figure*}[!ht]
   \centering
   \includegraphics[width=\linewidth]{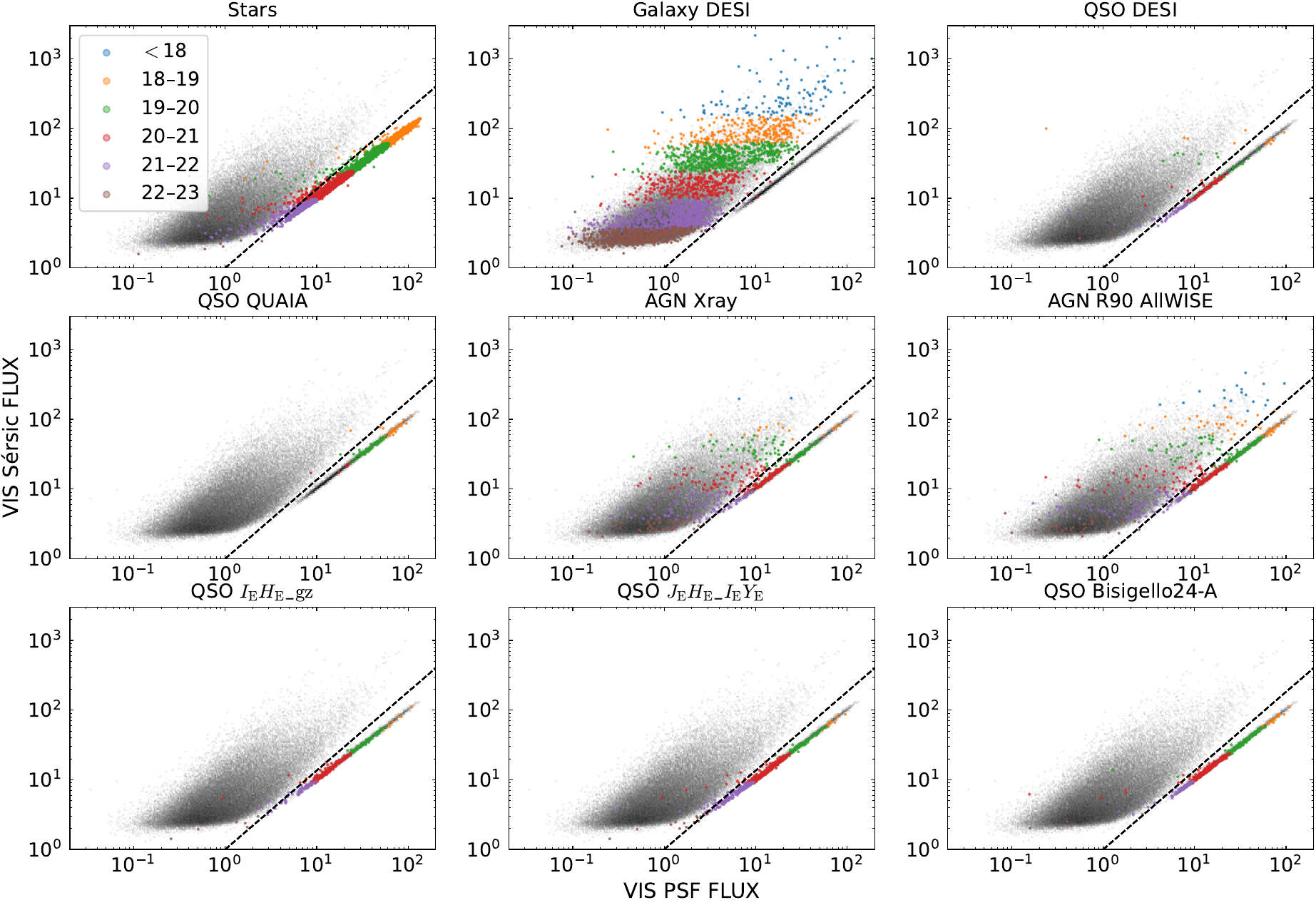}
   \caption{Comparing the PSF total flux with the Sérsic flux provides a clear separation between the galaxy class and the majority of the non-galaxy selection. This allows for a simple linear boundary for classifying potential AGN, with the majority being point-like objects. However, it is most apparent with selections from AllWISE R90 and X-ray-selected AGN that many sources would not be selected using these features. The coloured dots represent the VIS \IE magnitude for each positive candidate for the respective selections. The grey dots show the scores of all the images in the dataset. }
   \label{fig:scatter_psf_sersic_flux_train}
\end{figure*}

\begin{figure*}
   \centering
   \includegraphics[width=\linewidth]{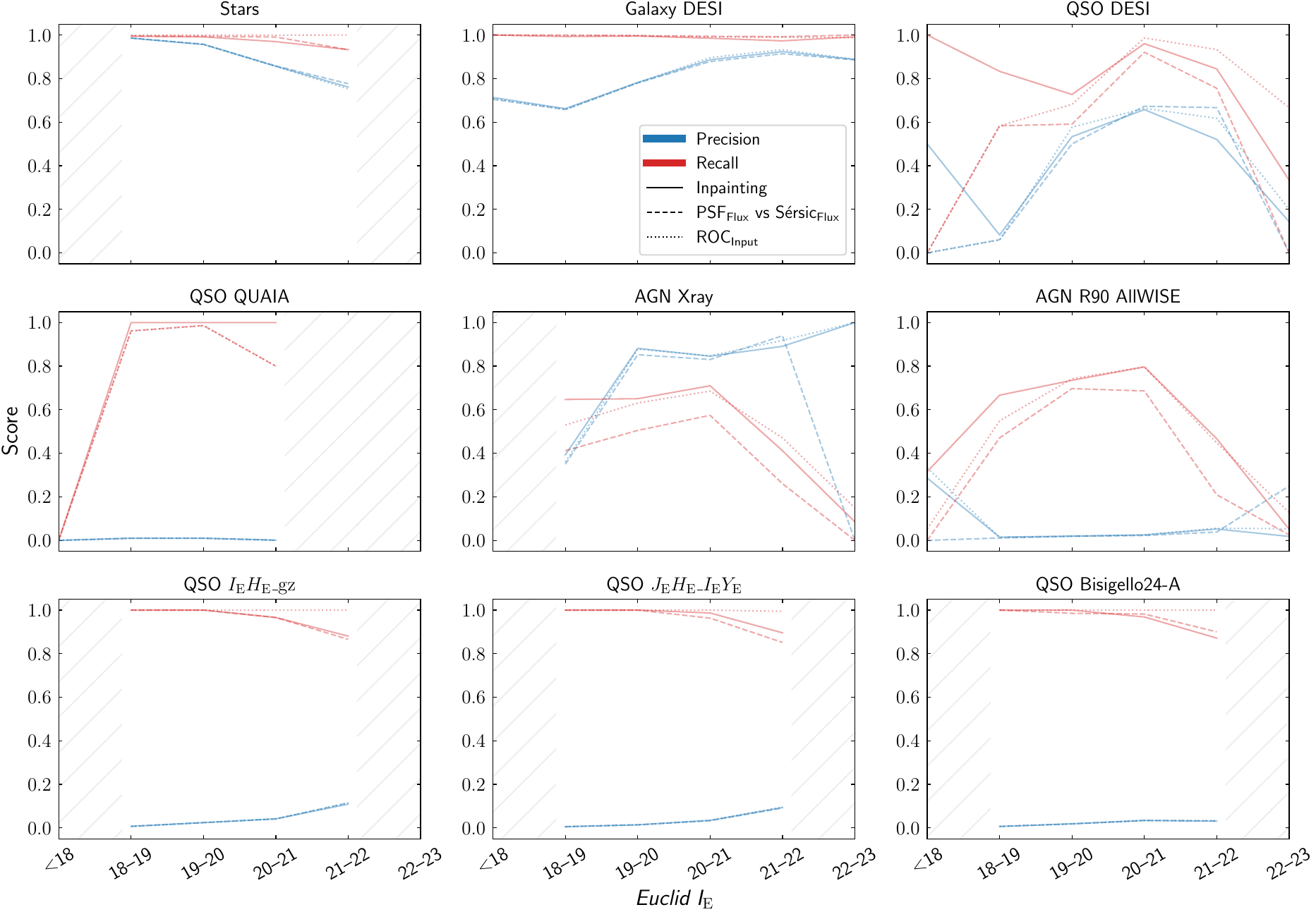}
   \caption{Precision and recall scores for the diffusion-based AGN predictions (solid line). As the various selections we compare against are not complete, sufficiently high recall allows us to see which traditional methods our selection overlaps with, indicating the types of sources our method is most appropriate for. } 
   \label{fig:scores_test}
\end{figure*}

\subsection{Comparing to other classifiers}
\label{sec:Scores_comparisons}

It is important to compare our AGN and QSO predictions to more traditional methods that are also able to utilise only the image data.

\subsubsection{Sérsic and PSF flux measurements}

To benchmark our non-parametric inpainting method against a more traditional approach, we constructed a classifier based on parametric modelling of galaxy light profiles. We utilised morphological parameters sourced directly from the Euclid MER Q1 catalogue \citep{Q1-TP004}. Specifically, we used the provided \texttt{flux\_vis\_sersic\_euclid} and \texttt{flux\_vis\_psf\_euclid} measurements.
These parameters were derived by the MER pipeline using the SourceExtractor++ software \citep{2020ASPC..527..461B,2020ASPC..527...29K}. For each detected source, the pipeline performs two separate and independent model fits. The \texttt{flux\_vis\_psf\_euclid }value represents the total flux obtained by fitting a model of the instrumental PSF to the source image. The \texttt{flux\_vis\_sersic\_euclid} value represents the total flux from fitting a two-dimensional Sérsic profile, which is designed to model the extended light distribution of a galaxy.
The underlying assumption for a classifier based on these parameters is that sources with a dominant, unresolved central component (such as a QSO or a Type-1 AGN) will be described well by a PSF model, resulting in a high \texttt{flux\_vis\_psf\_euclid} value. Conversely, normal resolved galaxies will be better described by the extended Sérsic profile, leading to a high \texttt{flux\_vis\_sersic\_euclid} value. By comparing the relative strength of these two flux measurements (see \cref{fig:background_noise_snr_levels}), we can construct a simple linear classifier to distinguish between point-like and extended sources. This baseline method is computationally inexpensive as it relies on pre-computed catalogue values, but it is fundamentally limited to this simple morphological distinction and may struggle with identifying AGN in complex host galaxies or at lower relative luminosities.

The distribution of sources with these parameters is shown in \cref{fig:scatter_psf_sersic_flux_train}. The equation of this linear boundary is $y=x^{1.13}$.

\subsubsection{ROC metric}

Our second comparison is with the original ROC metric used in \cref{sec:Data_expectations}. As a clear distinction between the galaxy class and the others is visible, it is important to verify how such a metric performs as a classifier. This is especially true given that such a metric can be generated from the raw image, and no machine-learning model or precomputed values are required. The boundary for this classifier is $x=0.75$.

\subsubsection{Performance against other candidates}

Given the variety and differences in the selection methods, there is an inherent incompleteness and potential for contamination. Therefore, our analysis focuses primarily on achieving sufficiently high recall. For an initial candidate selection method, such as the one proposed here, the primary goal is often to compile a large, inclusive sample for potential follow-up observations or more detailed multi-wavelength analysis. In this context, high recall is typically prioritised over high precision, which measures the purity of the selected sample. The scientific cost of missing a potentially interesting source (a false negative) is often considered higher than the operational cost of including a contaminant (a false positive) that can be filtered out in subsequent analysis stages. Therefore, our primary goal is to demonstrate the capability of our model to achieve high recall across a wide variety of AGN and QSO candidates which have been selected via different, more conventional methods and physical mechanisms.

Precision and recall are defined as follows:
\begin{equation}
   \label{eq:prec_rec}
   \qquad \mathrm{precision} = \frac{\mathrm{TP}}{\mathrm{TP}+\mathrm{FP}}
   \qquad \mathrm{recall} = \frac{\mathrm{TP}}{\mathrm{TP}+\mathrm{FN}}\;.
\end{equation}
For a given classification task (e.g. identifying DESI QSO objects):
\begin{itemize}
   \item TP (true positives) is the number of objects that belong to the class and are correctly predicted as belonging to it by the model.
   \item FP (false positives) is the number of objects that do not belong to the class but are incorrectly predicted as belonging to it by the model.
   \item FN (false negatives) is the number of objects that belong to the class but are not predicted as belonging to it by the model.
\end{itemize}

The test set scores for each respective selection are shown in \cref{fig:scores_test}, with the collated results for stars, galaxies, AGN, and QSOs shown in \cref{fig:scores_test_combined}. All predictions are made from images that were neither used during training nor used in the creation of any of the decision boundaries for the classifiers.
Due to differing constraints for each selection, we split the analysis to allow for a more representative comparison.

\subsubsection{Galaxies}
Our galaxy selection was determined by the DESI spectral type classification \texttt{SPECTYPE=GALAXY} \citep{DESI-2024AJ....168...58D}, with AGN contaminants removed following the procedure detailed in \citet{Q1-SP027}, hereafter called \citetalias{Q1-SP027}. Negative instances were formed from the remaining spectroscopically determined labels \texttt{SPECTYPE=STAR} and \texttt{SPECTYPE=QSO} \citep{DESI-2024AJ....168...58D} along with the contaminants removed from the original \texttt{SPECTYPE=GALAXY} selections.

The predictions from each classifier for whether a source is a normal galaxy are simply the inverse of the boundary's original predictions. As \cref{fig:scores_test} shows, all classifiers capture nearly all galaxies across every magnitude bin. This behaviour is expected, as the selection criteria used by each classifier are explicitly designed to favour AGN-like features, thereby excluding typical galaxies by construction. The precision scores also match extremely closely between classifiers, indicating that similar contaminants prove difficult, especially for the brighter sources. Such contaminants can be seen in \cref{fig:scatter_mse_maxratio_train,fig:scatter_psf_sersic_flux_train}, where stars and DESI-selected QSO fall outside the boundaries.

The ability to separate normal galaxies from other extragalactic or stellar objects benefits both groups of data. Being able to create a pure galaxy sample is vital for the precise measurements needed for cosmology, as well as to further our understanding of dark matter and dark energy, a primary object of the \Euclid mission \citep{EuclidSkyOverview}.

\subsubsection{Stars}
The star selection is made up of the positive instances of \Euclid selected (\texttt{star\_candidate\_euclid}, \citetalias{Q1-SP027}), DESI-selected \citep[SPECTYPE=STAR,][]{DESI-2024AJ....168...58D}, and \textit{Gaia}-selected stars. Due to the assumptions made for this method, the incidental capture of stars within our classifications is an expected outcome. The near-perfect recall shows that our classifier is able to capture stars well. This performance is beneficial when attempting to capture a high-purity galaxy sample, as was discussed in the previous section. However, as other works such as \citet{Q1-SP027} have shown, there are metrics to mitigate the contamination of stars. As a significant number of stars are likely to remain in our sample, for all subsequent selections we applied the cut \texttt{phz\_star\_prob\_phz\_class}$< 0.3$ to predictions of all classifiers to minimise the contamination of stellar sources.

\subsubsection{\Euclid-detected QSOs}
The \Euclid photometric-based selections for QSOs include $I_{\text{E}}H$\_$gz$ and $JH$\_$I_{\text{E}}Y$, as is described in \citetalias{Q1-SP027}, and the Bisigello24-A 2-colour selection \citep{EP-Bisigello}. This subset considers only sources with \texttt{MUMAX\_MINUS\_MAG} $\leq -2.6$, a parameter that indicates point-likeness \citep{Q1-TP004}. As these selections have yet to go through a follow-up validation, they remain candidates and as such should not be seen as the complete ground truth.

For the $I_{\text{E}}H$\_$gz$ and $JH$\_$I_{\text{E}}Y$ selections, we capture around 90\% of the candidate QSOs. This coverage remains consistent across the available magnitudes. The performance compared to the other classifiers is similar, with ROC capturing a small percentage more sources, but not different enough to be significant.

The two-colour selection proved harder to match for all classifiers, with recall dropping to 0.6 for the faintest objects. Again, no significant differences are seen between the classifiers.

\subsubsection{DESI-selected QSOs}
As these sources are assumed to be point-like, applying the above \texttt{MUMAX\_MINUS\_MAG} limit would be reasonable. However, due to the improvements in spatial resolution from DESI, at about $1\arcsec$, to \Euclid's now 0\farcs2, some sources that appeared point-like through DESI now appear extended. Therefore, to ensure all potential sources were found, we did not apply the restriction to this subset.

This selection provides the first evidence of significant differences in the suggested candidates for each classifier. The inpainting classifier is able to capture a significantly higher selection of lower-magnitude sources compared to the other classifiers. Throughout all magnitude bands, the PSF and Sérsic flux classifier has worse recall. The ROC classifier matched more candidates in the fainter sources.

\subsubsection{\textit{Gaia}-detected QSOs}
For the \textit{Gaia}-detected QSOs, we used the Quaia selection \citep{Storey-Fisher2024}. The subset consists of positive instances (\texttt{qso\_quaia}, \citetalias{Q1-SP027}), with negative instances from \textit{Gaia}-detected galaxies (\texttt{in\_galaxy\_candidates\_gaia}, \citetalias{Q1-SP027}) and stars (\texttt{star\_candidate\_gaia}, \citetalias{Q1-SP027}). We again applied the \texttt{MUMAX\_MINUS\_MAG} $\leq -2.6$ restriction to the sources.

After the initial magnitude bin, our classifier matched with a high proportion of the Quaia-selected QSOs. The other two classifiers achieved equal recall scores, but in the faintest detected sources our classifier was able to match with an additional 20\% of sources.

The particularly poor precision for this selection across all the classifiers is due to the considerable imbalance between positive and negative predictions. With only 0.5\% of the \textit{Gaia}-detected sources being Quaia candidates, selecting more than just the candidates is very likely, significantly impacting the precision.

\subsubsection{AllWISE R90 AGN}
Utilising WISE-detected sources allows us to verify our ability to select candidates that may have been obscured by dust. Using the selection criteria derived by \citet{Assef_2018_2018ApJS..234...23A}, we compared our candidates to the 90\% reliability (R90) selection, a high-purity AGN sample.

The \citet{Assef_2018_2018ApJS..234...23A} selection provides a more significant test for the classifiers due to no restrictions on point-like sources, as well as fewer contaminants that may have unintentionally improved the overlap of candidates. Similarly to the DESI-selected QSOs, our classifier shows a higher coverage than the other decision boundaries on the brightest sources. In the fainter sources, the diffusion and ROC classifiers are matched. Given the purity of the R90 selection, a large sample of suspected AGN remained unmatched by any of the classifiers, with the most matched magnitude bin achieving 80\%.

Much like the \textit{Gaia}-detected QSOs, the small fraction of candidates compared to the full sample of possible sources leads to many of the classifiers selecting sources outside the R90 selection. With a completeness of only 17\% \citep{Assef_2018_2018ApJS..234...23A} there will be many mid-infrared AGN in the negative sample, which the classifiers are likely finding.

\begin{table*}
\centering
\caption{Recall, precision, and area under the curve (AUC) scores for each classifier.}
\label{tab:all_scores}
\begin{tabular}{cccccccccccccc}
\hline\hline
 \noalign{\vskip 1pt}
& & & \multicolumn{3}{c}{Recall} & & \multicolumn{3}{c}{Precision} & & \multicolumn{3}{c}{AUC} \\
Selection & 18 $<$ \IE $<$ 22.5 & & Ours & PSF--S & ROC & & Ours & PSF--S & ROC & & Ours & PSF--S & ROC\\
\hline
 \noalign{\vskip 1pt}
\multirow{ 2}{*}{Stars} & $<20$ & & 0.99 & 0.99 & 1.00 & & 0.96 & 0.97 & 0.96 & & 0.50 & 0.51 & 0.50 \\

& $\geq20$ & & 0.97 & 0.98 & 1.00 & & 0.84 & 0.85 & 0.84 & & 0.51 & 0.53 & 0.50 \\
 \noalign{\vskip 6pt}
\multirow{ 2}{*}{Galaxy DESI} & $<20$ & & 1.00 & 1.00 & 1.00 & & 0.74 & 0.73 & 0.73 & & 0.64 & 0.63 & 0.64 \\

 & $\geq20$ & & 0.98 & 1.00 & 0.99 & & 0.90 & 0.90 & 0.91 & & 0.73 & 0.72 & 0.75 \\

 \noalign{\vskip 6pt}

\multirow{ 2}{*}{QSO DESI} & $<20$ & & \textbf{0.77} & 0.57 & 0.63 & & 0.18 & 0.14 & 0.15 & & \textbf{0.84} & 0.75 & 0.77 \\

 & $\geq20$ & & 0.90 & 0.84 & \textbf{0.96} & & 0.59 & \textbf{0.67} & 0.62 & & 0.93 & 0.91 & 0.96 \\
 \noalign{\vskip 6pt}
\multirow{ 2}{*}{QSO QUAIA} & $<20$ & & 1.00 & 0.98 & 0.98 & & 0.01 & 0.01 & 0.01 & & 0.59 & 0.58 & 0.58 \\

 & $\geq20$ & & \textbf{1.00} & 0.80 & 0.80 & & $<$0.01 & $<$0.01 & $<$0.01 & & \textbf{0.51} & 0.41 & 0.40\\

 \noalign{\vskip 6pt}

\multirow{ 2}{*}{AGN X-ray} & $<20$ & & 0.64 & 0.48 & 0.61 & & 0.74 & 0.73 & 0.74 & & 0.63 & 0.59 & 0.62 \\

 & $\geq20$ & & 0.51 & 0.38 & 0.53 & & 0.86 & 0.86 & 0.87 & & 0.53 & 0.51 & 0.55 \\

 \noalign{\vskip 6pt}

\multirow{ 2}{*}{AGN R90 AllWISE} & $<20$ & & \textbf{0.68} & 0.57 & 0.63 & & 0.02 & 0.02 & 0.02 & & 0.68 & 0.63 & 0.66 \\

 & $\geq20$ & & 0.59 & 0.45 & 0.60 & & 0.03 & 0.02 & 0.03 & & 0.75 & 0.68 & 0.75 \\

 \noalign{\vskip 6pt}

\multirow{ 2}{*}{QSO $I_{\text{E}}H$\_$gz$} & $<20$ & & 1.00 & 1.00 & 1.00 & & 0.02 & 0.02 & 0.02 & & 0.50 & 0.50 & 0.50 \\

 & $\geq20$ & & 0.94 & 0.92 & \textbf{1.00} & & 0.05 & 0.05 & 0.05 & & 0.49 & 0.47 & 0.50 \\

 \noalign{\vskip 6pt}

\multirow{ 2}{*}{QSO $JH$\_$I_{\text{E}}Y$} & $<20$ & & 1.00 & 1.00 & 1.00 & & 0.01 & 0.01 & 0.01 & & 0.50 & 0.50 & 0.50 \\

 & $\geq20$ & & 0.95 & 0.91 & \textbf{1.00} & & 0.04 & 0.04 & 0.04 & & 0.50 & 0.46 & 0.50 \\

 \noalign{\vskip 6pt}

\multirow{ 2}{*}{QSO Bisigello24-A} & $<20$ & & 1.00 & 0.99 & 1.00 & & 0.02 & 0.02 & 0.02 & & 0.50 & 0.50 & 0.50 \\

 & $\geq20$ & & 0.96 & 0.97 & 1.00 & & 0.03 & 0.03 & 0.03 & & 0.50 & 0.50 & 0.50 \\
\hline
\end{tabular}
\tablefoot{
Each selection is split into brighter and fainter sources, \IE $< 20$ and \IE $\geq 20$, respectively. Bold scores indicate significant improvements over the other classifiers (an increase of at least 0.05). The $\mathrm{PSF}_{\mathrm{FLUX}}$ versus\ S\'{e}rsi$\mathrm{c}_{\mathrm{FLUX}}$ results have been shortened to PSF--S.
}
\end{table*}

\subsubsection{X-ray-selected AGN}
We compared our candidates to the X-ray selection from the work by \citet{Q1-SP003}. The positive instances are when the calculated probability that the source is a galaxy is sufficiently low (\texttt{Gal\_proba\_roster} $<0.2$, \citetalias{Q1-SP027}). The negative instances are sources with \texttt{Gal\_proba\_roster} $\geq0.2$. Only sources with detection in X-rays are included in the results.

The main difference with the results of the X-ray-detected AGN and the other selections is that the precision outperforms the recall across nearly every magnitude bin. This is due to the low number of X-ray sources available. The recall performance between the diffusion model and the ROC classifier is very similar, with the PSF and Sérsic flux classifier again struggling to match as many candidates as the other two. 

Each of the selection criteria we compare captures distinct subsets of the AGN and QSO populations, influenced by specific observational biases and methodological constraints. Therefore, it is only when collating these selections and identifying trends and overlaps in the labels that we can be confident that we sufficiently cover the spectrum of possible AGN.

\cref{tab:all_scores} summarises the scores shown in \cref{fig:scores_test}. We combine the results into two \IE magnitude bins 18--20 and 20--22.5, separating into brighter and fainter sources, respectively. Due to similar scores in many of the selections, we highlight in bold any scores that outperform the other classifiers by at least 0.05.

The results show that the diffusion method achieves high recall scores across all tested selections, most notably outperforming the other classifiers in the bright QSO DESI, faint QSO QUAIA, and bright R90 selections. In other selections, it performs comparably to the ROC classifier, with both classifiers often significantly outperforming the $\mathrm{PSF}_{\mathrm{FLUX}}$ versus \ S\'{e}rsi$\mathrm{c}_{\mathrm{FLUX}}$ classifier.

\subsection{Decomposition of the AGN component}
\label{sec:decomposition}
Given our model's ability to classify suspected AGN, it is not unreasonable for the reader to assume that the residuals of the repainting process and the output image can independently represent the AGN and galaxy components. Although utilising inpainting for decomposition is a worthwhile future research direction, the validity of the residuals in accurately representing the AGN component for further scientific analysis has not been tested and, as such, cannot be used. A more significant exploration of which mask to use will also be vital for such a use case. A mask that adapts its size and shape for each input image will likely be necessary. For a deep-learning approach applying AGN decomposition to Euclid data, readers are referred to \citet{Q1-SP015}, hereafter called \citetalias{Q1-SP015}, whose method provides AGN fraction ($F_{\text{AGN}}$) measurements that we use to analyse our AGN candidates.

The subset of sources from the Q1 catalogue~\citep{Q1-TP004} used within this work and the subset tested in \citetalias{Q1-SP015} feature an overlap of 425\,781 sources. The confusion matrix for AGN predictions between the diffusion classifier and sources with an $F_{\text{AGN}} \geq 0.2$ (AGN candidate for \citetalias{Q1-SP015}) are shown in \cref{tab:fAGN_confusion}. The majority of sources ($\sim95\%$) are agreed by both methods not to be AGN candidates. For the remaining sources, only $\sim10\%$ are agreed upon, with over five times more AGN candidates selected by \citetalias{Q1-SP015}. When comparing all the AGN candidates selected using their respective subsets of data, this number reduces down to 3.5 times more candidates selected by \citetalias{Q1-SP015} than the diffusion model (\cref{tab:fAGN_confusion2}).

\begin{figure}[!t]
   \centering
   \includegraphics[width=\linewidth]{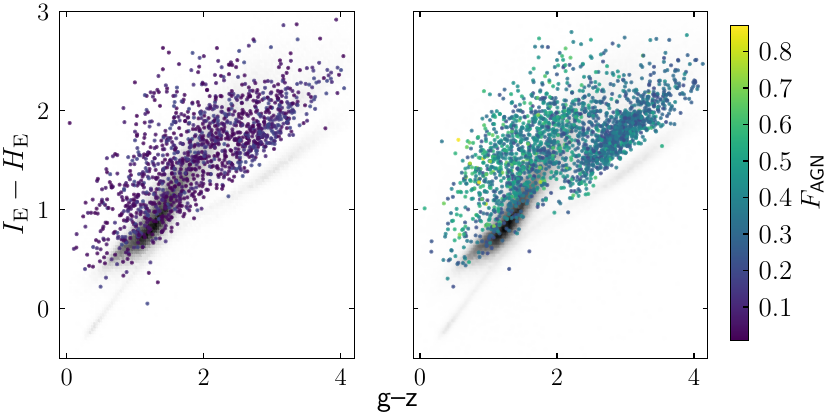}
   \caption{Diffusion model AGN candidates, colour-coded according to the AGN fraction of \citetalias{Q1-SP015}. \textit{Left:} Not identified as AGN by \citetalias{Q1-SP015} ($F_{\text{AGN}} < 0.2$), and \textit{right:} AGN candidates selected by \citetalias{Q1-SP015} ($F_{\text{AGN}} \geq 0.2$).}
   \label{fig:agn_fraction}
\end{figure}

The left panel of \cref{fig:agn_fraction} shows the AGN candidates selected by the diffusion classifier but predicted to have a low AGN fraction by the model in \citetalias{Q1-SP015}. The right panel shows the AGN fraction of sources that are agreed upon between the two methods. \cref{fig:fAGN_mse_max_ratio} shows where the \citetalias{Q1-SP015} selections are located on the inpainting metrics. As is shown in \cref{tab:fAGN_confusion}, many of the \citetalias{Q1-SP015} sources are not located above the diffusion threshold. Of the sources that lie above the threshold, the majority show an $F_{\text{AGN}} > 0.5$. This implies that the $F_{\text{AGN}}$ selection is more sensitive to fainter AGN.

\begin{table}
\centering
\caption{AGN candidate comparison between this work and the AGN fractions predicted by \citetalias{Q1-SP015}.}
\label{tab:fAGN_confusion}
\begin{tabular}{ccc}
\hline\hline
 \noalign{\vskip 1pt}
 Matched sources & \citetalias{Q1-SP015} AGN & \citetalias{Q1-SP015} non-AGN \\ \hline
 \noalign{\vskip 1pt}
 Diffusion selected & \phantom{0}2\,263 & \phantom{00}1\,331 \\
 Diffusion non-selected & 19\,758 & 402\,429 \\ \hline
 
\end{tabular}

\end{table}

\begin{table}
\centering
\caption{Total number of AGN candidates from each morphology method and their respective datasets.}
\label{tab:fAGN_confusion2}
\begin{tabular}{ccc}
\hline\hline
 \noalign{\vskip 1pt}
 Method & Diffusion & \citetalias{Q1-SP015} \\ \hline
 Total AGN candidates & 16\,053 & \phantom{0}57\,874 \\ \hline
 
\end{tabular}

\end{table}

\begin{figure}[!t]
   \centering
   \includegraphics[width=\linewidth]{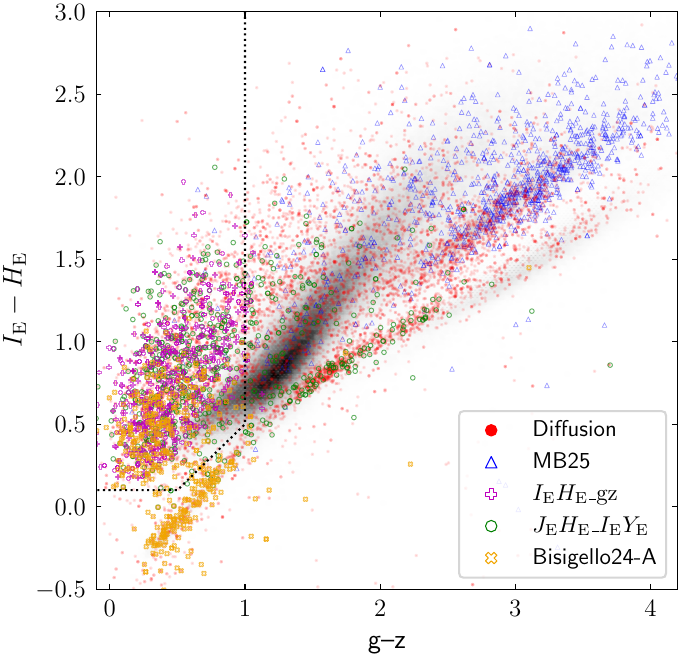}
   \caption{Comparison of our diffusion model selections with those from photometry-based selections \citep[\citetalias{Q1-SP027};][]{EP-Bisigello} and the selections from \citetalias{Q1-SP015}. Although all the diffusion selected sources are plotted, to improve clarity, only a representative subset of the other selections is shown. The photometry-based selections are limited to much bluer sources, highlighted by the dotted line showing the cut introduced in \citetalias{Q1-SP027}.}
   \label{fig:selections_compare}
\end{figure}

Comparing the predictions for each of the \Euclid-based selections in \cref{fig:selections_compare}, we can see that our diffusion model covers areas of the colour space that the photometry-based selections do not. Being able to select both bluer galaxies captured by the photometry selections \citep[\citetalias{Q1-SP027};][]{EP-Bisigello} and the much redder sources captured by \citetalias{Q1-SP015} shows the versatility of the method, highlighting its effectiveness across different galaxy populations.

\begin{figure*}[!ht]
   \centering
   \includegraphics[width=\linewidth]{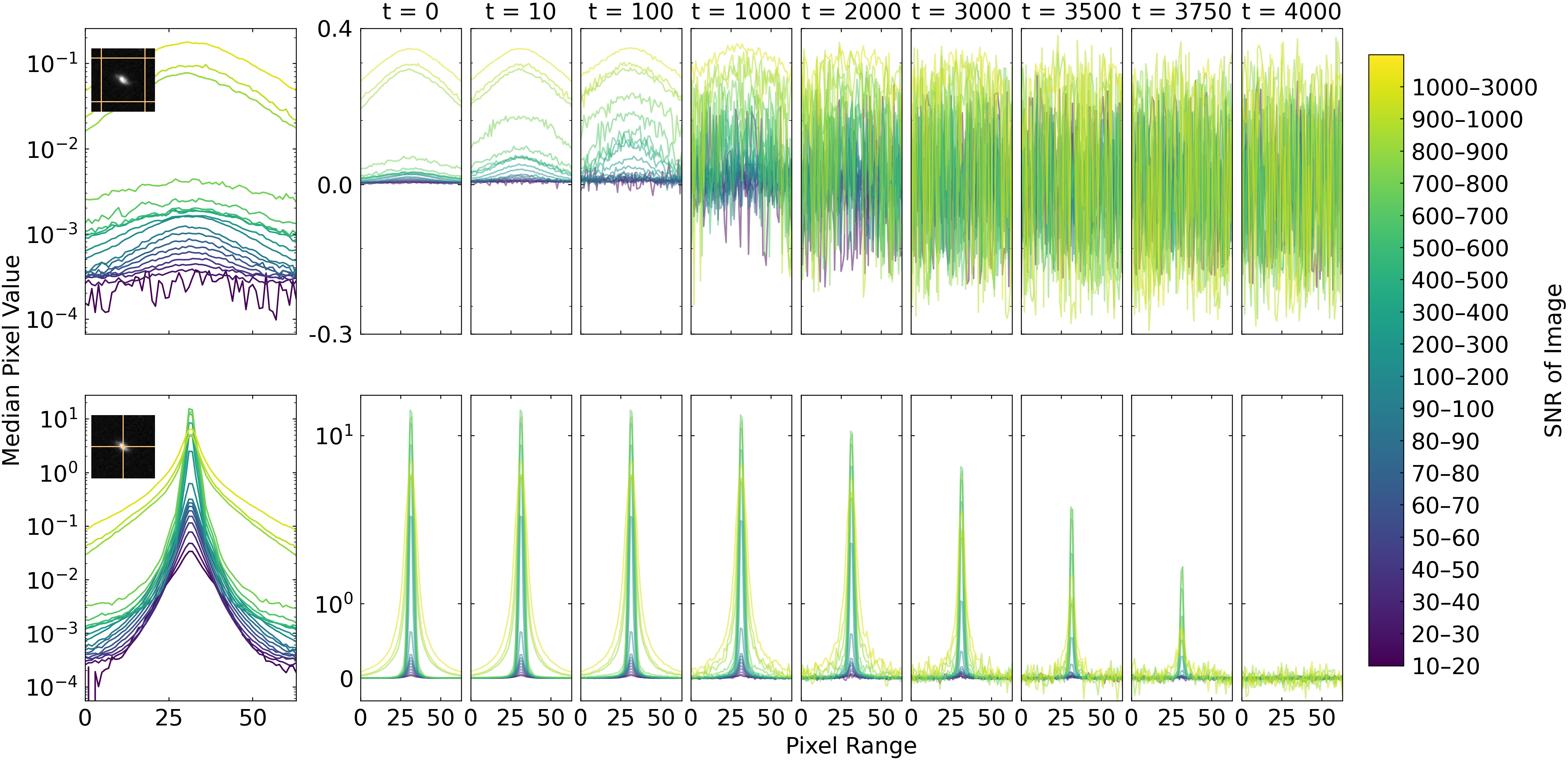}
   \caption{Impact of the diffusion noise scheduler on the median pixel values of the VIS images. The pixels covered in the median pixel calculation are shown as orange regions overlaid on the example source image in the leftmost plots. The top row shows how the additional noise impacts areas of the image that are more typically background noise, whereas the bottom row focuses on pixels that will be heavily influenced by the source. The binning of images according to their S/N shows how different levels of image clarity are affected by the gradual increase in introduced noise. Images with higher S/N can retain more of the source detail for more timesteps in the scheduler.}
   \label{fig:background_noise_snr_levels}
\end{figure*}

\begin{figure*}[!ht]
   \centering
   \includegraphics[width=\linewidth]{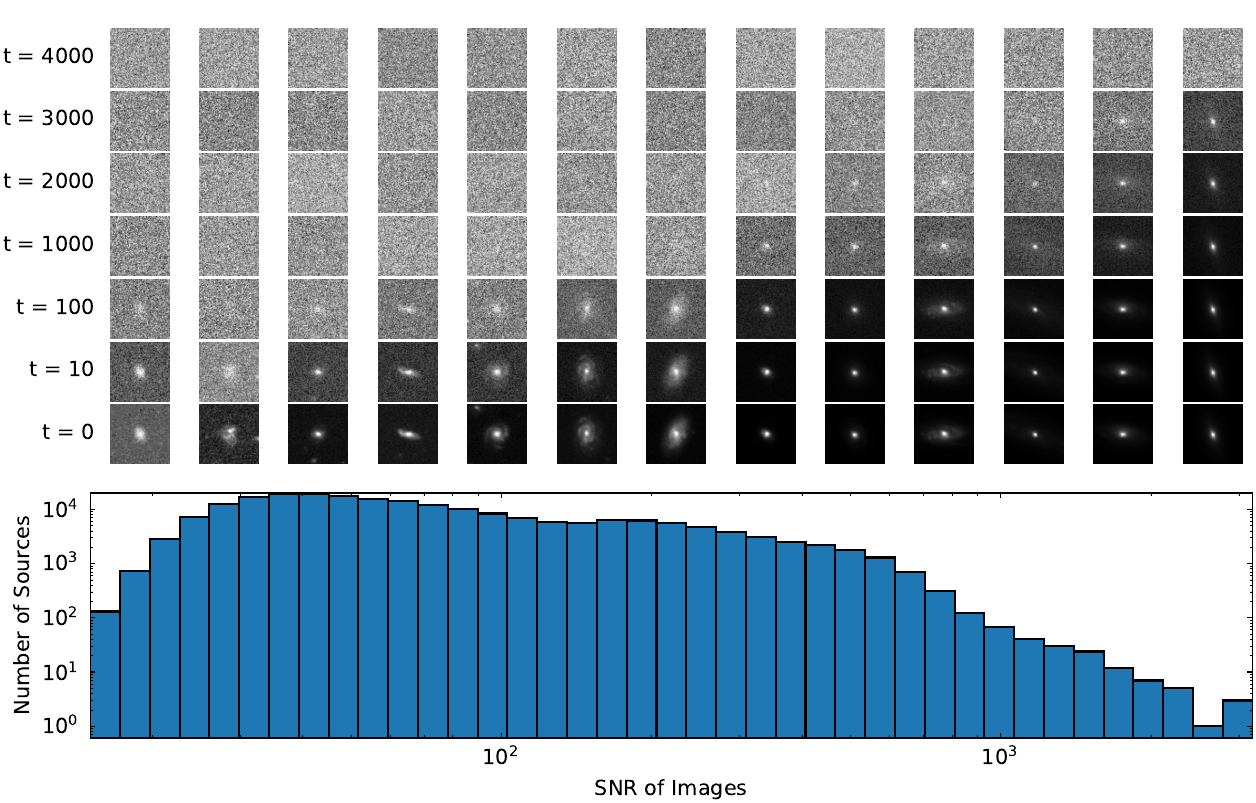}
   \caption{\textit{Top:} Respective noised images produced by the cosine-beta schedule at different timesteps. Each image is a sample from the respective S/N bin directly below it. Due to the scales of pixel values, the introduced noise has a more significant impact on the typically fainter, low-S/N images, leading to the images converging to Gaussian noise much sooner into the forward process. The relationship between the S/N and the rate of convergence results in the entirety of the top left of the grid of images being pure noise, indicating inefficient training for lower-S/N images. This highlights the difficulty in applying off-the-shelf pipelines to the complexities of real-world astronomical data that feature a high dynamic range and varying quality over images. \textit{Bottom:} Distribution of S/N of galaxy images. Even though the sample is dominated by lower-S/N images, a non-negligible number of sources with S/N>1000 remains in the training set.}
   \label{fig:diffusion_noise_images}
\end{figure*}

\section{Discussion}
\subsection{The effect of signal-to-noise}
\label{sec:Astro_noise_vs_diffusion}

In machine-learning research, the improvements of generative models are often demonstrated with datasets that are of a high resolution and typically noise-free. This provides a level of consistency across the full dataset, ensuring the benchmarks reflect the model's performance under ideal conditions. Whether through loss functions (as is discussed in \cref{sec:Training_objective}) or in the practical application of the pipelines, the model's response to one part of the data will be similar to another.

However, the difficulty in learning the dynamics of our data is not consistent throughout the images. Higher S/Ns and lower magnitudes provide higher contrast between the source features and the background noise, enabling the model to discern these features more accurately and faster than the low-S/N counterparts. In the context of diffusion models, where the aim throughout training is to learn the dynamics of the introduced noise at different timesteps, the varying S/N of the images could again cause inconsistencies. This can be seen in both the classifier results shown in \cref{fig:scores_test}, where the fainter objects typically have lower recall, as well as the increased variance during inpainting of fainter images shown in \cref{fig:boxplots_mse}.

To explore how the noise schedule impacts the \Euclid images, \cref{fig:background_noise_snr_levels} shows the variability of pixel values as they progress through timesteps. We take a row and column of pixels (a slice) through each image, bin them by S/N, and compute the median pixel value at each index within the slice. This allows us to observe how the values change across the image. Since pixels affected by the source have much higher values than those near the edges of the image, \cref{fig:background_noise_snr_levels} compares slices through the outer regions (\textit{top}) with slices through the centre of the image (\textit{bottom}). 

In the top row, we see that low-S/N images are more significantly affected by the introduction of noise, and this effect appears earlier in the scheduler. The background noise converges to the Gaussian distribution across all S/N bins between halfway ($t = 2000$) and three quarters ($t = 3000$) of the total noise steps. In contrast, for pixels dominated by the source (bottom row), high-S/N images remain relatively unaffected through much of the schedule. The brightest images do not show noticeable changes until around $t = 2000$, and only converge to the Gaussian distribution within the final 250 steps. It is evident that both the image's S/N level and the position of the pixels within the image determine the number of timesteps needed for the noise to resemble the target Gaussian distribution.

To visualise the true effect of this differing behaviour, \cref{fig:diffusion_noise_images} shows a sample of images throughout different timesteps. Each image shown is a sample from the respective S/N bins below. It is clear that the noise scheduling does not perform uniformly across the images.

To improve the performance of a diffusion model, whether to better recreate a dataset or to model more complex data, the simplest approach is to increase the number of unique denoising steps. The emphasis here is on uniqueness, as simply adding more timesteps can lead to redundancy, especially in the later stages when successive steps become too similar, add little value, and increase overhead. This is precisely why \citet{nichol2021improved} introduced the cosine-beta scheduler (discussed in \cref{sec:scheduler}); the original linear scheduler caused images to degrade into pure noise too early into the process.

The images in \cref{fig:diffusion_noise_images} show that for a large fraction of our dataset, the model spends the majority of its time and computational effort learning the trivial task of denoising pure Gaussian noise, which provides no useful learning signal for galaxy morphology. This is a fundamental inefficiency in the implementation of diffusion-based models that not only affects \Euclid images, but will apply to any scientific dataset that exhibits varying levels of S/N. Although this has implications in regards to the efficiency and optimisation of the pipeline, it highlights the bigger issue of the data quality and variety assumptions implied when using off-the-shelf deep learning methods to the complex, heterogeneous data typical in large astronomical surveys and the wider scientific community. While a detailed implementation is beyond the scope of this work, this finding highlights a critical direction for future research: the development of adaptive noise schedulers that are conditioned on image properties such as S/N. Such an advance would significantly improve both the efficiency and performance of diffusion models in scientific applications.

\subsection{Considerations for training and inference costs}

Our DDPM inpainting model required approximately 50 hours of training using a single Nvidia A6000 GPU, which aligns with expectations for models of a similar scale and architecture. However, the inference pipeline is heavily constrained by memory rather than computation, as is demonstrated by its near-linear scaling -- the inpainting of 1024 and 8192 images took roughly the same amount of time when sufficient GPU memory was available. This memory constraint is due to the need to store and iteratively manipulate large intermediate tensors during inference. Training, although more computationally expensive, has a more structured process whereby gradients are computed and accumulated in a predictable manner.

With expanded computational resources (4$\times$ Nvidia H100 GPUs), we can inpaint just over 2.5 million images per day. As we have shown in \cref{sec:Training,sec:scheduler}, our model architecture and hyperparameters remain largely unmodified from their standard configurations. Optimisations tailored explicitly for astronomical-based data, especially in regards to the noise scheduler discussed in \cref{sec:Astro_noise_vs_diffusion}, will significantly reduce the number of iterations and inference time required to process each image. For example, \cref{fig:diffusion_noise_images} shows an opportunity for a two- to four-fold speed-up in timesteps when correcting for the high number of pure Gaussian noise images alone.

The results from \cref{sec:Scores_comparisons} show that although our predictions have a significant overlap with many of the selections presented, it also highlights that the ROC classifier produces a similar overlap performance. Given that this boundary is formed by a simple calculation from the data, rather than requiring a complex model and therefore near-zero overhead to calculate, it may present itself as the preferred method. However, the ROC metric is static and its performance is fixed by its mathematical definition. It must be reiterated that the current pipeline for the diffusion model has had minimum optimisations for the data and the classification task and so can be expected to improve with larger training sets, more advanced network architectures, or the future incorporation of multi-wavelength data as conditioning information. The ROC classifier is also limited to outputting a single number. The generative capabilities of the diffusion model, where it outputs a reconstructed image, provides the potential for a much richer set of scientific applications beyond simple classification such as decomposition as was discussed in \cref{sec:decomposition}. Therefore, although it has shown good performance, its limited approach restricts the ROC classifier to only serve as a fundamental baseline for comparison in our study.

In regards to the PSF and Sérsic flux classifier, although there are no computational costs to the user, traditional parametric fitting is a serial, per-object process that becomes a bottleneck for survey-scale datasets~\citep{tuccillo2018deep}. Having parallelisable and scalable alternative methods, even if they feature a large one-time training cost, allows us to make a strategic investment as we cover more of the extragalactic sky. For AGN identification specifically, the simple parametric models may not be well suited to handle complex or merged morphologies. This could explain the worse AGN-based recall seen for this classifier in \cref{fig:scores_test}.

Future enhancements to the diffusion model are expected to not only improve the accuracy of identifying AGN candidates but also refine our reconstruction error maps for better analysis of prospective AGN components. The other classifiers do not offer the same potential for improvement, making the diffusion method the most promising moving forward.

\section{Conclusion}
\label{sec:Conclusions}

In this paper, we have used a diffusion-based inpainting model to identify AGN and QSOs using VIS images from the Q1 data \citep{Q1cite} from the \Euclid telescope. Our approach focuses on inpainting the brightest pixels within the centre of a galaxy and employs a novel thresholding approach based on the reconstruction errors to differentiate AGN and QSOs from the rest of the galaxy population. Using only VIS images, our method can generate a large, reliable sample of optically selected AGN without requiring AGN labels ahead of time. Utilising a standard training and inference pipeline with minimal modifications for astronomical data, our method demonstrates high recall and significant overlap with AGN candidates obtained through more traditional methods, across various wavelengths including optical, near-infrared (NIR), mid-infrared, and X-ray.

Our key contributions and conclusions include the following:

\begin{itemize}
\item Compared to traditional colour-selection criteria, the approach attains high coverage in AGN candidates while achieving near-perfect separation of galaxies not featuring an AGN component. Reliably separating these different sources allows for the creation of subsets with high purity, which are vital for the cosmological objectives of the \Euclid mission.
\item By demonstrating that VIS images alone can generate reliable and extensive samples of AGN and QSOs, our method showcases the impressive capabilities of \Euclid's imaging. Given that the future of the mission will cover $\mathrm{14\,000
\;deg^2}$, the ability for AGN identification using only VIS allows for the possibility of some science aims to be met without the immediate need to integrate multiple data sources. This can significantly reduce the time and complexity of processing and analysing the data.
\item The application of diffusion models with minimal adaptations for astronomical imaging highlights the potential for the use of more advanced machine-learning tools and pipelines for future \Euclid data, as well as other scientific datasets. The ability to leverage models that can process large volumes of data efficiently, without costly and extensive specialised tuning, will be vital as scientific datasets continue to increase in size and complexity.
\item Although our model performs well without significant optimisation for our use case, we provide insights into the behaviour of the model given the challenges presented with astronomical imaging. We show how the assumptions of machine-learning literature, particularly regarding fixed noise schedules, can result in non-optimal training on data with a high dynamic range and large variations in S/N. By highlighting these inefficiencies, we identify a clear direction for future progress in both astronomy and machine-learning domains through the development of adaptive, data-aware training strategies.
\item Because this is the first use of these pipelines for such an application, we have routinely provided guidance on improvements and next steps to refine this work. Enhancing the model's precision, improving the efficiency of both training and inference and expanding its capabilities will only increase its impact. 
\end{itemize}

In conclusion, our results demonstrate the feasibility and effectiveness of using diffusion-based inpainting for AGN and QSO identification. The ability to use VIS images exclusively for selecting these candidates provides a potential strategic flexibility to the \Euclid mission, while reducing reliance on complex multi-spectral data integration. By proving the utility of applying our pipelines to the task of AGN identification, we believe it can be expanded and can significantly contribute to a broader range of astronomical applications, such as identifying gravitational lensing or finding transient objects.

\begin{acknowledgements}

The authors would like to thank the referee for their very insightful and constructive feedback during the review process.

\AckQone

\AckEC

Based on data from UNIONS, a scientific collaboration using three Hawaii-based telescopes: CFHT, Pan-STARRS, and Subaru (\url{www.skysurvey.cc}\,).

Based on data from the Dark Energy Camera (DECam) on the Blanco 4-m Telescope at CTIO in Chile (\url{https://www.darkenergysurvey.org}\,).

This work uses results from the ESA mission {\it Gaia}, whose data are being processed by the Gaia Data Processing and Analysis Consortium (\url{https://www.cosmos.esa.int/gaia}\,).

This publication makes use of data products from the Wide-field Infrared Survey Explorer, which is a joint project of the University of California, Los Angeles, and the Jet Propulsion Laboratory/California Institute of Technology, funded by the National Aeronautics and Space Administration.

DESI construction and operations is managed by the Lawrence Berkeley National Laboratory. This research is supported by the U.S. Department of Energy, Office of Science, Office of High-Energy Physics, under Contract No. DE–AC02–05CH11231, and by the National Energy Research Scientific Computing Center, a DOE Office of Science User Facility under the same contract. Additional support for DESI is provided by the U.S. National Science Foundation, Division of Astronomical Sciences under Contract No. AST-0950945 to the NSF’s National Optical-Infrared Astronomy Research Laboratory; the Science and Technology Facilities Council of the United Kingdom; the Gordon and Betty Moore Foundation; the Heising-Simons Foundation; the French Alternative Energies and Atomic Energy Commission (CEA); the National Council of Science and Technology of Mexico (CONACYT); the Ministry of Science and Innovation of Spain, and by the DESI Member Institutions. The DESI collaboration is honored to be permitted to conduct astronomical research on Iolkam Du’ag (Kitt Peak), a mountain with particular significance to the Tohono O’odham Nation.

\\Grant Stevens acknowledges financial support from the UKRI for an EPSRC Doctoral Prize Fellowship at the University of Bristol (EP/W524414/1).

This work has benefited from the support of Royal Society Research Grant RGS{\textbackslash}R1\textbackslash231450.
This research was supported by the International Space Science Institute (ISSI) in Bern, through ISSI International Team project \#23-573 ``Active Galactic Nuclei in Next Generation Surveys''.

``ELSA: Euclid Legacy Science Advanced analysis tools'' (Grant Agreement no. 101135203) is 
funded by the European Union. Views and opinions expressed are however those of the 
author(s) only and do not necessarily reflect those of the European Union or Innovate UK. 
Neither the European Union nor the granting authority can be held responsible for them. UK 
participation is funded through the UK Horizon guarantee scheme under Innovate UK grant 
10093177.

The authors acknowledge the use of computational resources from the parallel computing cluster of the Open Physics Hub (https://site.unibo.it/openphysicshub/en) at the Physics and Astronomy Department in Bologna.

The authors acknowledge the use of resources provided by the Isambard-AI National AI Research Resource (AIRR). Isambard-AI is operated by the University of Bristol and is funded by the UK Government’s Department for Science, Innovation and Technology (DSIT) via UK Research and Innovation; and the Science and Technology Facilities Council [ST/AIRR/I-A-I/1023].

\end{acknowledgements}

\bibliographystyle{aa}
\bibliography{refs}

\begin{thebibliography}{57}
\expandafter\ifx\csname natexlab\endcsname\relax\def\natexlab#1{#1}\fi

\bibitem[{{Adam} {et~al.}(2022){Adam}, {Coogan}, {Malkin}, {Legin}, {Perreault-Levasseur}, {Hezaveh}, \& {Bengio}}]{adam2022posterior}
{Adam}, A., {Coogan}, A., {Malkin}, N., {et~al.} 2022, arXiv preprint arXiv:2211.03812

\bibitem[{Adam {et~al.}(2025)Adam, Stone, Bottrell, Legin, Hezaveh, \& Perreaul-Levasseur}]{adam2023echoes}
Adam, A., Stone, C., Bottrell, C., {et~al.} 2025, \aj, 169, 254

\bibitem[{{Assef} {et~al.}(2018){Assef}, {Stern}, {Noirot}, {Jun}, {Cutri}, \& {Eisenhardt}}]{Assef_2018_2018ApJS..234...23A}
{Assef}, R.~J., {Stern}, D., {Noirot}, G., {et~al.} 2018, \apjs, 234, 23

\bibitem[{{Bertin} {et~al.}(2020){Bertin}, {Schefer}, {Apostolakos}, {{\'A}lvarez-Ayll{\'o}n}, {Dubath}, \& {K{\"u}mmel}}]{2020ASPC..527..461B}
{Bertin}, E., {Schefer}, M., {Apostolakos}, N., {et~al.} 2020, in Astronomical Society of the Pacific Conference Series, Vol. 527, Astronomical Data Analysis Software and Systems XXIX, ed. R.~{Pizzo}, E.~R. {Deul}, J.~D. {Mol}, J.~{de Plaa}, \& H.~{Verkouter}, 461

\bibitem[{{Davis} {et~al.}(2018){Davis}, {Graham}, \& {Cameron}}]{Davis2018ApJ...869..113D}
{Davis}, B.~L., {Graham}, A.~W., \& {Cameron}, E. 2018, ApJ, 869, 113

\bibitem[{{Davis} {et~al.}(2019){Davis}, {Graham}, \& {Cameron}}]{Davis2019ApJ...873...85D}
{Davis}, B.~L., {Graham}, A.~W., \& {Cameron}, E. 2019, ApJ, 873, 85

\bibitem[{{DESI Collaboration} {et~al.}(2024){DESI Collaboration}, {Adame}, {Aguilar}, {Ahlen}, {Alam}, {Aldering}, {Alexander}, {Alfarsy}, {Allende Prieto}, {Alvarez}, {Alves}, {Anand}, {Andrade-Oliveira}, {Armengaud}, {Asorey}, {Avila}, {Aviles}, {Bailey}, {Balaguera-Antol{\'\i}nez}, {Ballester}, {Baltay}, {Bault}, {Bautista}, {Behera}, {Beltran}, {BenZvi}, {Beraldo e Silva}, {Bermejo-Climent}, {Berti}, {Besuner}, {Beutler}, {Bianchi}, {Blake}, {Blum}, {Bolton}, {Brieden}, {Brodzeller}, {Brooks}, {Brown}, {Buckley-Geer}, {Burtin}, {Cabayol-Garcia}, {Cai}, {Canning}, {Cardiel-Sas}, {Carnero Rosell}, {Castander}, {Cervantes-Cota}, {Chabanier}, {Chaussidon}, {Chaves-Montero}, {Chen}, {Chen}, {Chuang}, {Claybaugh}, {Cole}, {Cooper}, {Cuceu}, {Davis}, {Dawson}, {de Belsunce}, {de la Cruz}, {de la Macorra}, {Della Costa}, {de Mattia}, {Demina}, {Demirbozan}, {DeRose}, {Dey}, {Dey}, {Dhungana}, {Ding}, {Ding}, {Doel}, {Doshi}, {Douglass}, {Edge}, {Eftekharzadeh}, {Eisenstein}, {Elliott}, {Ereza}, {Escoffier},
  {Fagrelius}, {Fan}, {Fanning}, {Fawcett}, {Ferraro}, {Flaugher}, {Font-Ribera}, {Forero-Romero}, {Forero-S{\'a}nchez}, {Frenk}, {G{\"a}nsicke}, {Garc{\'\i}a}, {Garc{\'\i}a-Bellido}, {Garcia-Quintero}, {Garrison}, {Gil-Mar{\'\i}n}, {Golden-Marx}, {Gontcho A Gontcho}, {Gonzalez-Morales}, {Gonzalez-Perez}, {Gordon}, {Graur}, {Green}, {Gruen}, {Guy}, {Hadzhiyska}, {Hahn}, {Han}, {Hanif}, {Herrera-Alcantar}, {Honscheid}, {Hou}, {Howlett}, {Huterer}, {Ir{\v{s}}i{\v{c}}}, {Ishak}, {Jacques}, {Jana}, {Jiang}, {Jimenez}, {Jing}, {Joudaki}, {Joyce}, {Jullo}, {Juneau}, {Kara{\c{c}}ayl{\i}}, {Karim}, {Kehoe}, {Kent}, {Khederlarian}, {Kim}, {Kirkby}, {Kisner}, {Kitaura}, {Kizhuprakkat}, {Kneib}, {Koposov}, {Kov{\'a}cs}, {Kremin}, {Krolewski}, {L'Huillier}, {Lahav}, {Lambert}, {Lamman}, {Lan}, {Landriau}, {Lang}, {Lange}, {Lasker}, {Leauthaud}, {Le Guillou}, {Levi}, {Li}, {Linder}, {Lyons}, {Magneville}, {Manera}, {Manser}, {Margala}, {Martini}, {McDonald}, {Medina}, {Medina-Varela}, {Meisner}, {Mena-Fern{\'a}ndez},
  {Meneses-Rizo}, {Mezcua}, {Miquel}, {Montero-Camacho}, {Moon}, {Moore}, {Moustakas}, {Mueller}, {Mundet}, {Mu{\~n}oz-Guti{\'e}rrez}, {Myers}, {Nadathur}, {Napolitano}, {Neveux}, {Newman}, {Nie}, {Nikutta}, {Niz}, {Norberg}, {Noriega}, {Paillas}, {Palanque-Delabrouille}, {Palmese}, {Pan}, {Parkinson}, {Penmetsa}, {Percival}, {P{\'e}rez-Fern{\'a}ndez}, {P{\'e}rez-R{\`a}fols}, {Pieri}, {Poppett}, {Porredon}, \& {Pothier}}]{DESI-2024AJ....168...58D}
{DESI Collaboration}, {Adame}, A.~G., {Aguilar}, J., {et~al.} 2024, \aj, 168, 58

\bibitem[{Dhariwal \& Nichol(2021)}]{dhariwal2021diffusion}
Dhariwal, P. \& Nichol, A. 2021, Advances in neural information processing systems, 34, 8780

\bibitem[{Dia {et~al.}(2025)Dia, Yantovski-Barth, Adam, Bowles, Perreault-Levasseur, Hezaveh, \& Scaife}]{dia2025iris}
Dia, N., Yantovski-Barth, M., Adam, A., {et~al.} 2025, arXiv preprint arXiv:2501.02473

\bibitem[{{Dressler}(1989)}]{1989IAUS..134..217D}
{Dressler}, A. 1989, in IAU Symposium, Vol. 134, Active Galactic Nuclei, ed. D.~E. {Osterbrock} \& J.~S. {Miller}, 217

\bibitem[{{Euclid Collaboration: Aussel} {et~al.}(2024){Euclid Collaboration: Aussel}, {Kruk}, {Walmsley}, {et~al.}}]{EP-Aussel}
{Euclid Collaboration: Aussel}, B., {Kruk}, S., {Walmsley}, M., {et~al.} 2024, \aap, 689, A274

\bibitem[{{Euclid Collaboration: Bisigello} {et~al.}(2024){Euclid Collaboration: Bisigello}, {Massimo}, {Tortora}, {et~al.}}]{EP-Bisigello}
{Euclid Collaboration: Bisigello}, L., {Massimo}, M., {Tortora}, C., {et~al.} 2024, \aap, 691, A1

\bibitem[{{Euclid Collaboration: Cropper} {et~al.}(2025){Euclid Collaboration: Cropper}, {Al-Bahlawan}, {Amiaux}, {et~al.}}]{EuclidSkyVIS}
{Euclid Collaboration: Cropper}, M., {Al-Bahlawan}, A., {Amiaux}, J., {et~al.} 2025, A\&A, 697, A2

\bibitem[{{Euclid Collaboration: Margalef-Bentabol} {et~al.}(2025){Euclid Collaboration: Margalef-Bentabol}, {Wang}, {La Marca}, {et~al.}}]{Q1-SP015}
{Euclid Collaboration: Margalef-Bentabol}, B., {Wang}, L., {La Marca}, A., {et~al.} 2025, A\&A, accepted (Euclid Q1 SI), arXiv:2503.15318

\bibitem[{{Euclid Collaboration: Matamoro Zatarain} {et~al.}(2025){Euclid Collaboration: Matamoro Zatarain}, {Fotopoulou}, {Ricci}, {et~al.}}]{Q1-SP027}
{Euclid Collaboration: Matamoro Zatarain}, T., {Fotopoulou}, S., {Ricci}, F., {et~al.} 2025, A\&A, in press (Euclid Q1 SI), \url{https://doi.org/10.1051/0004-6361/202554619}, arXiv:2503.15320

\bibitem[{{Euclid Collaboration: Mellier} {et~al.}(2025){Euclid Collaboration: Mellier}, {Abdurro'uf}, {Acevedo~Barroso}, {et~al.}}]{EuclidSkyOverview}
{Euclid Collaboration: Mellier}, Y., {Abdurro'uf}, {Acevedo~Barroso}, J., {et~al.} 2025, A\&A, 697, A1

\bibitem[{{Euclid Collaboration: Romelli} {et~al.}(2025){Euclid Collaboration: Romelli}, {K\"ummel}, {Dole}, {et~al.}}]{Q1-TP004}
{Euclid Collaboration: Romelli}, E., {K\"ummel}, M., {Dole}, H., {et~al.} 2025, A\&A, in press (Euclid Q1 SI), \url{https://doi.org/10.1051/0004-6361/202554586}, arXiv:2503.15305

\bibitem[{{Euclid Collaboration: Roster} {et~al.}(2025){Euclid Collaboration: Roster}, {Salvato}, {Buchner}, {et~al.}}]{Q1-SP003}
{Euclid Collaboration: Roster}, W., {Salvato}, M., {Buchner}, J., {et~al.} 2025, A\&A, accepted (Euclid Q1 SI), arXiv:2503.15316

\bibitem[{{Euclid Collaboration: Walmsley} {et~al.}(2025){Euclid Collaboration: Walmsley}, {Huertas-Company}, {Quilley}, {et~al.}}]{Q1-SP047}
{Euclid Collaboration: Walmsley}, M., {Huertas-Company}, M., {Quilley}, L., {et~al.} 2025, A\&A, submitted (Euclid Q1 SI), arXiv:2503.15310

\bibitem[{{Euclid Quick Release Q1}(2025)}]{Q1cite}
{Euclid Quick Release Q1}. 2025, \url{https://doi.org/10.57780/esa-2853f3b}

\bibitem[{Feng {et~al.}(2023)Feng, Smith, Rubinstein, Chang, Bouman, \& Freeman}]{feng2023score}
Feng, B.~T., Smith, J., Rubinstein, M., {et~al.} 2023, in Proceedings of the IEEE/CVF International Conference on Computer Vision, 10520--10531

\bibitem[{{Ferrarese} \& {Merritt}(2000)}]{2000ApJ...539L...9F}
{Ferrarese}, L. \& {Merritt}, D. 2000, \apjl, 539, L9

\bibitem[{{Gaia Collaboration: Bailer-Jones} {et~al.}(2023){Gaia Collaboration: Bailer-Jones}, {Teyssier}, {Delchambre}, {Ducourant}, {Garabato}, {Hatzidimitriou}, {Klioner}, {Rimoldini}, {Bellas-Velidis}, {Carballo}, {Carnerero}, {Diener}, {Fouesneau}, {Galluccio}, {Gavras}, {Krone-Martins}, {Raiteri}, {Teixeira}, {Brown}, {Vallenari}, {Prusti}, {de Bruijne}, \& et~al.}]{GaiaCollaboration_2023_2023A&A...674A..41G}
{Gaia Collaboration: Bailer-Jones}, C.~A.~L., {Teyssier}, D., {Delchambre}, L., {et~al.} 2023, \aap, 674, A41

\bibitem[{{Gaia Collaboration: Vallenari} {et~al.}(2023){Gaia Collaboration: Vallenari}, {Brown}, {Prusti}, {de Bruijne}, {Arenou}, {Babusiaux}, {Biermann}, {Creevey}, {Ducourant}, {Evans}, {Eyer}, {Guerra}, {Hutton}, {Jordi}, {Klioner}, {Lammers}, {Lindegren}, \& et~al.}]{GaiaCollaboration_2023_2023A&A...674A...1G}
{Gaia Collaboration: Vallenari}, A., {Brown}, A.~G.~A., {Prusti}, T., {et~al.} 2023, \aap, 674, A1

\bibitem[{Goodfellow {et~al.}(2014)Goodfellow, Pouget-Abadie, Mirza, Xu, Warde-Farley, Ozair, Courville, \& Bengio}]{goodfellow2014generative}
Goodfellow, I., Pouget-Abadie, J., Mirza, M., {et~al.} 2014, Advances in neural information processing systems, 27

\bibitem[{Goodfellow {et~al.}(2020)Goodfellow, Pouget-Abadie, Mirza, Xu, Warde-Farley, Ozair, Courville, \& Bengio}]{goodfellow2020generative}
Goodfellow, I., Pouget-Abadie, J., Mirza, M., {et~al.} 2020, Communications of the ACM, 63, 139

\bibitem[{Graikos {et~al.}(2022)Graikos, Malkin, Jojic, \& Samaras}]{graikos2022diffusion}
Graikos, A., Malkin, N., Jojic, N., \& Samaras, D. 2022, Advances in Neural Information Processing Systems, 35, 14715

\bibitem[{Harrison \& Ramos~Almeida(2024)}]{harrison2024observational}
Harrison, C.~M. \& Ramos~Almeida, C. 2024, Galaxies, 12, 17

\bibitem[{Ho {et~al.}(2020)Ho, Jain, \& Abbeel}]{ho2020denoising}
Ho, J., Jain, A., \& Abbeel, P. 2020, Advances in neural information processing systems, 33, 6840

\bibitem[{{Kingma} \& {Welling}(2013)}]{kingma2013auto}
{Kingma}, D.~P. \& {Welling}, M. 2013, arXiv e-prints, arXiv:1312.6114

\bibitem[{{Kormendy} \& {Richstone}(1995)}]{1995ARA&A..33..581K}
{Kormendy}, J. \& {Richstone}, D. 1995, \araa, 33, 581

\bibitem[{{K{\"u}mmel} {et~al.}(2020){K{\"u}mmel}, {Bertin}, {Schefer}, {Apostolakos}, {{\'A}lvarez-Ayll{\'o}n}, \& {Dubath}}]{2020ASPC..527...29K}
{K{\"u}mmel}, M., {Bertin}, E., {Schefer}, M., {et~al.} 2020, in Astronomical Society of the Pacific Conference Series, Vol. 527, Astronomical Data Analysis Software and Systems XXIX, ed. R.~{Pizzo}, E.~R. {Deul}, J.~D. {Mol}, J.~{de Plaa}, \& H.~{Verkouter}, 29

\bibitem[{Lugmayr {et~al.}(2022)Lugmayr, Danelljan, Romero, Yu, Timofte, \& Van~Gool}]{lugmayr2022repaint}
Lugmayr, A., Danelljan, M., Romero, A., {et~al.} 2022, in IEEE/CVF conference on computer vision and pattern recognition (IEEE), 11461--11471

\bibitem[{Lupton {et~al.}(2004)Lupton, Blanton, Fekete, Hogg, O’Mullane, Szalay, \& Wherry}]{lupton2004preparing}
Lupton, R., Blanton, M.~R., Fekete, G., {et~al.} 2004, PASP, 116, 133

\bibitem[{{Magorrian} {et~al.}(1998){Magorrian}, {Tremaine}, {Richstone}, {Bender}, {Bower}, {Dressler}, {Faber}, {Gebhardt}, {Green}, {Grillmair}, {Kormendy}, \& {Lauer}}]{1998AJ....115.2285M}
{Magorrian}, J., {Tremaine}, S., {Richstone}, D., {et~al.} 1998, \aj, 115, 2285

\bibitem[{Nguyen {et~al.}(2017)Nguyen, Clune, Bengio, Dosovitskiy, \& Yosinski}]{nguyen2017plug}
Nguyen, A., Clune, J., Bengio, Y., Dosovitskiy, A., \& Yosinski, J. 2017, in Proceedings of the IEEE conference on computer vision and pattern recognition, 4467--4477

\bibitem[{Nichol \& Dhariwal(2021)}]{nichol2021improved}
Nichol, A.~Q. \& Dhariwal, P. 2021, in Proceedings of Machine Learning Research, Vol. 139, Proceedings of the 38th International Conference on Machine Learning, ed. M.~Meila \& T.~Zhang (PMLR), 8162--8171

\bibitem[{Pathak {et~al.}(2016)Pathak, Krahenbuhl, Donahue, Darrell, \& Efros}]{pathak2016context}
Pathak, D., Krahenbuhl, P., Donahue, J., Darrell, T., \& Efros, A.~A. 2016, in IEEE conference on computer vision and pattern recognition (IEEE), 2536--2544

\bibitem[{Radford {et~al.}(2021)Radford, Kim, Hallacy, Ramesh, Goh, Agarwal, Sastry, Askell, Mishkin, Clark, {et~al.}}]{radford2021learning}
Radford, A., Kim, J.~W., Hallacy, C., {et~al.} 2021, in Proceedings of Machine Learning Research, Vol. 139, Proceedings of the 38th International Conference on Machine Learning, ed. M.~Meila \& T.~Zhang (PMLR), 8748--8763

\bibitem[{{Ramesh} {et~al.}(2022){Ramesh}, {Dhariwal}, {Nichol}, {Chu}, \& {Chen}}]{ramesh2022hierarchical}
{Ramesh}, A., {Dhariwal}, P., {Nichol}, A., {Chu}, C., \& {Chen}, M. 2022, arXiv e-prints, arXiv:2204.06125

\bibitem[{Remy {et~al.}(2023)Remy, Lanusse, Jeffrey, Liu, Starck, Osato, \& Schrabback}]{remy2023probabilistic}
Remy, B., Lanusse, F., Jeffrey, N., {et~al.} 2023, A\&A, 672, A51

\bibitem[{Saharia {et~al.}(2022)Saharia, Chan, Saxena, Li, Whang, Denton, Ghasemipour, Gontijo~Lopes, Karagol~Ayan, Salimans, Ho, Fleet, \& Norouzi}]{saharia2022photorealistic}
Saharia, C., Chan, W., Saxena, S., {et~al.} 2022, in Advances in Neural Information Processing Systems, ed. S.~Koyejo, S.~Mohamed, A.~Agarwal, D.~Belgrave, K.~Cho, \& A.~Oh, Vol.~35 (Curran Associates, Inc.), 36479--36494

\bibitem[{{Sahu} {et~al.}(2019){Sahu}, {Graham}, \& {Davis}}]{Sahu2019ApJ...876..155S}
{Sahu}, N., {Graham}, A.~W., \& {Davis}, B.~L. 2019, ApJ, 876, 155

\bibitem[{{Sampson} {et~al.}(2024){Sampson}, {Melchior}, {Ward}, \& {Birmingham}}]{2024A&C....4900875S}
{Sampson}, M.~L., {Melchior}, P., {Ward}, C., \& {Birmingham}, S. 2024, Astronomy and Computing, 49, 100875

\bibitem[{{Smith} {et~al.}(2022){Smith}, {Geach}, {Jackson}, {Arora}, {Stone}, \& {Courteau}}]{2022MNRAS.511.1808S}
{Smith}, M.~J., {Geach}, J.~E., {Jackson}, R.~A., {et~al.} 2022, \mnras, 511, 1808

\bibitem[{Sohl-Dickstein {et~al.}(2015)Sohl-Dickstein, Weiss, Maheswaranathan, \& Ganguli}]{sohl2015deep}
Sohl-Dickstein, J., Weiss, E., Maheswaranathan, N., \& Ganguli, S. 2015, in International conference on machine learning, pmlr, 2256--2265

\bibitem[{Spagnoletti {et~al.}(2024)Spagnoletti, Boucaud, Kabalan, Biswas, {et~al.}}]{spagnoletti2024bayesian}
Spagnoletti, A., Boucaud, A., Kabalan, W., Biswas, B., {et~al.} 2024, in 38th conference on Neural Information Processing Systems

\bibitem[{Stone \& Courteau(2019)}]{stone2019intrinsic}
Stone, C. \& Courteau, S. 2019, ApJ, 882, 6

\bibitem[{Stone {et~al.}(2021)Stone, Courteau, \& Arora}]{stone2021intrinsic}
Stone, C., Courteau, S., \& Arora, N. 2021, ApJ, 912, 41

\bibitem[{{Storey-Fisher} {et~al.}(2024){Storey-Fisher}, {Hogg}, {Rix}, {Eilers}, {Fabbian}, {Blanton}, \& {Alonso}}]{Storey-Fisher2024}
{Storey-Fisher}, K., {Hogg}, D.~W., {Rix}, H.-W., {et~al.} 2024, \apj, 964, 69

\bibitem[{Thanh-Tung \& Tran(2020)}]{thanh2020catastrophic}
Thanh-Tung, H. \& Tran, T. 2020, in 2020 international joint conference on neural networks (ijcnn), 1--10

\bibitem[{Tuccillo {et~al.}(2018)Tuccillo, Huertas-Company, Decenci{\`e}re, Velasco-Forero, Dom{\'\i}nguez~S{\'a}nchez, \& Dimauro}]{tuccillo2018deep}
Tuccillo, D., Huertas-Company, M., Decenci{\`e}re, E., {et~al.} 2018, MNRAS, 475, 894

\bibitem[{Walmsley {et~al.}(2020)Walmsley, Smith, Lintott, Gal, Bamford, Dickinson, Fortson, Kruk, Masters, Scarlata, {et~al.}}]{walmsley2020galaxy}
Walmsley, M., Smith, L., Lintott, C., {et~al.} 2020, MNRAS, 491, 1554

\bibitem[{Wang {et~al.}(2018)Wang, Tao, Qi, Shen, \& Jia}]{wang2018image}
Wang, Y., Tao, X., Qi, X., Shen, X., \& Jia, J. 2018, Advances in neural information processing systems, 31

\bibitem[{Yeh {et~al.}(2017)Yeh, Chen, Yian~Lim, Schwing, Hasegawa-Johnson, \& Do}]{yeh2017semantic}
Yeh, R.~A., Chen, C., Yian~Lim, T., {et~al.} 2017, in IEEE conference on computer vision and pattern recognition (IEEE), 5485--5493

\bibitem[{Yu {et~al.}(2018)Yu, Lin, Yang, Shen, Lu, \& Huang}]{yu2018generative}
Yu, J., Lin, Z., Yang, J., {et~al.} 2018, in IEEE conference on computer vision and pattern recognition (IEEE), 5505--5514

\bibitem[{Zhang {et~al.}(2023)Zhang, Rao, \& Agrawala}]{zhang2023adding}
Zhang, L., Rao, A., \& Agrawala, M. 2023, in IEEE/CVF International Conference on Computer Vision (IEEE), 3836--3847

\end{thebibliography}

\begin{appendix}
\onecolumn
\section{Masks}

\begin{figure*}[ht!]
   \centering
   \includegraphics[width=\linewidth]{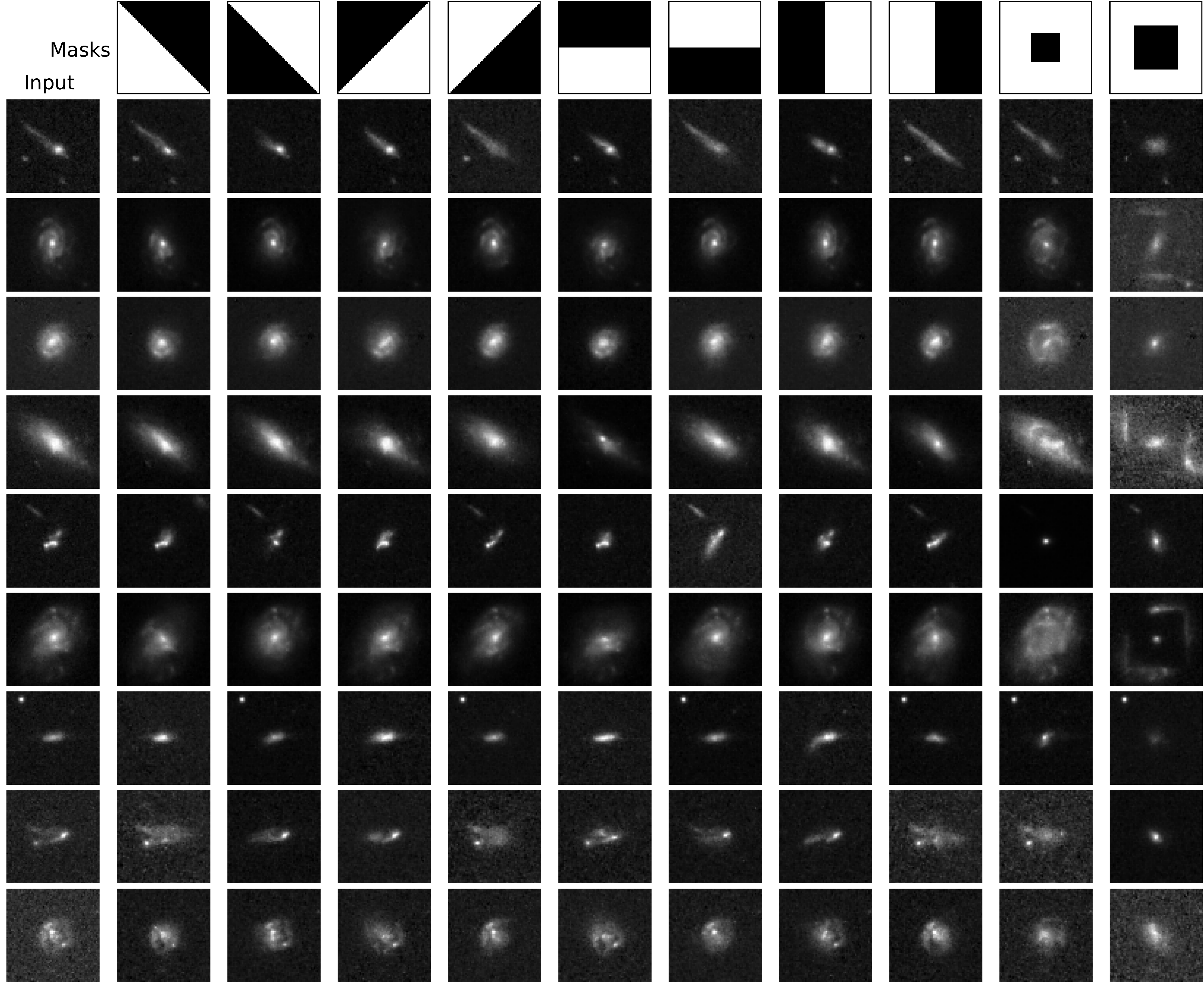}
   \caption{The output of applying the inpainting method to a selection of images with example masks (top row). The input image (leftmost column) has the pixels in the white area of each mask fixed, while the diffusion-based inpainting regenerates the black area. Although this model's aim is not explicitly large-scale image generation, applying masks that cover large selections of the image showcases the model's ability to create realistic galaxies. Ensuring that the generated images preserve a shape and brightness consistent with the fixed pixels shows that the model has a good internal representation of the dynamics of galaxy structure.}
   \label{fig:masks_output}
\end{figure*}

To explore the ability of the model at recreating galaxy images, we apply a variety of different masks to a selection of unseen images. \Cref{fig:masks_output} shows the original image on the leftmost column, followed by the regenerated images in subsequent columns. With each mask shown on the top row, the white area shows the pixels that are fixed from the original image, whereas the black areas are generated using the diffusion model.

As our model's aim is not to generate large collections of synthetic data, the variety and scientific validity of the produced galaxies are not put under high scrutiny. Instead, we only check that there are no obvious catastrophic mistakes. In the last two columns, we can see that due to the large mask and fewer pixels to guide the reconstruction, some images produce a visible boundary around the image where the combination of fixed and generated pixels is not as seamless as the other images. Given that the masks used for AGN detection are significantly smaller than those presented in \cref{fig:masks_output}, the fraction of pixels that the inpainting is conditioned on will be much higher, providing tighter constraints on the output compared to these examples. Improvements to the scheduler, as discussed in \cref{sec:Astro_noise_vs_diffusion}, would also reduce such artefacts.

\section{Alternative loss}

\begin{SCfigure}[1.5][ht!]
   \centering
   \includegraphics[width=0.66\textwidth]{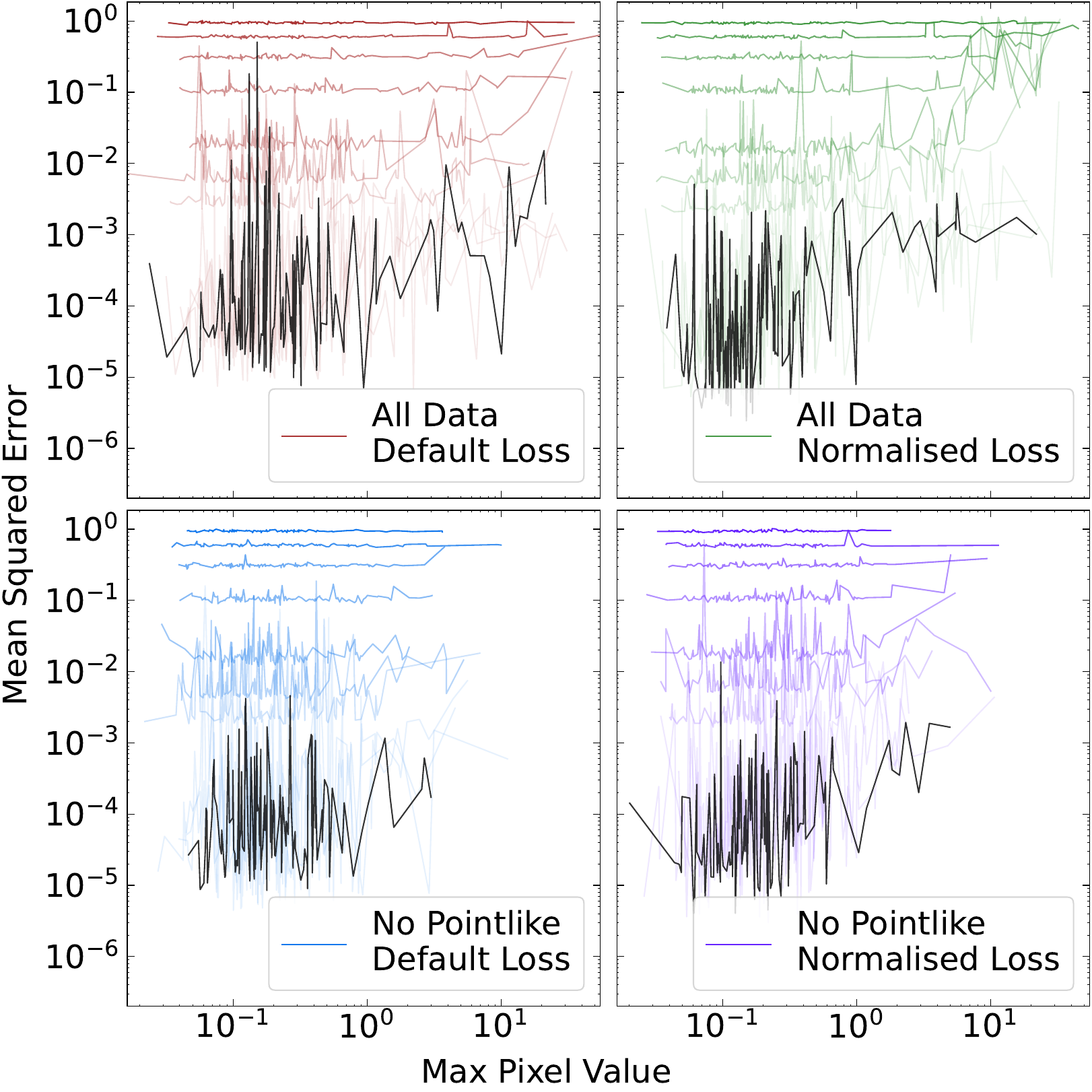}
   \caption{MSE of a random batch at different epochs throughout training. The opacity of each line indicates how far into the training epochs the measurement is, with fainter lines at higher epochs. Each training pair exhibits different behaviour as training progresses, indicating the weight each puts on an image's pixel brightness. Because each model was trained on more than solely MSE, these measurements are only to highlight the raw pixel performance of image recreation at each stage of training. The black line indicates the final step during training.}
   \label{fig:mse_max_pixel_losses}
\end{SCfigure}

Making use of standard metrics whose primary use-case is data on a consistent scale, such as 0--255 pixels, may not function optimally with the high dynamic range of VIS data. Specifically, since MSE is not scale invariant, having images that feature pixels that are an order of magnitude brighter, results in relative errors increasing by a square of that difference. This is typically a benefit as it forces outlier predictions to be severely penalised. However, with our data, this results in brighter images being disproportionately penalised compared to fainter ones, even if the predictions are both the same relative distance away from the true prediction.

As shown in \cref{sec:Training_objective}, we applied a normalisation to $L_\mathrm{MSE}$ by dividing each image's error by its respective maximum pixel value, which could be from a brighter external source. To visualise the impact of amending the loss function, with respect to an images brightness, \cref{fig:mse_max_pixel_losses} shows how the error from the unmodified MSE changes as the model is trained for longer. Each line represents the error for a collection of images in a random batch, with the increased transparency indicating the passage of time where the model had been trained on more epochs. The final epoch of each model is presented by a black line. The effect of removing point-like sources, as discussed in \cref{sec:sample_selection}, is also explored.

Although all four models arrive at a similar minimum loss, and begin the training process near-identically, their respective behaviour throughout training differs significantly. For the full dataset with the default loss (top left of \cref{fig:mse_max_pixel_losses}), training prioritises brighter images, leading to some significantly low errors towards the rightmost data. However, this comes at the expense of fainter data, where wild fluctuations in error show that the model's output is not very robust when it comes to precise predictions. This happens even when MSE might suggest low error, since pixel values below 1.0 result in minimal losses despite the predictions being relatively incorrect.

For the normalised loss with the full dataset (top right), since the penalisation is shared more fairly across pixel values, it is the density of images that has the most influence. Early on in training we can see that as the majority of data reduces in error, the brightest images remain at near-starting performance. This continues until the errors of the fainter images become so low that model is forced to prioritise the bright images. The key difference here is this only happens after the rest of the images are at a sufficient standard. The final outcome is fainter images performing well and the entire scale of images having the same maximum error.

The two models which feature the reduced training set (bottom row) share similar dynamics during training, regardless of the loss used. This is largely due to many point-like sources also being some of the brightest images and therefore the scale of the dataset is reduced. As the most problematic data are removed, the impact of disproportionate penalisation is significantly reduced, allowing the fainter sources to be the focus during training.

\twocolumn
\section{Reconstruction metrics}

\begin{figure}[ht!]
   \centering
   \includegraphics[width=\linewidth]{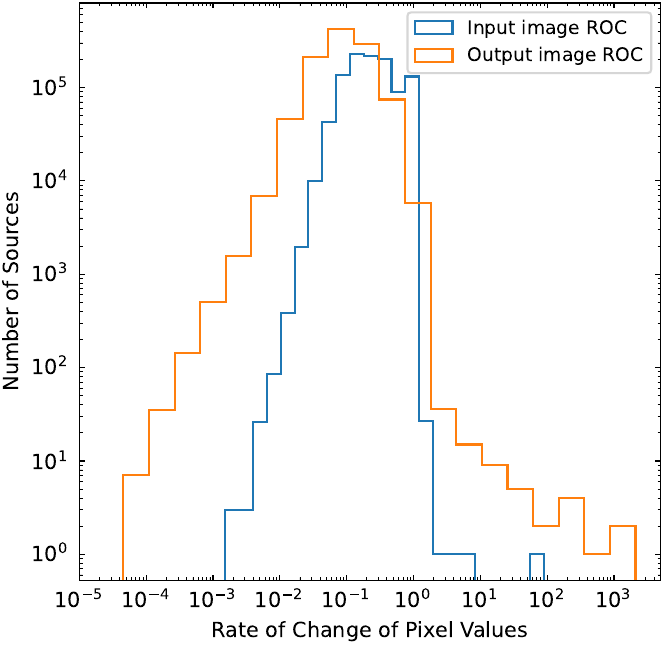}
   \caption{Comparison of the ROC between the original images and the inpainted images. The similarity in the peak of the distributions shows that the model is able to accurately recreate the distribution of the data.} 

   \label{fig:ROC_input_output}
\end{figure}

\Cref{fig:ROC_input_output} shows the differences in the ROC measurement between input images and the inpainting output.

\onecolumn
\section{Extreme examples}
\begin{figure*}[ht!]
   \centering
   \includegraphics[width=\linewidth]{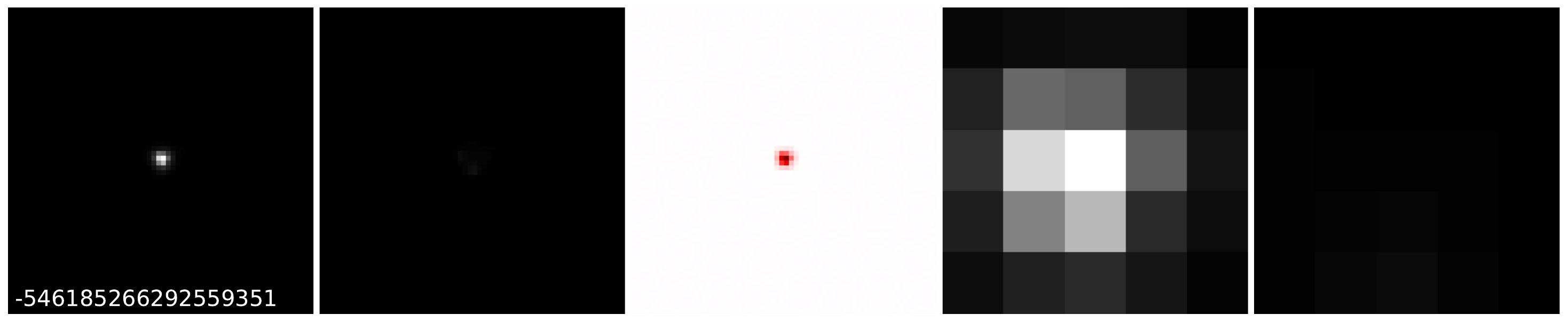}
   \includegraphics[width=\linewidth]{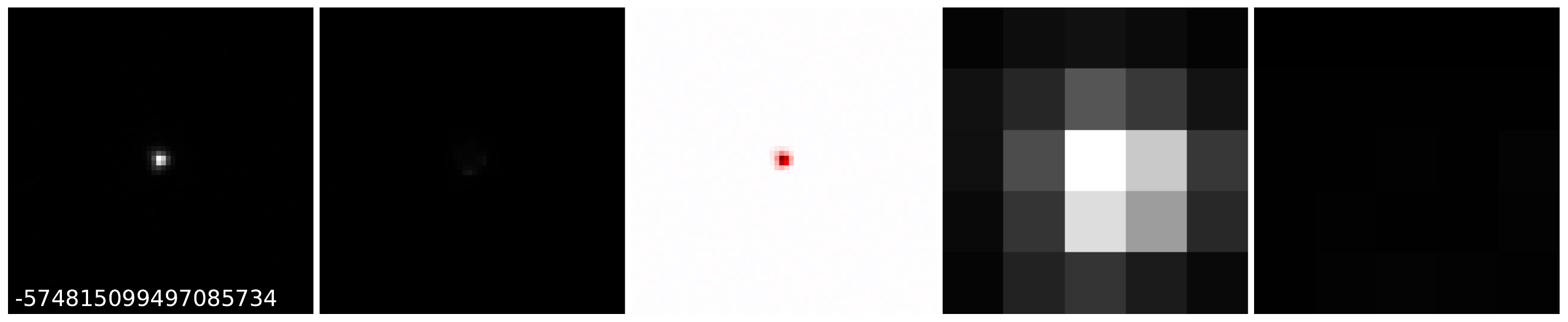}
   \includegraphics[width=\linewidth]{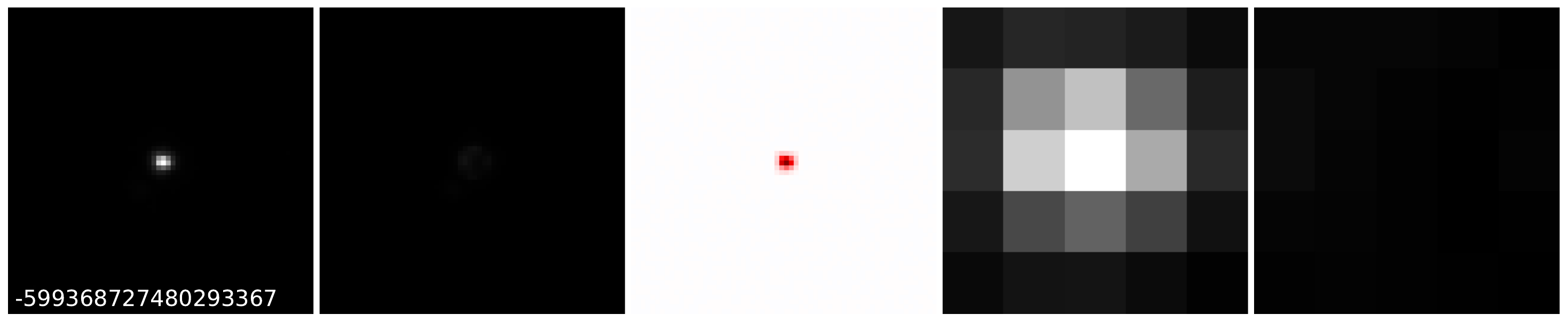}
   \includegraphics[width=\linewidth]{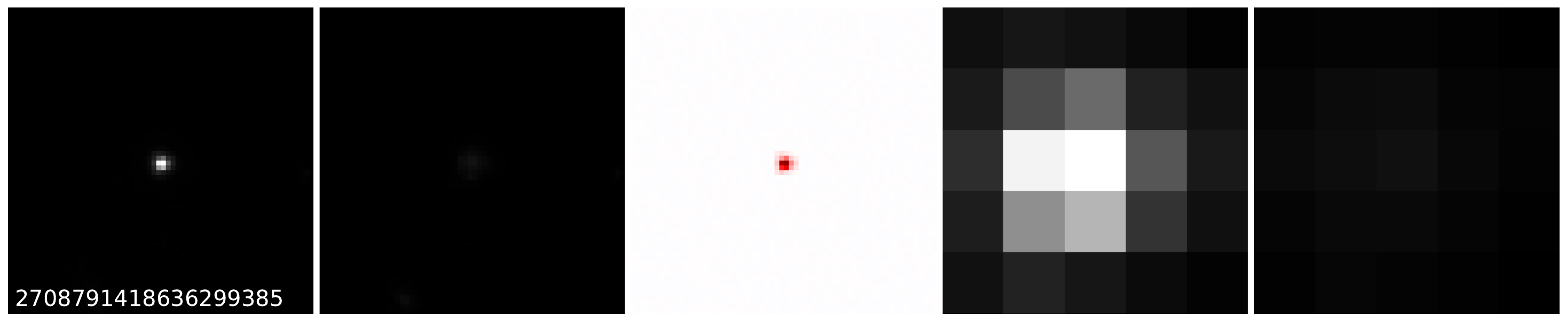}
   \includegraphics[width=\linewidth]{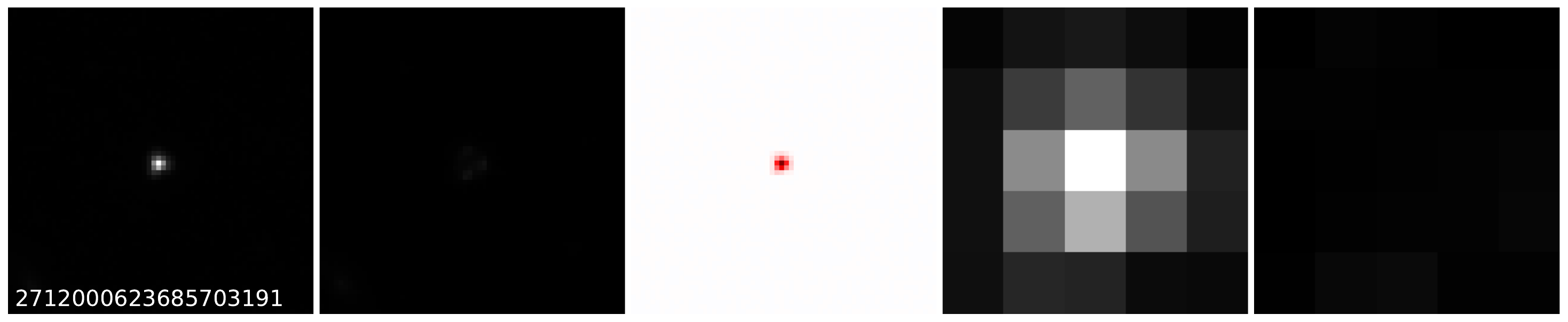}
   
   \caption{Examples of sources that achieved a very high maximum pixel difference ratio.}
   \label{fig:outputs_removed}
\end{figure*}

\begin{figure*}
   \includegraphics[width=\linewidth]{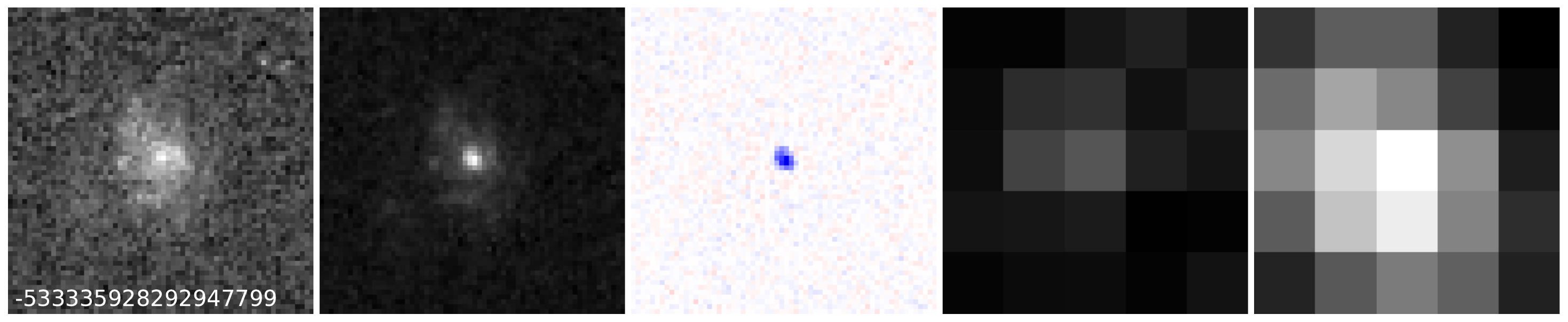}
   \includegraphics[width=\linewidth]{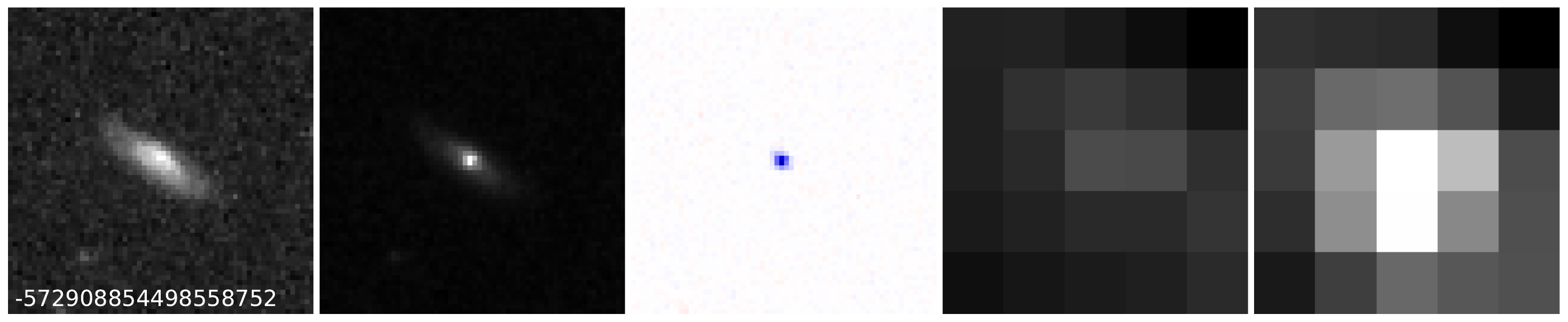}
   \includegraphics[width=\linewidth]{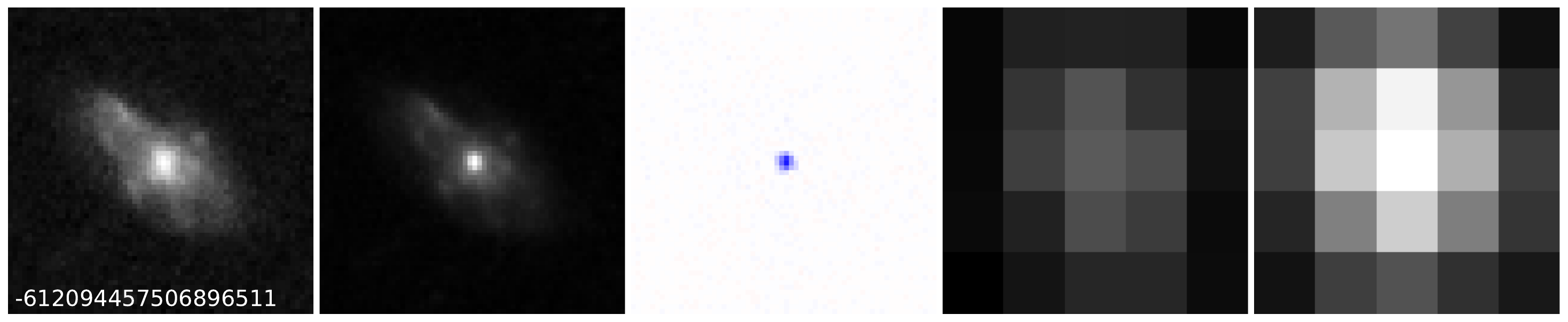}
   \includegraphics[width=\linewidth]{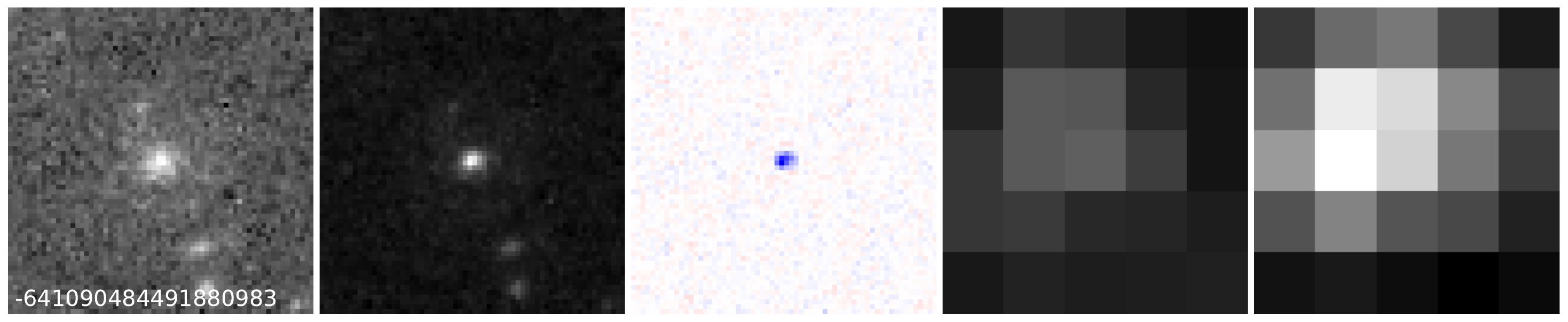}
   \includegraphics[width=\linewidth]{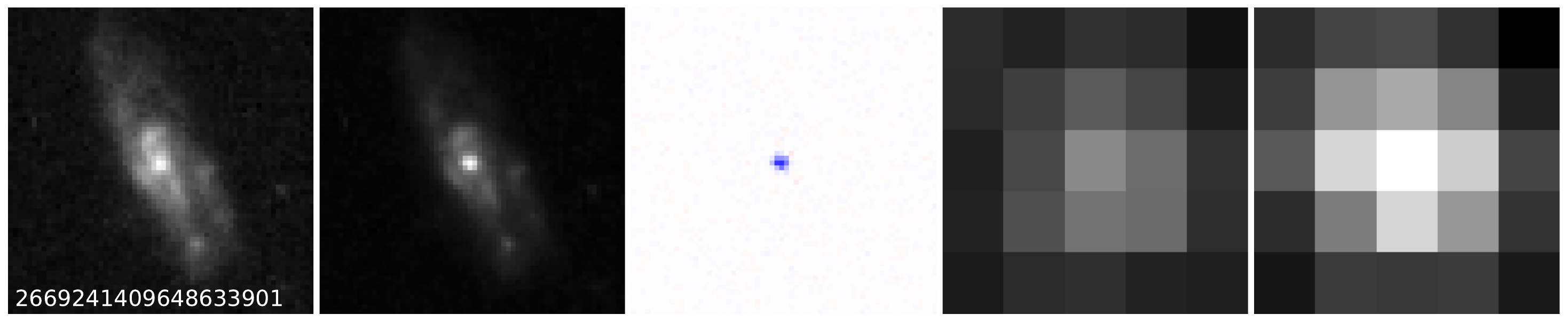}
   \includegraphics[width=\linewidth]{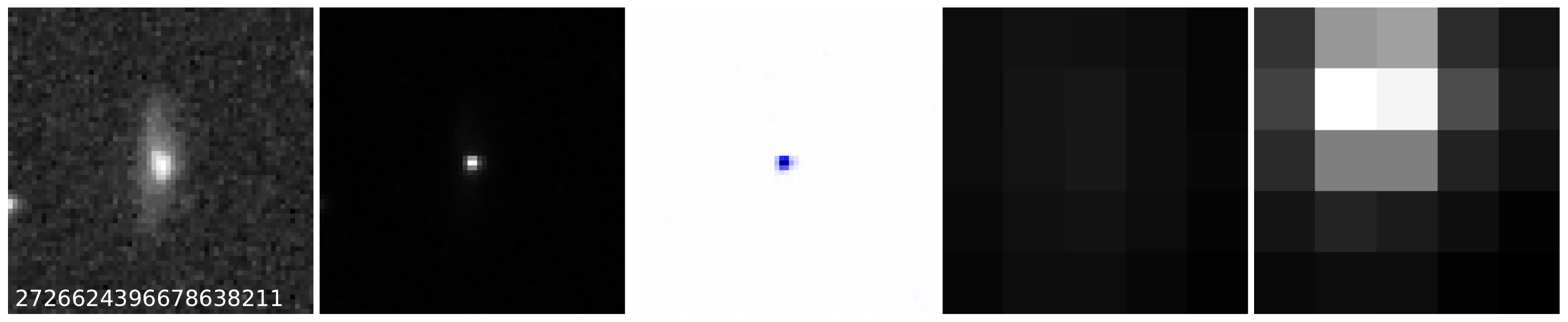}
   
   \caption{Examples of sources that achieved a very low maximum pixel difference ratio.}
   \label{fig:outputs_brighter}
\end{figure*}

We present output examples from the inpainting pipeline where pixel values deviate significantly from the expected input pixels. \Cref{fig:outputs_removed} highlights much fainter pixels, while \Cref{fig:outputs_brighter} shows much brighter ones.

\onecolumn
\section{Morphology examples}
\begin{figure}[ht!]
   \centering
   \includegraphics[width=\linewidth]{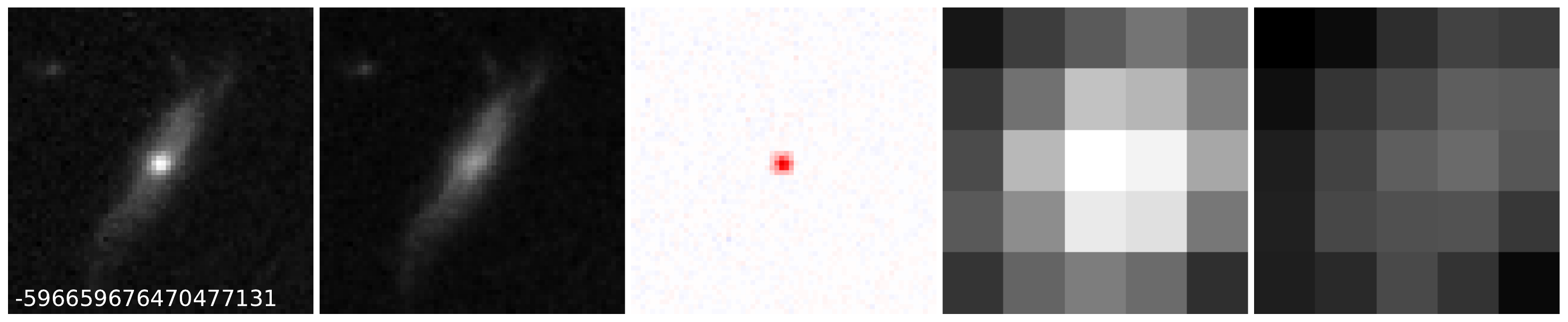}
   \includegraphics[width=\linewidth]{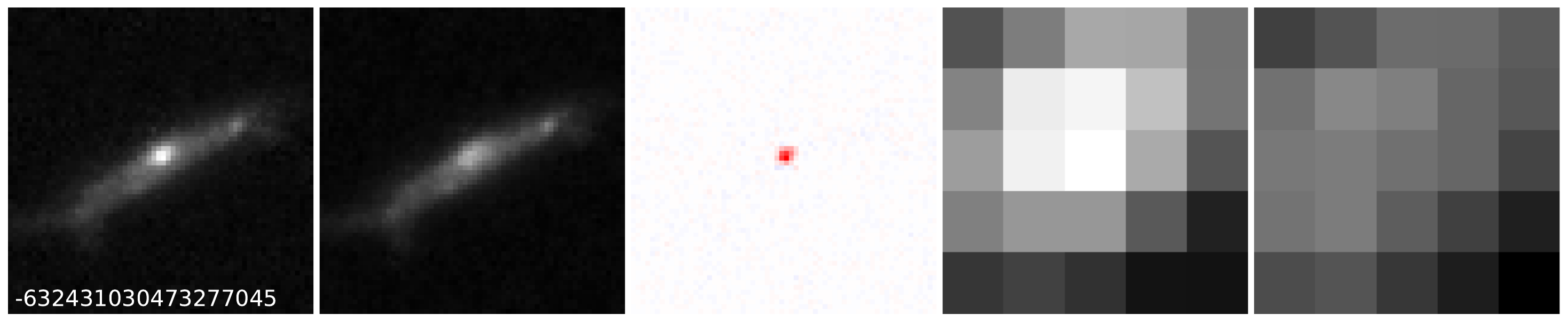}
   \includegraphics[width=\linewidth]{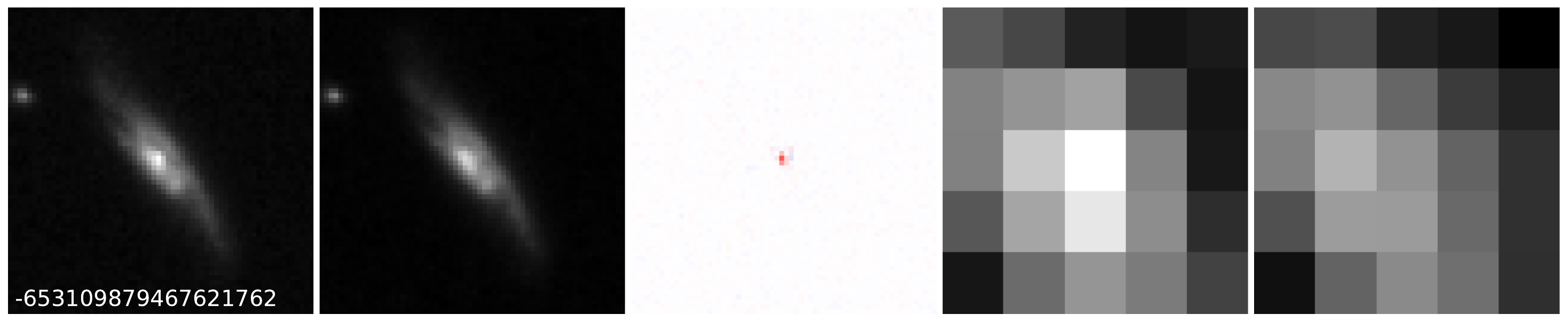}
   \includegraphics[width=\linewidth]{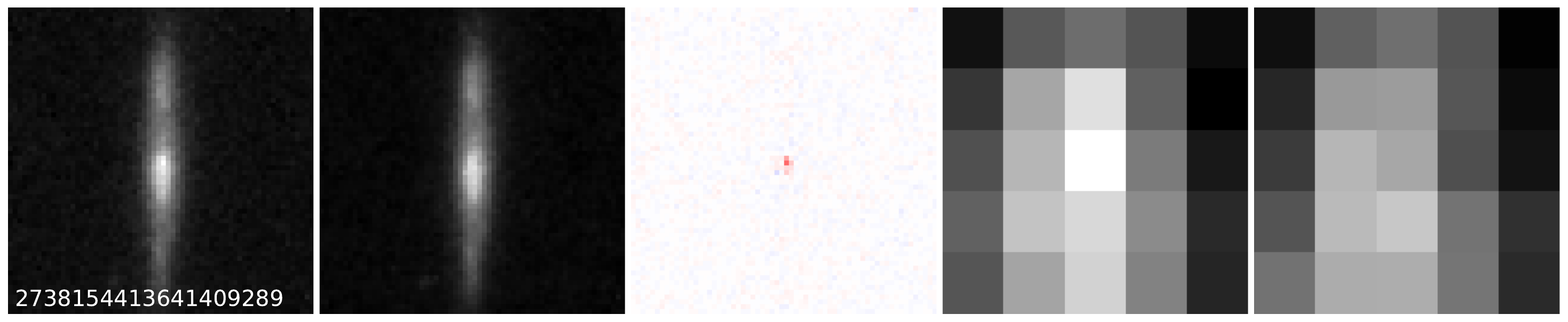}
   \includegraphics[width=\linewidth]{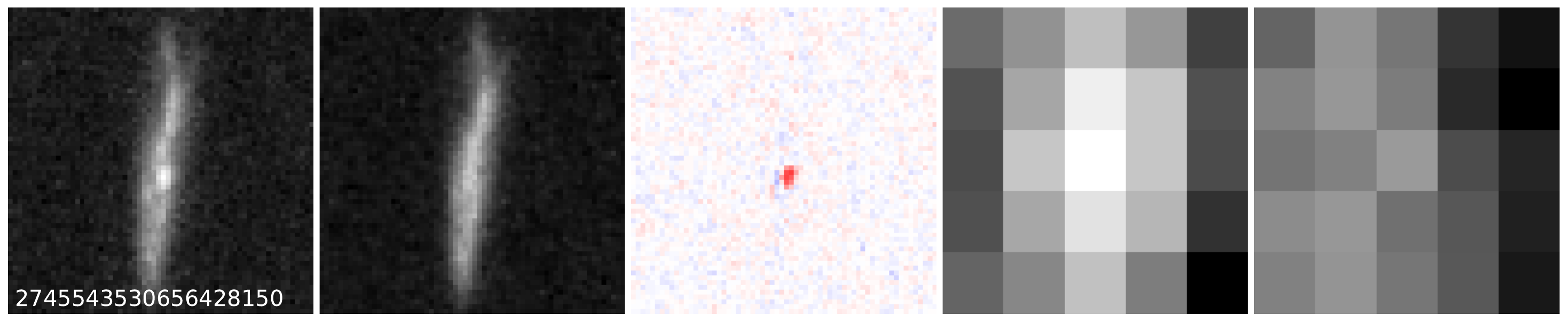}
   
   \caption{Examples of inpainting on edge-on sources.}
   \label{fig:morphology_edge_on}
\end{figure}

\begin{figure}[ht!]
   \centering
   \includegraphics[width=\linewidth]{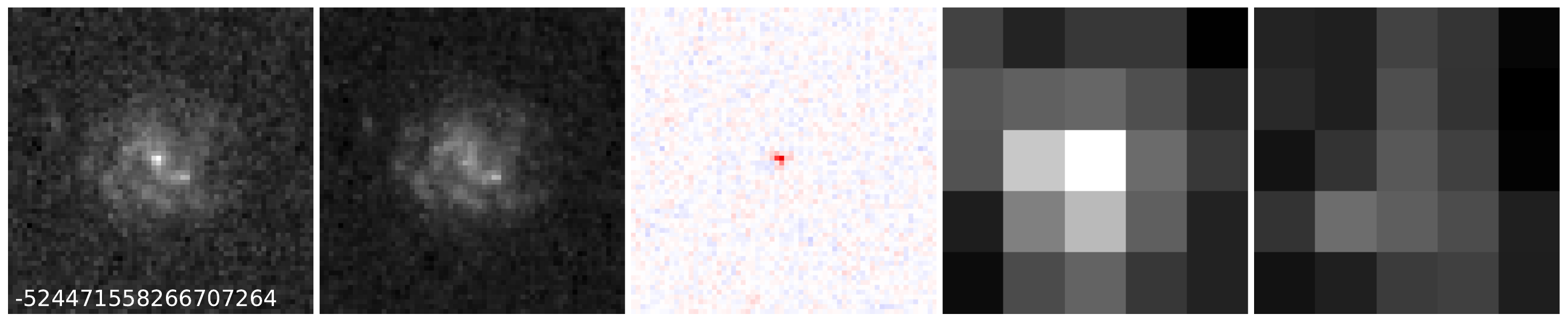}
   \includegraphics[width=\linewidth]{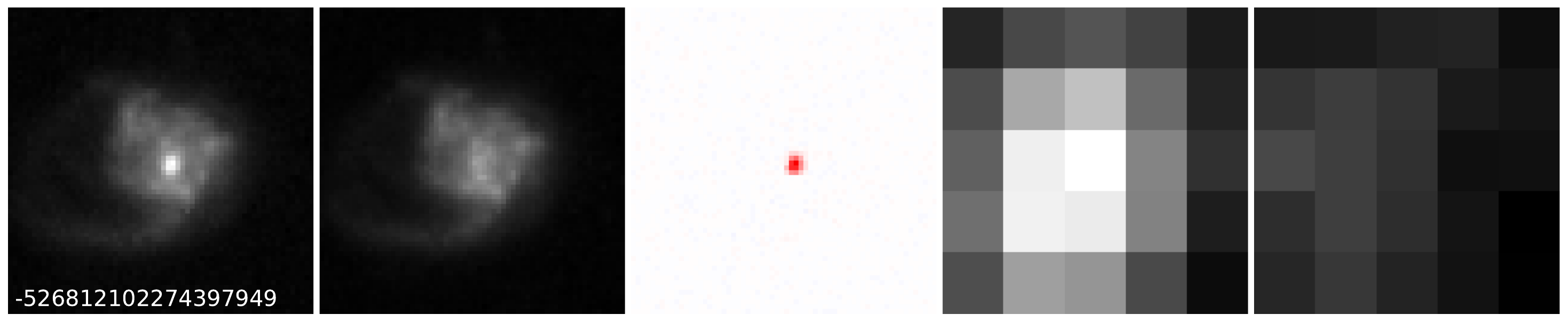}
   \includegraphics[width=\linewidth]{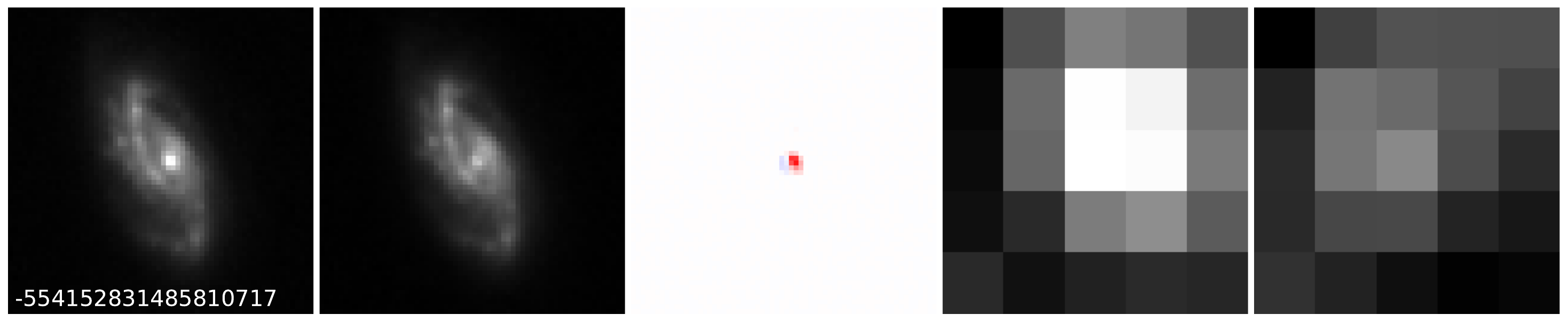}
   \includegraphics[width=\linewidth]{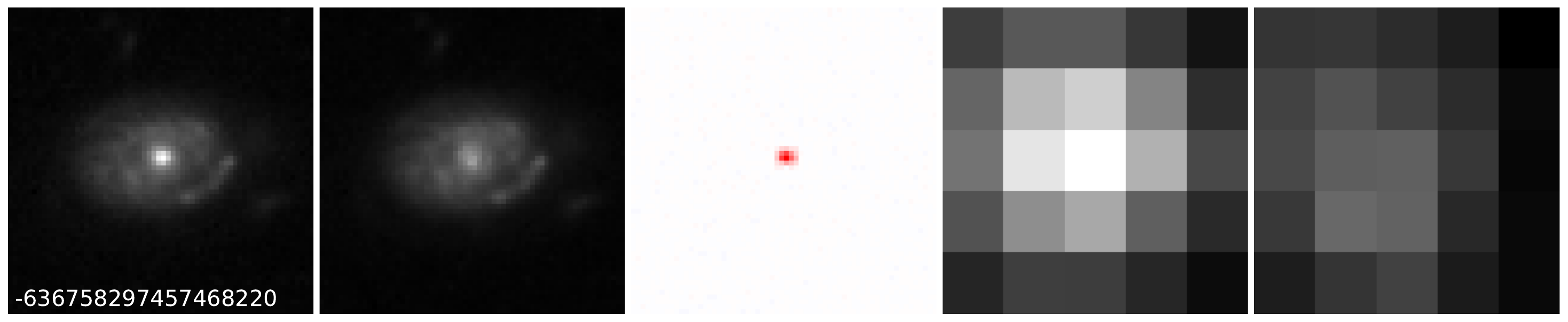}
   \includegraphics[width=\linewidth]{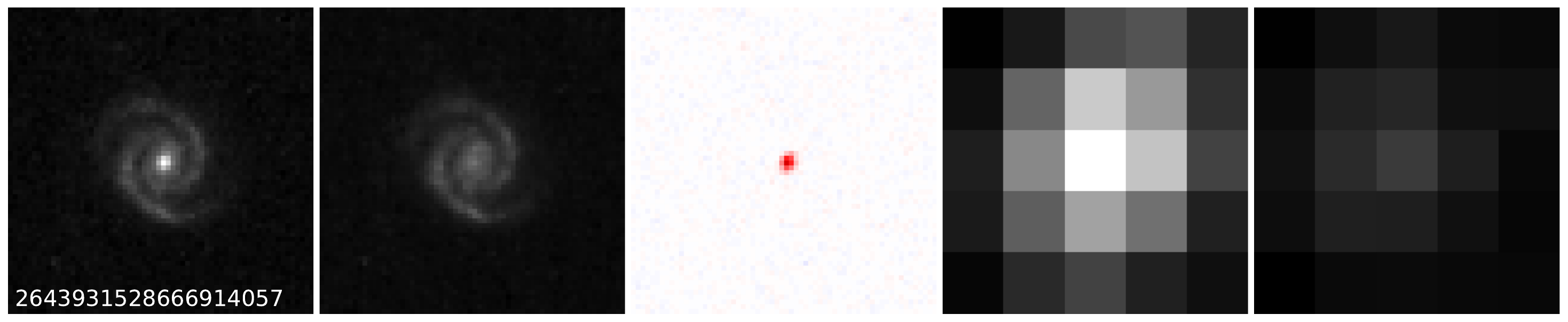}
   \includegraphics[width=\linewidth]{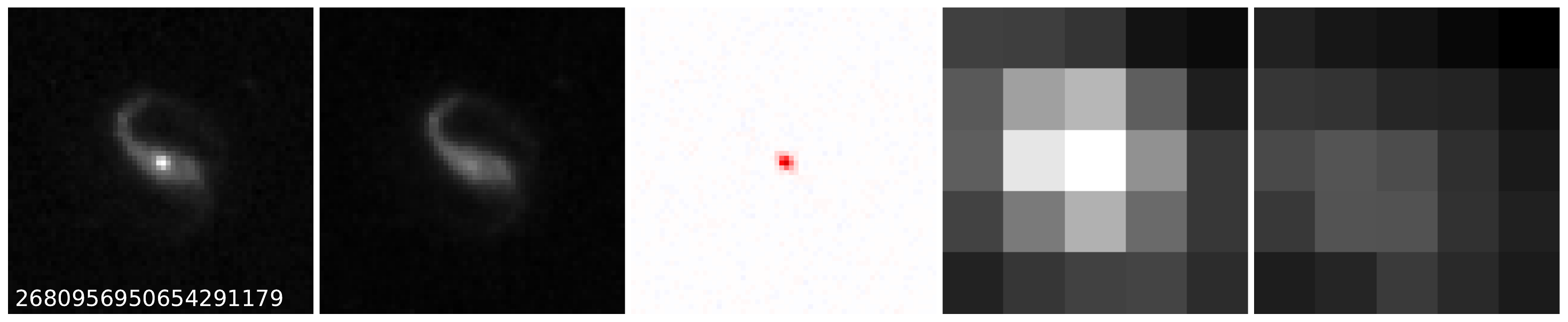}
   
   \caption{Examples of inpainting on spiral sources.}
   \label{fig:morphology_spiral}
\end{figure}

\begin{figure}[ht!]
   \centering
   \includegraphics[width=\linewidth]{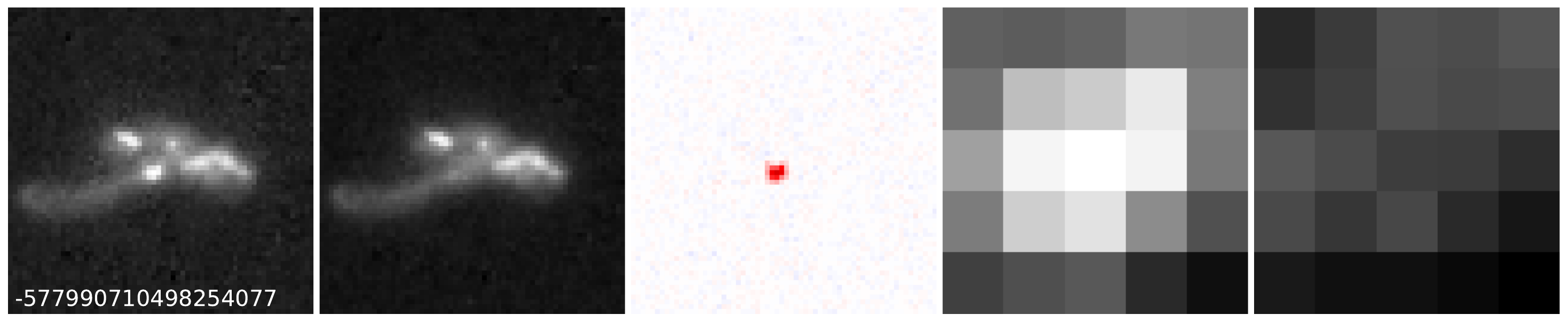}
   \includegraphics[width=\linewidth]{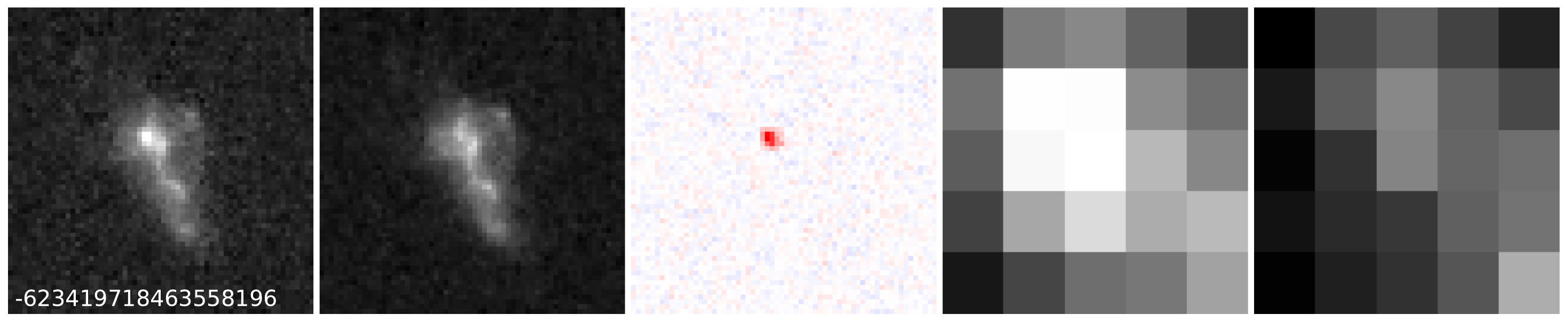}
   \includegraphics[width=\linewidth]{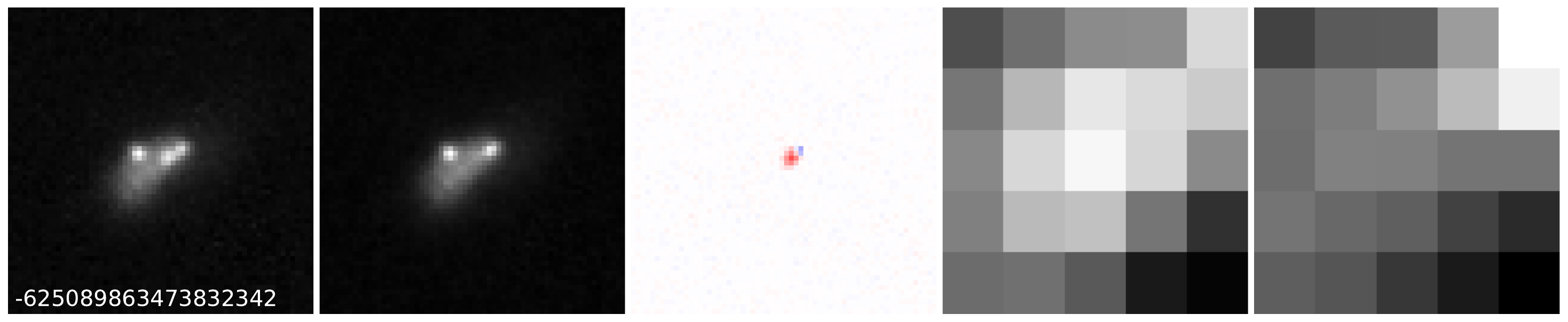}
   \includegraphics[width=\linewidth]{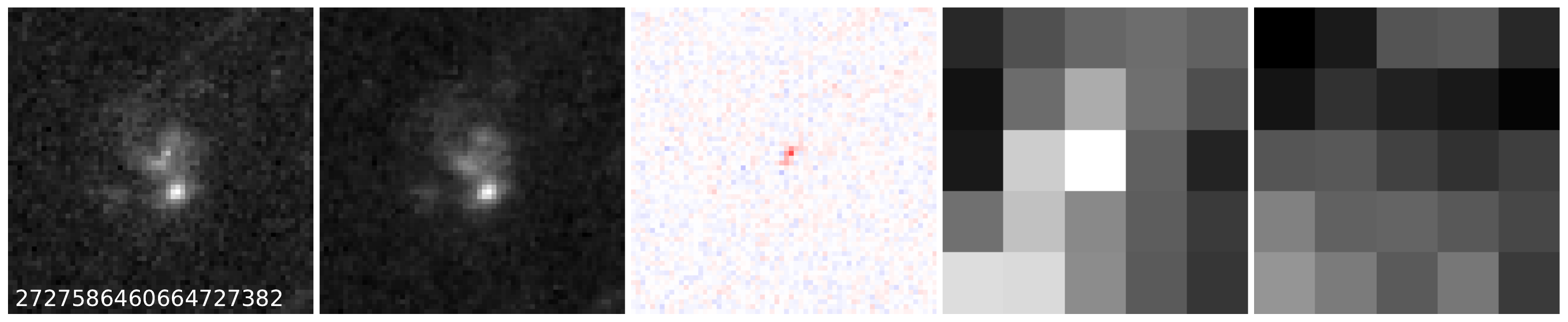}
   
   \caption{Examples of inpainting on mergers.}
   \label{fig:morphology_merger}
\end{figure}

We show output examples from various morphology types to showcase how the diffusion model adapts to different galaxy shapes. Edge-on galaxies, spirals and mergers are shown in \cref{fig:morphology_edge_on,fig:morphology_spiral,,fig:morphology_merger}, respectively.

\twocolumn
\section{Additional results}
\begin{figure}[ht!]
   \centering
   \includegraphics[width=\linewidth]{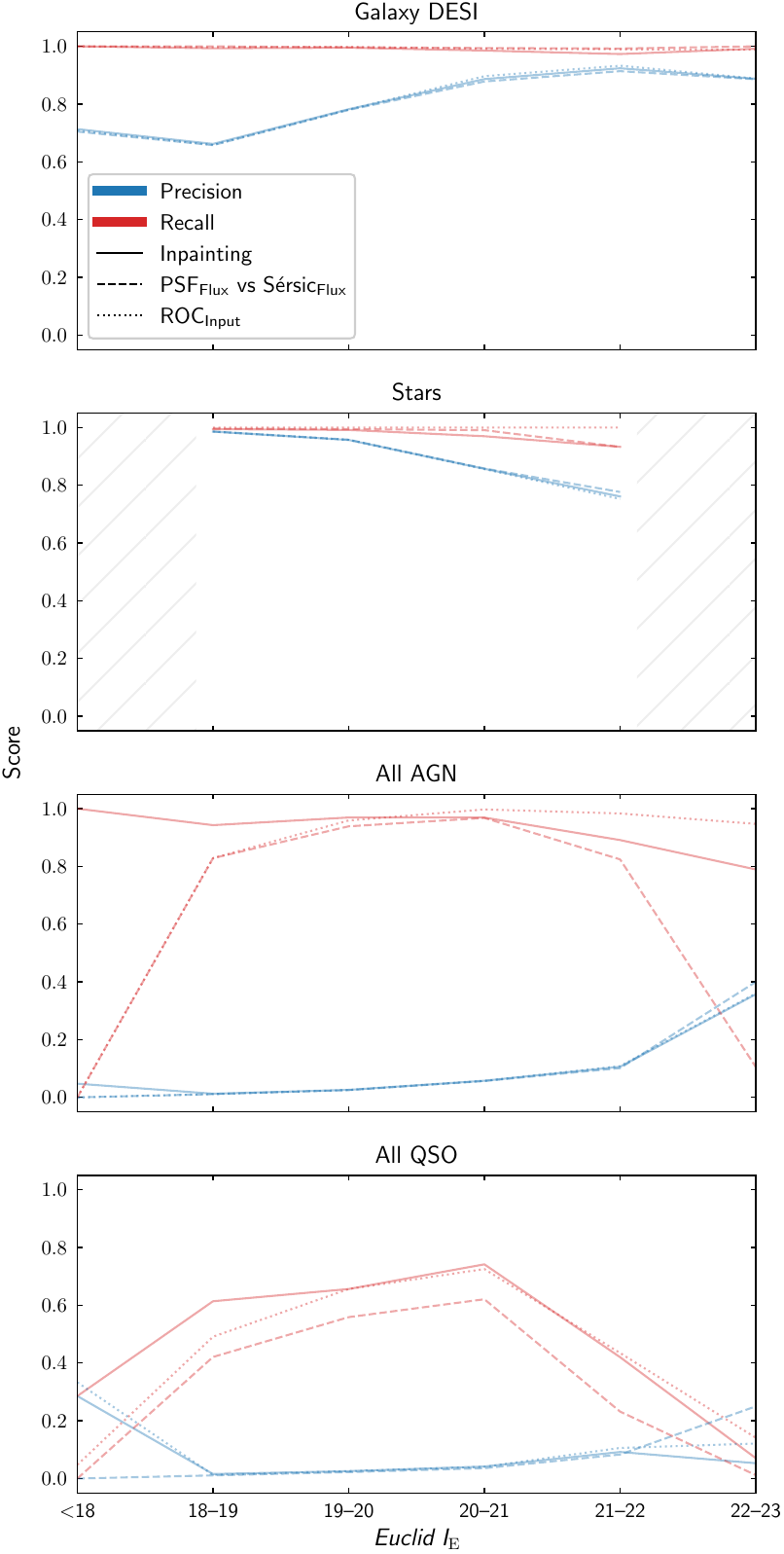}
   \caption{ Precision and recall scores for the diffusion-based AGN predictions (solid line). All of the respective selection are collated into their common classification types.} 

   \label{fig:scores_test_combined}
\end{figure}

\begin{figure}[ht!]
   \centering
   \includegraphics[width=\linewidth]{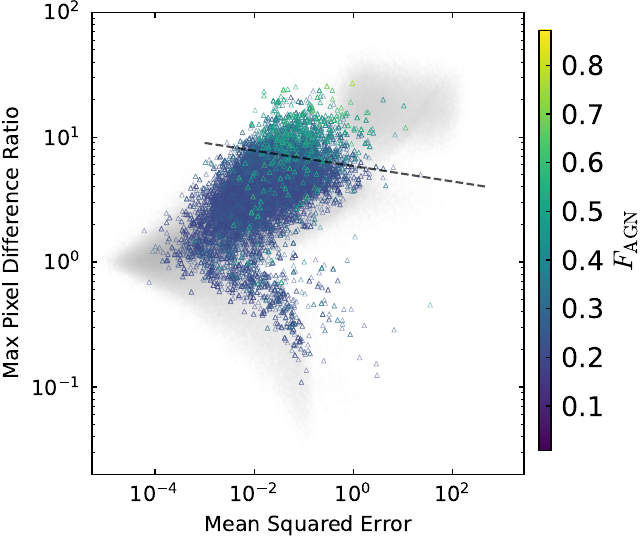}
   \caption{ \citetalias{Q1-SP015} selection on our inpainting metrics. The $F_{\text{AGN}}$ values are provided by \citetalias{Q1-SP015}. All shown sources have an $F_{\text{AGN}} > 0.2$. A large selection of sources within the \citetalias{Q1-SP015} selection are not located within our boundary. Of the sources that are within our boundary, many exhibit higher $F_{\text{AGN}}$.} 

   \label{fig:fAGN_mse_max_ratio}
\end{figure}

We present collated results for each of the star, galaxy, AGN and QSO selections in \cref{fig:scores_test_combined}. In \cref{fig:fAGN_mse_max_ratio}, the MSE and max ratio scores of all the \citetalias{Q1-SP015} selections as shown, indicating our diffusion model is more sensitive to selecting brighter $F_{\text{AGN}}$ sources.

\end{appendix}

\end{document}